\documentclass[rmp, superscriptaddress, twocolumn, 10pt]{revtex4}  
\usepackage{amsmath,amsthm,amssymb}
\usepackage{bbm}
\usepackage{graphicx}
\usepackage{color}
\usepackage[colorlinks,linkcolor=blue,citecolor=blue,urlcolor=blue]{hyperref}

\makeatletter
\newcommand{\oast}{\mathbin{\mathpalette\make@circled\ast}}
\newcommand{\make@circled}[2]{%
  \ooalign{$\m@th#1\smallbigcirc{#1}$\cr\hidewidth$\m@th#1#2$\hidewidth\cr}%
}
\newcommand{\smallbigcirc}[1]{%
  \vcenter{\hbox{\scalebox{0.77778}{$\m@th#1\bigcirc$}}}%
}
\makeatother

\newcommand{\hc}{\hat{c}}

\newcommand{\hH}{\hat{H}}

\newcommand{\hn}{\hat{n}}

\newcommand{\eqq}[1]{\begin{align} #1 \end{align}}

\begin{document}

\author{Yuta Murakami}
\thanks{All authors contributed equally.}
\affiliation{Center for Emergent Matter Science, RIKEN, Wako, Saitama 351-0198, Japan}

\author{Denis Gole\v z}
\thanks{All authors contributed equally.}
\affiliation{Jozef Stefan Institute, Jamova 39, SI-1000 Ljubljana, Slovenia}
\affiliation{Faculty of Mathematics and Physics, University of Ljubljana, Jadranska 19, 1000 Ljubljana, Slovenia}

\author{Martin Eckstein}
\thanks{All authors contributed equally.}
\affiliation{Institute of Theoretical Physics, University of Hamburg, 20355 Hamburg, Germany}

\author{Philipp Werner}
\thanks{All authors contributed equally.}
\affiliation{Department of Physics, University of Fribourg, 1700 Fribourg, Switzerland}

\title{Photo-induced nonequilibrium states in Mott insulators}

\date{October 2023}

\hyphenation{}

\begin{abstract}

The study of nonequilibrium phenomena in interacting lattice systems can provide new perspectives on correlation effects, and information on metastable states of matter. Mott insulators are a promising class of systems for nonequilibrium studies, since they exhibit exotic phenomena and complex phase diagrams upon doping, and because a large Mott gap provides  protection against fast thermalization and heating after photo-excitations. We can thus expect the emergence of interesting transient states and photo-induced phases in Mott systems. This review presents the current understanding of the mechanisms which control the time evolution  of photo-doped charge carriers and the properties of photo-induced metastable states. We focus on recent theoretical progress, identify the relevant underlying concepts, and link them to experimental observations. The review starts with a general discussion of field-induced nonequilibrium setups and an overview of key experiments which revealed characteristic properties of photo-excited Mott states, proceeds with a compact overview of the theoretical tools which have been developed to investigate these strongly correlated nonequilibrium states, and then analyzes Mott insulators driven out of equilibrium by static electric fields, periodic fields, and short laser pulses. We also discuss the appearance of nonthermal electronic orders in photo-excited Mott systems, including nonthermal spin and orbital orders, $\eta$ pairing states, and novel types of excitonic orders.

\end{abstract}

\maketitle
\tableofcontents

%%%%%%%%%%%%%%%%%%%%%%%%%%%%%%%%%%%%%%%%%%%%%%%%%%%%%%%%%%%%%%%%%%%%%%%%
\section{Introduction}
\label{sec:introduction}
%%%%%%%%%%%%%%%%%%%%%%%%%%%%%%%%%%%%%%%%%%%%%%%%%%%%%%%%%%%%%%%%%%%%%%%%

The manipulation of material properties by intense laser pulses or other nonequilibrium protocols has lead to surprising discoveries with potential for future technological applications, as discussed in several recent review articles \cite{Basov2017,Giannetti2016,oka2019,Sentef2021,Koshihara2022,boschini2023time}. Mott insulators, or strongly correlated systems in proximity to a Mott phase, are particularly interesting platforms in this regard: They combine a high sensitivity to small perturbations -- present already in equilibrium (Sec.~\ref{sec:introduction_equi}) -- with a robust energy gap that prevents trivial heating by the laser. 
While fascinating nonequilibrium phenomena have been reported in these systems, both in experiments and in model studies, a full understanding of the underlying physics is often lacking.  
The aim of this review is to identify generic concepts governing the nonequilibrium physics of Mott systems, to explain their current theoretical interpretation, and to link them to prototypical experimental results. 

\subsection{(Doped) Mott insulators in equilibrium} 
\label{sec:introduction_equi}

Mott insulators are materials which should be metals according to band theory, but are insulating because strong electron-electron interactions suppress charge fluctuations. 
If mobile electron- or hole-like charge carriers are introduced into these systems, many fascinating phenomena can be induced, ranging from non-Fermi liquid behavior to unconventional superconductivity \cite{Imada1998}. Mott insulating phases exist in different classes of compounds, including transition metal oxides (NiO, La$_2$CuO$_4$, 
\dots) and molecular crystals (Cs$_3$C$_{60}$, $\kappa$-(BEDT-TTF)$_2$Cu[N(CN)$_2$]Cl, \dots). These materials exhibit partially filled narrow bands near the Fermi level, while Coulomb interactions comparable or larger than the bandwidth act on the corresponding $d$, $f$ or molecular orbitals. Mott insulators often exhibit magnetic and/or orbital orders at low temperature~\cite{Dagotto2005, Tokura2000}, but in contrast to weakly-correlated systems with long-range order, such as Slater antiferromagnets, a charge gap persists even above the ordering temperature. 

In multi-orbital systems, Mott insulating behavior can be found for each (open shell) integer filling. The stability of these phases depends sensitively on the degeneracy of the levels, the crystal field splittings, and the Hund coupling \cite{Georges2013}. In an orbitally degenerate system, a rough estimate of the interaction strength for which Mott insulating behavior is expected can be obtained by comparing the charge excitation gap in the atomic limit, $E_{N+1}+E_{N-1}-2E_N$, to the bandwidth. Here, $E_N$ denotes the energy for a filling of $N$ electrons. Such an analysis reveals that in the usual case of a positive Hund coupling, the half-filled Mott phase with dominant high-spin configurations is the most stable one. The fullerides A$_3$C$_{60}$ however exhibit an effectively negative Hund coupling, due to the dynamical Jahn-Teller effect \cite{Fabrizio1997}, which leads to qualitatively different properties, such as dominant low-spin states and a destabilization of the half-filled Mott phase \cite{Capone2009}. 

In some materials, such as transition metal oxides, ligand states can be energetically close to the partially filled narrow bands, so that after the splitting of the narrow bands into lower and upper Hubbard bands, these states may overlap with or even appear above the lower Hubbard band. In such a situation, the lowest-energy charge excitation is not from the lower Hubbard band to the upper Hubbard band, but from the ligand band to the upper Hubbard band, and the material is classified as a charge transfer insulator \cite{Zaanen1985}. 

\begin{figure}[t]
\begin{center}
\includegraphics[angle=0, width=\columnwidth]{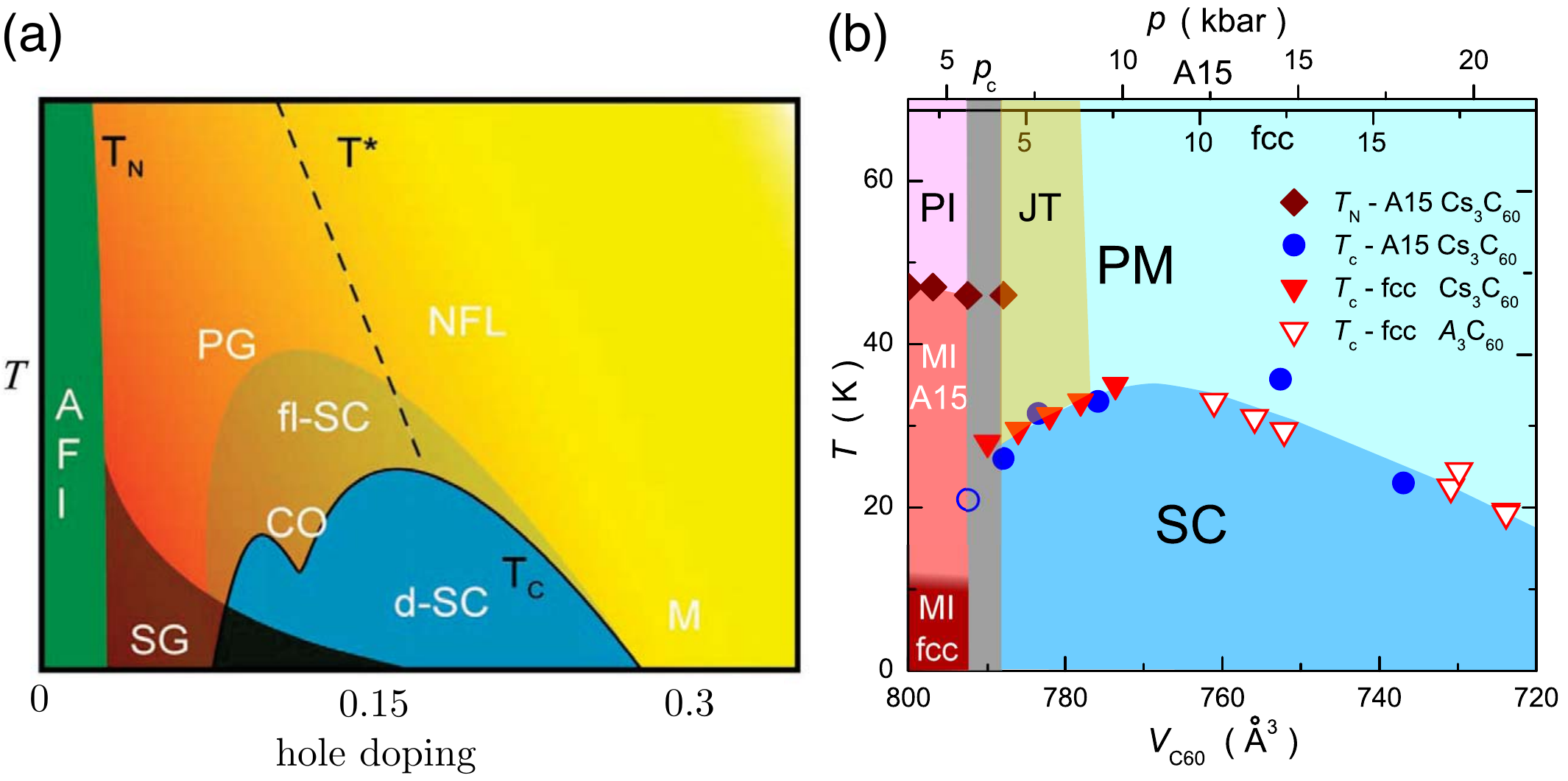}
\caption{
(a) Schematic phase diagram of hole doped cuprates (adapted from \cite{Shen2008}), featuring an antiferromagnetic (Mott) insulating phase (AFI), spin glass phase (SG), $d$-wave superconducting phase (d-SC), charge ordered phase (CO), in addition to a pseudo-gapped metal (PG) which crosses over into a non-Fermi liquid metal (NFL) and conventional Fermi liquid like metal (M). The N\'eel temperature $T_N$, superconducting critical temperature $T_c$ and pseudogap crossover temperature $T^*$ are also indicated.   
(b) Phase diagram of fulleride compounds, with the bandwidth-controlled Mott transition marked by the vertical gray line (adapted from \cite{Ihara2011}). Magnetic insulating (MI), paramagnetic insulating (PI), paramagnetic metallic (PM) and superconducting (SC) phases are shown. We also sketch the Jahn-Teller metal phase (JT) reported by \onlinecite{Zadik2015}.}
\label{fig_expmott}
\end{center}
\end{figure}

In equilibrium, the Mott insulator to metal transition can be studied experimentally by changing the carrier concentration through chemical doping, or by varying the bandwidth via the application of pressure. 
A sketch of the phase diagram of hole-doped cuprates is shown in Fig.~\ref{fig_expmott}(a), while the pressure dependent phase diagram of fulleride compounds is illustrated in Fig.~\ref{fig_expmott}(b). These examples demonstrate a close connection between Mott insulating behavior, magnetism and superconductivity, as well as the appearance of other complex phases of matter, such as spin glass states or Jahn-Teller metals \cite{Zadik2015} in the vicinity of the Mott phase.  
Close to a Mott phase, one typically observes an incoherent metallic state, resulting from barely itinerant electrons \cite{Imada1998} and the freezing of magnetic or orbital moments. Strong interactions between the charge carriers and spin, orbital or lattice degrees of freedom can lead to the formation of polarons with heavy masses and non-Fermi liquid behavior~\cite{lee2006}.
Further away from the Mott phase in the filling-versus-interaction plane, this incoherent metal crosses over into a more conventional Fermi liquid metal \cite{Medici2011}. 
Especially in the case of multi-orbital systems, this crossover region is characterized by distinct non-Fermi liquid properties \cite{Werner2008}, and the enhanced fluctuations of local moments can lead to unconventional forms of superconductivity or excitonic order \cite{Kunes2015,Hoshino2016}, which may compete with magnetic, orbital, and charge orders.

%%%%%%%%%%%%%%%%%%%%%%%%%%%%%
%%%%%%%%%%%%%%%%%%%%%%%%%%%%%

\begin{figure*}[t]
\begin{center}
\includegraphics[angle=0, width=2.0\columnwidth]{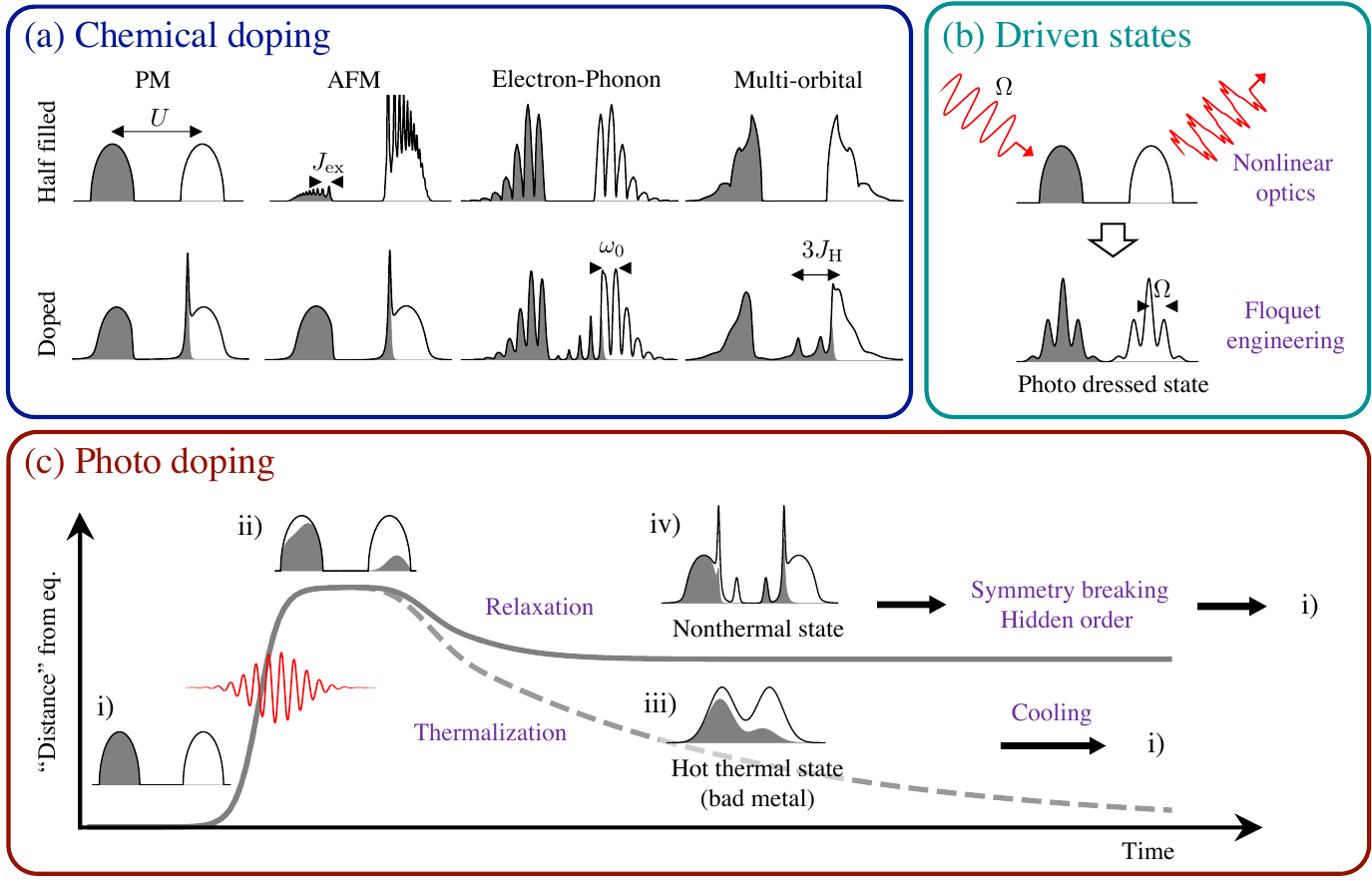} 
\caption{
Single-particle spectral functions of Mott systems. 
(a) Spectral functions of undoped and doped Mott insulators. The panels in the top row show the spectral functions (black line) and occupations (grey shading) of the paramagnetic~(PM) Hubbard model, the minority-spin component of the antiferromagnetic~(AFM) Hubbard model, the Holstein-Hubbard model and the two-orbital Hubbard model at half filling. The panels in the bottom row show the spectra and occupations for doped systems. 
The arrows indicate how the different energy scales (Hubbard $U$, antiferromagnetic exchange $J_\text{ex}$, Hund coupling $J_{\rm H}$, phonon frequency $\omega_0$) are imprinted in the spectrum.
(b) Modification of the spectrum of a Mott insulator by periodic driving with frequency $\Omega$. (c) Illustration of different stages i)-iv) in the excitation and relaxation process of photo-doped Mott system.  In large-gap Mott insulators~(stage iv), the intraband relaxation leads to distinct nonthermal quasiparticle distributions. 
In small-gap Mott insulators~(stage iii), fast thermalization to a bad metallic state with elevated temperature is expected.
}
\label{fig_mott}
\end{center}
\end{figure*}

Several important properties  of Mott insulators can be deduced from the single-particle spectral function, which shows the states accessible by adding or removing an electron.  A good starting point for the theoretical analysis is the atomic limit. For example, in the most basic model for correlated electron systems, the single-band Hubbard model \cite{Gutzwiller1963,Hubbard1963}, the local Hamiltonian has the form $H_\text{loc}=U n_\uparrow n_\downarrow -\mu(n_\uparrow+n_\downarrow)$, with $n_\sigma$ measuring the number of electrons with spin $\sigma$, $U$ the local interaction and $\mu$ the chemical potential. In the particle-hole symmetric case ($\mu=U/2$) this becomes $H_\text{loc}=U(n_\uparrow-\tfrac12)(n_\downarrow-\tfrac12)$, up to an irrelevant constant. The eigenenergies are $E_0=E_2=U/4$ for the empty and doubly occupied states and $E_\uparrow=E_\downarrow=-U/4$ 
for the singly occupied states. Hence, in the half-filled ground state, the electron addition spectrum exhibits a peak at $\omega=E_2-E_\sigma=U/2$, while the energy removal spectrum has a peak at $\omega=E_\sigma-E_0=-U/2$, corresponding to a gap of size $U$. If a small hopping between the lattice sites is turned on, these delta-function peaks broaden into Hubbard bands. The resulting local spectral function is sketched for a paramagnetic (PM) system in the top row of Fig.~\ref{fig_mott}(a), where the gray shaded area indicates the electron removal spectrum (filled states). The filled lower Hubbard band and the empty upper Hubbard band represent the addition and removal of an electron to or from the Mott ground state, respectively. As the ground state consists of predominantly singly occupied sites, up to virtual charge fluctuations, adding (removing) electrons leads to  excitations which are mainly of doubly occupied (empty) character. These mobile excitations, which also play a key role in photo-excited systems, will therefore be called doublons and holes (or holons) in the following. 

If the chemical potential is shifted into the upper Hubbard band, a certain population of doublons is produced by chemical doping. In the PM state, these doublons can move in the background of half-filled sites, which turns the Mott insulator into a correlated metal. As is shown in the bottom row of Fig.~\ref{fig_mott}(a)
the doublons form a narrow quasi-particle band at the edge of the Hubbard band, while the gap to the lower Hubbard band persists. This doped Mott insulator state can exhibit Fermi liquid properties, albeit with a strongly enhanced effective mass of the charge carriers~\cite{lee2006, Werner2007}. 

Below the N\'eel temperature, the Hubbard bands split up into spin-polaron bands \cite{dagotto1994,Sangiovanni2006}, as illustrated in the second column of Fig.~\ref{fig_mott}(a), which shows the minority-spin spectral function of the half-filled system. These features (which are absent in $D=1$ due to spin-charge separation) are the result of the strong spin-charge coupling in dimensions $D\ge 2$:
If a doublon or holon moves in an antiferromagnetic (AFM) spin background, each hop breaks several AFM bonds, at an energy cost proportional to the spin exchange $J_\text{ex}$. This leads to a large effective mass and near-localization of the charge carriers. 
The properties of the spin-polaron peaks have been worked out in detail for a single hole in the $t$-$J$ model~\cite{Strack1992,dagotto1994,jaklivc2000,kane1989,Trugman1988}. 
The fine structures with characteristic energy $J_\text{ex}$ 
can be understood as signatures of a particle moving in a linear potential (produced by so-called string states), and they indicate a large polaron mass.
Further doping leads to intricate low-energy phenomena which require a proper description of short-ranged correlations, and which have been analyzed in previous reviews~\cite{lee2006,dagotto1994,Maier2005}. 
At high enough doping, the AFM spin background is reduced or completely suppressed,
resulting in a spectral function similar or identical to that of the doped PM system. 

%%%%%%%%%%%%%%%%%%%%%%%%%%%%%
%%%%%%%%%%%%%%%%%%%%%%%%%%%%%

Substructures in the Hubbard bands can also result from a strong coupling to (optical) phonons, 
or other bosonic excitations such as collective charge excitations (plasmons), as shown in the third row of Fig.~\ref{fig_mott}(a). In the former case, the separation between the polaronic subbands is given by the 
phonon frequency $\omega_0$. 
For very large coupling, the entire electronic band can split up into sidebands
with exponentially decaying weights at high energies. In the case of the upper (lower) Hubbard band, the high-energy subbands correspond to an electron insertion (removal) with multiple phonon emissions.
(If the temperature is of order $\omega_0$, phonons become thermally occupied and sidebands associated with phonon absorption upon electron addition would show up in the gap region.)
Also chemically doped doublons form dressed polarons. Hence, the removal of a doublon can partially release the phonon cloud and  leave the Mott state with additional phonon excitations, as manifested by a series of new ingap sidebands in the electron removal spectrum, which extends from the upper Hubbard band into the gap region. 
At least in simple model calculations, the phonon sidebands are much more robust against doping (and heating) than the spin-polaron sidebands. 

Finally, multi-orbital Hubbard models exhibit rich multiplet structures~\cite{Kanamori1963,Georges2013}. 
For example, in the case of a two-orbital model with density-density interactions, the Hund coupling $J_H$ distinguishes the energies of the half filled configurations with two electrons in the same orbital ($E_\text{int}=U$), in both orbitals with opposite spin ($E_\text{int}=U-2J_H$), and in both orbitals with parallel spin ($E_\text{int}=U-3J_H$). At low temperature, almost all the sites of the half-filled Mott insulator are in the high-spin configuration, which results in Hubbard bands with a separation of approximately $U+J_H$. The higher-energy substructures seen in the spectral function in the fourth row of Fig.~\ref{fig_mott}(a)
can be associated with electron insertion or removal processes with additional high-spin/low-spin excitations on neighboring sites.  
Once the chemical potential is shifted into the upper Hubbard band, long-lived ``triplons" are inserted into the half-filled background of high-spin doublon states. Removing an electron from these triplons can produce any of the three possible half-filled configurations, which results in three occupied (side-)peaks near the upper Hubbard band. The partial occupation of the main Hubbard band is associated with the creation of a high-spin doublon with $E_\text{int}=U-3J_H$, the first side band with a low-spin doublon with $E_\text{int}=U-2J_H$ and the second sideband with a low-spin doublon with $E_\text{int}=U$. Similar in-gap features would also show up at half-filling, once the temperature becomes of the order of $J_H$. 
The existence of long-lived triplons in the doped case also leads to the appearance of high-energy features in the unoccupied part of the spectrum, which can be linked to the creation of quadruplons.

The above summary of how chemical doping  affects the spectral function of equilibrium Mott  insulators provides a useful guide to the discussion of photo-excited Mott insulators, which can sometimes be interpreted as systems with simultaneous electron and hole doping.

\subsection{Photo-excited Mott insulators: Overview} \label{sec:overview}

For the understanding of photo-excited Mott states, it is useful to distinguish two types of nonequilibrium settings: 
The first type involves states whose nonthermal properties are induced and supported 
by a persistent external drive, such as quasi-static or time-periodic fields, see Fig.~\ref{fig_mott}(b). 
In the second setting a pulsed field or other perturbation produces electronic excitations across the Mott gap, and triggers a sequence of characteristic relaxation steps,
see Fig.~\ref{fig_mott}(c), steps i)-iv). This can lead to thermalization at an elevated temperature [path iii)], or, under suitable conditions, to the realization of photo-induced phases with interesting properties [path iv)]. 
In the following subsections, we summarize important concepts and phenomena related to these settings. This overview is  meant to serve as a guide for the review; detailed discussions and additional links to the original literature can be found in the referenced sections. 

\subsubsection{Driven states}
\label{sec:driven_overview}
A widely studied class of nonequilibrium states encompasses (quasi) steady or time-periodic states emerging in the presence of an external drive (Fig.~\ref{fig_mott}(b)). The application of external fields can result in modified effective Hamiltonian parameters, 
which is the basis of Floquet engineering~\cite{Bukov2015,Eckardt2017,oka2019,Sentef2021}. 
A celebrated example is the realization of the Haldane model in a honeycomb lattice system subject to circularly polarized fields \cite{Oka2009,Kitagawa11Haldane}. This theoretical proposal has been realized in cold atom simulators \cite{Jotzu2014}, and in solids~\cite{McIver2020}, where the latter are generally more challenging due to the strong energy absorption from the drive.
Band structure engineering can also be achieved in Mott systems (Sec.~\ref{sec:Floquet}).
Here, unwanted heating can be suppressed if the driving frequency lies within the transparency window set by the Mott gap.
In addition to the band structure renormalization, the spectrum of the driven state can exhibit Floquet sidebands~(photon-dressed states)  \cite{Wang2013},  which can reduce the gap in Mott systems (Fig.~\ref{fig_mott}(b)). 
Moreover, Mott insulators also provide interesting opportunities for controlling the low energy physics of spins and orbitals using external driving (Sec.~\ref{sec:effective_j_ac}).

As in the case of Floquet engineering, the application of strong DC fields dresses the many-body ground state. This effect is 
relevant, e.g., in experiments with strong THz fields, 
which can be regarded as a quasistatic on the fs electronic timescale. 
Floquet  and strong field physics are closely related  in many ways:
Quasi-static fields can  give rise to sidebands (Wannier-Stark bands) (Sec.~\ref{eq:WS_local}), and they may be used to modify low energy Hamiltonians in a way analogous to Floquet engineering~(Sec.~\ref{sec:effective_jex}). 

Not only the laser field itself, but also coherent excitations in the material can produce a periodic driving. 
The most prominent  mechanism along these lines is nonlinear phononics~\cite{Foerst2011}, where the coherent oscillation of one phonon mode can -- in the presence of anharmonicity -- lead to a nonzero time-averaged displacive force on other phonon modes. Similarly, the nonlinear coupling to the coherently excited phonon can affect the electrons, e.g., through induced magnetic or pairing interactions, or the modification of the Hubbard $U$ (Sec.~\ref{sec:nonlinear_phononics}). An interesting question is whether related protocols enable efficient switching pathways between different phases. 

Finally, excitations with strong fields are accompanied by nonlinear optical effects such as high-harmonic generation (Sec.~\ref{sec:nonlinear}).
Nonlinear responses can be used to 
explore the energy landscape around equilibrium states, or to track 
the quasiparticle dynamics in driven states.
It is thus important to understand the mechanisms which control these nonlinear optical responses.
 
\subsubsection{Photo-doped states}

In this subsection we discuss the different  stages in the evolution of a Mott system after a pulsed excitation (Fig.~\ref{fig_mott}(c)). Strong electric fields acting on a Mott insulator can create mobile charge carriers, such as doublons and holons in the single-orbital case. This amounts to a transfer of occupied spectral weight from the lower to the upper Hubbard band, see step from i) to ii) in Fig.~\ref{fig_mott}(c). Although the total number of electrons remains unchanged, this process will be referred to as photo-doping. The formation of photo-doped states, their properties, lifetime and decay have been subject to numerous experimental (Sec.~\ref{sec:exp_probes}) and theoretical (Sec.~\ref{sec:photo_doped}) investigations, and the analysis of these processes and phenomena constitutes a main part of this review. 

{\it Excitation across the Mott gap ---}  First, the possible excitation pathways deserve a closer look, as their  detailed understanding may help to design tailored excitation protocols.
Both the creation of doublons and holons within the same band and the charge transfer between different bands correspond to optically active transitions, and can be achieved in linear absorption using frequencies resonant to the gap (Sec.~\ref{sec:photocarrier_generation}). Moreover, carriers can be generated via nonlinear (strong-field) processes, such as tunneling across the Mott gap (Sec.~\ref{sec:breakdown}) or multi-photon absorption (Sec.~\ref{sec:single_multi}). The crossover between the two nonlinear mechanisms is the many-body analog of the Keldysh crossover. 

{\it Electron thermalization ---}   The common assumption for an interacting electron system is that electron-electron scattering 
leads to fast thermalization, even before the energy is passed on to  phonons or other degrees of freedom. 
A thermal state is entirely characterized by the total electron number and the electronic temperature $T_e$, 
so that the total absorbed energy would be all that matters for the electronic excitation process.
Whether, and on what timescale, an interacting isolated quantum system shows such ergodic behavior is a fundamental question in many-body physics \cite{Polkovnikov2011RMP, Alessio2016}.

In laser excited metals, on timescales larger than a few $100$~fs, the formation of a quasi-thermal ``hot electron state'' is commonly assumed, even in good metals where the electron-electron interaction is weak. This provides the basis for the conventional few-temperature descriptions of photo-excited phases  \cite{Allen1987}. Importantly, 
however, increasing the interaction strength to the strong correlation regime does not necessarily lead to faster thermalization.
In particular, theoretical and experimental evidence shows that the Mott gap provides a bottleneck against rapid thermalization (Sec.~\ref{sec_recombination}),
which is a key ingredient for the realization of interesting nonequilibrium states (Sec.~\ref{sec:noneq_phases}). 
Fast thermalization, on the other hand, is still expected in the metal-insulator crossover regime and for small-gap Mott insulators (Fig.~\ref{fig_mott}(c), panel iii)). In this case, however, even a thermal hot-electron state can be highly nontrivial. For example, strong interactions can lead to unconventional  bad metallic  behavior at temperatures as high as the electronic bandwidth. In equilibrium, this regime lies beyond the melting temperature of most solids, and needs to be studied in model calculations or cold atom simulators. 
Ultrafast laser experiments can access this regime in transient states and help to  
clarify the properties of correlated electrons at high temperature, as was noted early on in the context of time-resolved APRES studies \cite{Perfetti2006}, see Sec.~\ref{sec:exp_probes}.  

{\it Electron relaxation ---}
If the Mott gap is  sufficiently large to prevent rapid thermalization, the system relaxes to a transient nonthermal state, see evolution from ii) to iv) in Fig.~\ref{fig_mott}(c). In this nonthermal state, the relative populations of the local states are no longer fixed by the electronic temperature, as they would be in a thermal state. This also implies that the thermalization process, which takes place on longer timescales, generically involves charge carrier production or recombination.

In the initial relaxation to the nonthermal state, photo-doped carriers lose kinetic energy due to their interactions with bosonic degrees of freedom, such as antiferromagnetically correlated spins, phonons, or plasmons (Sec.~\ref{sec:intraband_relax}).  This intra-Hubbard-band relaxation eventually leads to  doublon and holon populations with a thermal-like distribution, but separate chemical potentials for the doublons and holons. 
Theoretically, such (quasi-)steady states can often be described using methods specifically adapted to nonequilibrium stationary states or to quasi-equilibrium states (Sec.~\ref{sec:quasiseq}). 
Since effectively cold photo-doped states may exhibit nontrivial electronic orders, it is relevant to think about nonequilibrium protocols which actively cool the charge carrier distributions in the Hubbard bands (Sec.~\ref{sec_entropy}). 

{\it Nonequilibrium quasiparticles ---} 
Microscopic insights into the nonequilibrium response and properties of photo-doped systems can be obtained by analyzing the nature of photo-doped quasi-particles (Sec.~\ref{sec:photo_doped}).
Typically, charge carriers are initially produced with excess kinetic energy and get subsequently dressed through their interaction with phonons (Sec.~\ref{sec:electron-phonon}) and spins  (Sec.~\ref{sec:electron-spin}),
resulting in the cooling described above. After the intra-band relaxation, quasi-particle features or in-gap states can appear in the spectral function. They are similar to the case of chemically doped Mott insulators (Fig.~\ref{fig_mott}(a)), but now there are both features associated with electron- and hole-like charge excitations (compare the doped spectra of Fig.~\ref{fig_mott}(a) with panel iv) of Fig.~\ref{fig_mott}(c)). Moreover, quasi-particles in Mott insulators can be bound into excitons by the nonlocal Coulomb interaction, or by interactions mediated by bosonic fluctuations (Sec.~\ref{sec:MottExciton}). 

{\it Photo-induced phase transitions ---}
If nonthermal states of correlated electron systems are sustained for sufficiently long time, excitations beyond some threshold can drive the system into new ordered phases. Experimental observations include the switching to 
hidden insulating \cite{Ichikawa2011} and metallic \cite{Stojchevska2014} phases, or metastable superconducting-like states \cite{Cavalleri2018}. While none of these experimental observations are completely understood, several pathways leading to nonthermal phases have been proposed for Mott systems. 
On the one hand, in a Mott insulator or charge transfer insulator, excitations of electrons across the gap will induce changes in the electronic structure. This includes both band shifts resulting from the redistribution of electrons between different orbitals, and possible modifications of the local interaction parameters 
by changes in the screening environment  (Sec.~\ref{sec:dynamical}). Both effects play an important role in
photo-induced insulator-metal transitions. 
As a second mechanism, the nonthermal melting of long-range order (in particular in the presence of competing orders) can lead to the transient stabilization of nonthermal symmetry-broken phases (Sec.~\ref{hidden_magnetic_orbital}). Finally, an interesting playground for novel photo-induced phases  are the above-mentioned long-lived photo-doped states. For sufficiently cold charge carrier distributions, the liquid of photo-doped carriers may become unstable towards the formation of new orders. Theoretical predictions include hidden $\eta$-pairing states, 
as well as nonthermal magnetic, orbital, spin-orbital, and odd-frequency orders which cannot be stabilized under equilibrium conditions~(Sec.~\ref{sec:hidden_phases}).  Often, the photo-induced electronic orders strongly couple to the lattice, which may further stabilize a hidden state.

%%%%%%%%%%%%%%%%%%%%%%%%%%%%%
\subsection{Experiments on photo-excited Mott states}
\label{sec:exp_probes}

The above mentioned phenomena in nonequilibrium Mott systems have been experimentally explored both in materials and cold atom simulators. 
The latter provide a platform which allows to experimentally realize the models typically considered in theoretical studies. 
\cite{Bloch2008,Gross2017}.  
Materials investigations often involve pump-probe setups \cite{Giannetti2016,boschini2023time} in which one of the following spectroscopies is used to probe the electronic properties.

{\em Time-resolved optics ---}
Time-resolved optics measures the evolution of optical responses, such as the absorption, reflection and transmission of the probe field.
An insulating state is characterized by a gap in the absorption spectrum. From modifications of the linear optical spectrum 
one can detect the metallization of the system, renormalization of the gap, and emergence of photo-induced in-gap structures.
Moreover, nonlinear responses can provide additional insights.
For example, the second harmonic signal is a sensitive probe of inversion symmetry breaking.
The use of even higher order responses as a probe is a currently active research direction.

{\em Time-resolved photo-emission spectroscopy ---}
Photo-emission spectroscopy (PES) analyzes the electrons ejected from the solid by the probe light.
Time-resolved PES measures the evolution of the occupied part of the single-particle spectrum, and its angle-resolved version (ARPES) allows to map out the dispersion of energy bands~\cite{boschini2023time}. 
With this approach, one can directly track the relaxation dynamics of photo-carriers, as well as the renormalization of bands or gaps. 

{\em Time-resolved X-ray spectroscopy ---}
Time-resolved X-ray absorption spectroscopy (XAS) and resonant inelastic X-ray scattering (RIXS) involve charge neutral excitations and are sensitive to the unoccupied states, which makes them particularly useful tools for nonequilibrium investigations of insulators. The absorption peaks in XAS can be used to identify the positions of bands, and the (thermal or nonthermal) population of different local states.
In the RIXS process, a core electron is excited into a valence state, creating an intermediate state which lasts for a few femto-seconds, until the core hole is filled 
via emission of an X-ray photon \cite{Ament2011}. The analysis of the emitted radiation provides information on single-particle and collective excitations within the valence manifold. Because of the large momentum of X-ray photons, the dispersions of spin, orbital and charge excitations can be measured. Local transitions between multiplet states give rise to sharp excitation peaks, with energy splitting of the order of the Hund coupling, while inter-(Hubbard-)band excitations manifest themselves as broad charge transfer features at higher energies. Time-resolved RIXS measurements have recently become possible thanks to the development of X-ray free-electron lasers \cite{Bostedt2016}, which enable investigations of the dynamics of spin and charge correlations~\cite{Dean2016,Mitrano2019}.

In the following, we briefly discuss a selection of nonequilibrium experiments on Mott systems,
to  relate the measured phenomena to the concepts introduced in Sec.~\ref{sec:overview}. 
Additional experimental results are discussed in the subsequent sections.

\subsubsection{Periodically driven systems}

%%%%%%%%%%%%%%%%%%%%%%%%%%%%%
\begin{figure}[t]
\begin{center}
\includegraphics[angle=0, width=1.0\columnwidth]{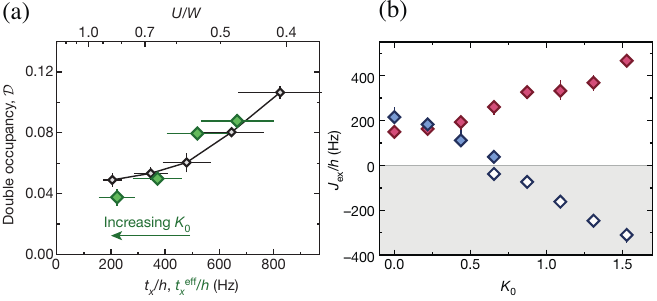}
\caption{Three-dimensional cold atom system with driving frequency $\Omega \gg U,t_x ,t_y,t_z$ and driving strength $K_0$. (a) Comparison of the double occupation $\mathcal{D}$ of the periodically driven system (open symbols) and the effective Floquet system (green symbols). 
 (b) Exchange coupling in the limit of isolated dimers with $U>\Omega$ (red) and $U<\Omega$ (blue), which becomes negative for stronger driving (open symbols). Here, $\Omega, U \gg t_x$, and the two dimer sites are along the $x$ direction. 
 (From \onlinecite{Gorg2018}.) }
\label{fig_Floqeuet_cold_atom}
\end{center}
\end{figure}   
%%%%%%%%%%%%%%%%%%%%%%%%%%%%%

An example of Floquet engineering of strongly correlated fermion systems was demonstrated with cold atom simulators. \onlinecite{Gorg2018} implemented the Hubbard model on a three dimensional lattice with anisotropic hoppings $t_i (\equiv v_{0i})$, $i\in\{x,y,z\}$, and, by shaking the optical lattice, mimicked the effect of a time-periodic electric field with frequency $\Omega$ along the $x$ direction. 
The double occupation $\mathcal{D}$ was measured in the driven state. 
In the off-resonant regime ($\Omega \gg U,t_x,t_y,t_z $), the Floquet Hamiltonian predicts an effective reduction of $t_x$.
This leads to a suppression of $\mathcal{D}$ (enhanced correlations), as shown in Fig.~\ref{fig_Floqeuet_cold_atom}(a), and the results indeed match the prediction from Floquet theory.
In the near-resonant regime with $\Omega, U \gg t_x,t_y,t_z$, the spin exchange coupling can either be enhanced, reduced or even change the sign, depending on $\Omega$ and the field strength $K_0$~\cite{Mentink2015} (Sec.~\ref{sec:effective_j_ac}).
The cold atom experiments observed hints of an enhancement of antiferromagnetic correlations for $U>\Omega$, while the sign of the spin correlations changed at sufficiently strong fields for $U<\Omega$. The exchange coupling of isolated two-site systems behaves in a similar way under periodic driving (Fig.~\ref{fig_Floqeuet_cold_atom}(b)).

\subsubsection{Photo-doped systems}
The photo-carrier generation and relaxation, and the associated modifications of physical properties, have been studied in solids using various time-resolved spectroscopic tools.

%%%%%%%%%%%%%%%%%%%%%%%%%%%%%
\begin{figure}[t]
\begin{center}
\includegraphics[angle=0, width=1.0\columnwidth]{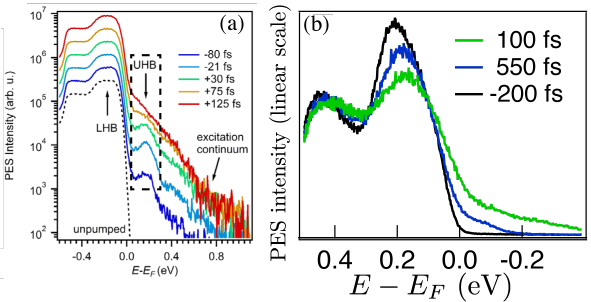}
\caption{
(a) Time-resolved PES signal in the weak excitation limit shown with arbitrary offsets from the unpumped system~(dashed) with fully occupied lower Hubbard band~(LHB). The partial occupation of the upper Hubbard band~(UHB) by doublons gives rise to the signal inside the dashed box. 
(b) Photo-excitation with a strong pulse results in a rapid (partial) filling of the gap.  
(Adapted from \onlinecite{Ligges2018} (a), and \onlinecite{Perfetti2006} (b).) 
}
\label{fig_ligges}
\end{center}
\end{figure}   
%%%%%%%%%%%%%%%%%%%%%%%%%%%%%

{\em Carrier dynamics and heating ---}
With time-resolved PES, one can monitor the dynamics of the photo-carriers as well as changes in the band structure.
An interesting example is the polaronic insulator 1$T$-TaS$_2$, which is often described as a small-gap single-orbital Mott insulator, although the precise nature of the insulating state is under debate~\cite{Butler2020,kratochvilova2017,Petocchi2022}.
After a relatively weak excitation, the PES signal reveals the partial occupation of the upper Hubbard band \cite{Ligges2018}, as illustrated by the feature inside the dashed box in Fig.~\ref{fig_ligges}(a). From the time evolution  of this signal, the doublon lifetime can be extracted.  
The surprisingly short ($<100$~fs) lifetime, compared to theoretical predictions for half-filled Mott insulators \cite{Eckstein2011thermalization}, led \onlinecite{Ligges2018} to conclude that the measured 1$T$-TaS$_2$ sample may have been intrinsically hole-doped. Metallic low-energy states enable an efficient doublon-holon recombination and remove the thermalization bottleneck characteristic of Mott insulators (Sec.~\ref{sec_recombination}). 
An alternative explanation for the short lifetime could be that 1$T$-TaS$_2$ is not a simple Mott insulator, but features a bonding/antibonding gap associated with bilayer structures, especially in the bulk region \cite{Petocchi2023}. 
Using a stronger pump pulse, \onlinecite{Perfetti2006} observed a fast ($< 100$~fs) 
partial filling of the Mott gap, see Fig.~\ref{fig_ligges}(b). 
This was attributed to strong heating,
which should induce a transition from the Mott insulator phase to the high-temperature incoherent metallic state in the metal-insulator crossover regime 
(thermalization in a hot metal state, see Fig.~\ref{fig_mott}(c)). This interpretation was supported by equilibrium dynamical mean field theory (DMFT) calculations at high temperatures. 

%%%%%%%%%%%%%%%%%%%%%%%%%%%%%
\begin{figure}[t]
\begin{center}
\includegraphics[angle=0, width=\columnwidth]{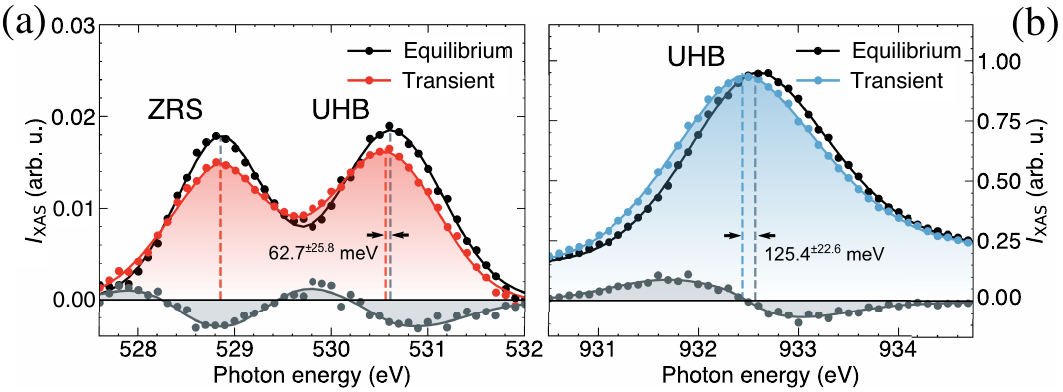}
\caption{Pump induced change in the XAS signal at the O $K$ edge (a) and Cu $L_3$ edge (b) in La$_{1.905}$Ba$_{0.095}$CuO$_4$. The black lines show the equilibrium spectra before the pump pulse, with ZRS and UHB referring to the Zhang-Rice singlet and upper Hubbard band feature, respectively. The colored lines show the XAS signals of the pumped system at pump-probe delay $\Delta t\approx 0$~ps and the gray lines the difference between the photo-excited and equilibrium spectra. (From~\onlinecite{Baykusheva2022}.) 
}
\label{fig_baykusheva}
\end{center}
\end{figure}
%%%%%%%%%%%%%%%%%%%%%%%%%%%%%

{\em Dynamical screening and band shifts ---} 
In nonequilibrium states, photo-induced charge carriers  can modify the electronic structure (Fig.~\ref{fig_mott}(c)). 
Time-resolved XAS provides valuable insights into these modifications. For example, \onlinecite{Baykusheva2022} studied a slightly hole-doped cuprate, La$_{1.905}$Ba$_{0.095}$CuO$_4$, and revealed an almost instantaneous red-shift of the feature associated with the upper Hubbard band, after a pump excitation with a 1.55 eV laser~($\Omega\approx$ gap), while the peak associated with the Zhang-Rice singlet barely moves, see Fig.~\ref{fig_baykusheva}. 
Similar results were also recently reported for above-gap excitations in NiO, where in addition to the shift of the upper Hubbard band, side peaks related to local many-body multiplet excitations could be resolved \cite{lojewski2023}.
The observed band shifts have been interpreted as evidence for a photo-induced change in the on-site Hubbard interaction \cite{Baykusheva2022}. While changes in dynamical screening can lead to reduced Mott gaps~\cite{Golez2015,Tancogne-Dejean2018}, spectral shifts of the type reported in Fig.~\ref{fig_baykusheva} could also result from photo-induced Hartree shifts, since the pump pulse redistributes charge between the orbitals (Sec.~\ref{sec:dynamical}). 

\begin{figure}[t]
\begin{center}
\includegraphics[angle=0, width=\columnwidth]{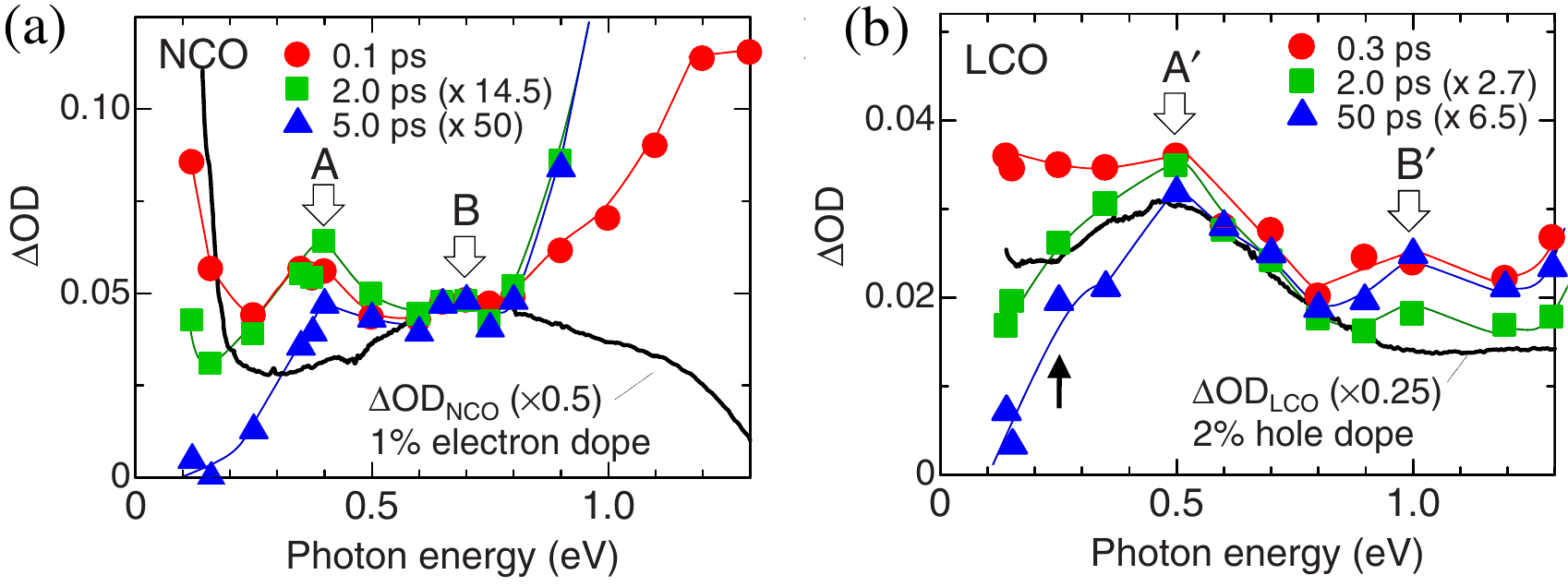}
\caption{Time-resolved change in the absorption spectra (OD: optical density) of photo-doped Nd$_2$CuO$_4$ (NCO, panel (a)) and La$_2$CuO$_4$ (LCO, panel (b)) with midgap absorption peaks marked by empty arrows. The black lines show the change in the equilibrium absorption upon chemical doping. (Adapted from \onlinecite{Okamoto2010}.) 
}
\label{fig_okamoto}
\end{center}
\end{figure}

{\em Metallization and in-gap states ---}
Photo-carriers turn a Mott insulator into a (nonthermal) metal, whose formation and decay can be investigated with time-resolved optics. Exemplarily, Fig.~\ref{fig_okamoto} shows the change in the absorption $\Delta $OD (OD stands for optical density) in Nd$_2$CuO$_4$ (NCO, panel (a)) and La$_2$CuO$_4$ (LCO, panel (b)) in the visible to IR region \cite{Okamoto2010}.
In both materials, a photo-induced metallic state with a Drude-like absorption peak is formed on a timescale much shorter than the 200 fs time resolution of the experiment, as can be seen from the low-energy up-turn of the signal measured at 0.1 ps in panel (a). This upturn is similar to the change in the absorption which is induced in equilibrium NCO by 1\% electron doping. Other noteworthy features are the midgap peaks indicated by A, B (A', B') in NCO (LCO). 
These form on a somewhat slower timescale than the Drude peak, and resemble the structures in chemically doped systems~\cite{uchida1991}.
\onlinecite{Okamoto2010} associated the midgap features with (spin-)polarons, but they could also be related to Mott excitons~\cite{jeckelmann2003} or nonthermal state populations.
The midgap features result from the intra-band relaxation and are a manifestation of a photo-doped quasi-steady state (Fig.~\ref{fig_mott}(c)), since they persists up to long times of the order of 10 ps. The Drude feature itself disappears after a few ps, presumably due to trapping of the mobile charge carriers.   

%%%%%%%%%%%%%%%%%%%%%%%%%%%%%
%%%%%%%%%%%%%%%%%%%%%%%%%%%%%

\subsubsection{Strongly correlated photo-induced phases}

The interest in nonthermal phases has been stimulated by remarkable experiments which indicate that long-lived states with novel properties can be photo-induced in certain correlated materials. Metastable states with nonthermal lattice structures were reported in manganites thin films~(Nd$_{0.5}$Sr$_{0.5}$MnO$_3$) \cite{Ichikawa2011}, LaTe$_3$~\cite{kogar2020} as well as Ca$_2$RuO$_4$ subject to epitaxial strain \cite{verma2023}.
The demonstration that the correlated insulator 1$T$-TaS$_2$ can be switched into a long-lived hidden metallic state by 35~fs laser pulses \cite{Stojchevska2014} or by slower voltage pulses \cite{Hollander2015} has triggered a large number of experimental studies and the development of first ultrafast memory devices  based on the switching between hidden and thermal states of matter \cite{Vaskivskyi2016}. A similar photo-induced transition from an insulating to a hidden metallic state was also observed in manganites films~\cite{zhang2016}. The community has furthermore been captivated by reports of light-induced superconducting-like states in laser driven cuprates \cite{Fausti11,Kaiser2014,Katsum2023PRB},  iron-based compounds \cite{Suzuki2019}, and strongly correlated organic compounds \cite{Mitrano2015,Buzzi2020}. While the mechanisms underlying these remarkable phenomena are not yet fully understood, these and related experiments have motivated the search for hidden phases, and in particular nonthermal electronic orders, in model systems.

%%%%%%%%%%%%%%%%%%%%%%%%%%%%%%%%%%%%%%%%%%%%%%%%%%%%%%%%%%%%%%%%%%%%%%%%
\section{Models}
\label{sec:models}
%%%%%%%%%%%%%%%%%%%%%%%%%%%%%%%%%%%%%%%%%%%%%%%%%%%%%%%%%%%%%%%%%%%%%%%%

\subsection{Models for strongly correlated lattice fermions}
\label{subsec:hamiltonians}

The theoretical discussions in this review are mostly based on simple but relevant models, which we list in this section. The most basic model for the study of Mott insulators is the single-band Hubbard model \cite{Gutzwiller1963,Hubbard1963},
\begin{align}
H_\text{Hub}
=&-\sum_{i,j,\sigma} v_{ij}c^\dagger_{i\sigma}c_{j\sigma}-\mu\sum_{i,\sigma}n_{i\sigma}+U\sum_i n_{i\uparrow}n_{i\downarrow},
\label{H_hubbard}
\end{align}
which describes electrons on a lattice with hopping amplitude $v_{ij}$, chemical potential $\mu$, and on-site repulsion $U$. Here, $c_{i\sigma}$ denotes the annihilation operator for an electron on site $i$ with spin $\sigma$ and $n_{i\sigma}=c^\dagger_{i\sigma} c_{i\sigma}$ is the density operator. 
In the large-$U$ regime, the spectral function of this model features upper and lower Hubbard bands separated by an energy $U$, as sketched in Fig.~\ref{fig_mott}(a). 

Relevant observables are the double occupation 
\begin{equation}
\mathcal{D}(t)=\langle n_\uparrow (t)n_\downarrow(t)\rangle,
\end{equation}
as well as the potential and kinetic energies,
\begin{align}
E_\text{pot}(t)&=N_\text{sites}U\mathcal{D}(t),
\\
E_\text{kin}(t)&=\sum_{{\bf k}\sigma} \epsilon_{\bf k}(t)n_{{\bf k}\sigma}(t), 
\end{align}
where the momentum occupation $n_{{\bf k}\sigma}$ (free-electron dispersion $ \epsilon_{\bf k}$) is the lattice Fourier transform of $n_{i\sigma}$ ($v_{ij}$).

In the limit of large $U$ and at half filling, the low energy physics of Eq.~\eqref{H_hubbard} is described by the Heisenberg model.
In the case of nearest-neighbor hopping $v_0$, it is given by
\begin{align}
H_J=&\frac{J_\text{ex}}{2}\sum_{\langle i,j\rangle} {\bf S}_{i}\cdot {\bf S}_{j},
\label{H_heisenberg}
\end{align}
where ${\bf S}_i=\sum_{\alpha,\beta}c^\dagger_{i\alpha}{\boldsymbol\sigma}_{\alpha,\beta} c_{i\beta}$ is the spin operator at site $i$, written with the vector ${\boldsymbol \sigma}$ of Pauli matrices.
$\langle i,j\rangle$ denotes nearest-neighbor sites and $J_\text{ex}=\frac{4v^2_0}{U}$ is the antiferromagnetic exchange coupling between neighboring spins induced by virtual doublon fluctuations.
In the hole-doped case, the  
low-energy physics is described by the 
$t$-$J$ model \cite{Chao1977,Gros1987} 
\begin{align}
H_{tJ}=&-v_0\sum_{\langle i,j\rangle ,\sigma} 
\tilde c^\dagger_{i\sigma}\tilde c_{j\sigma}-\mu\sum_{i,\sigma}\tilde n_{i\sigma}+\frac{J_\text{ex}}{2}\sum_{\langle i,j\rangle} {\bf S}_{i}\cdot {\bf S}_{j}.
\label{H_tj}
\end{align}
Here, doubly occupied sites are projected out, and  $\tilde c^\dagger_{i\sigma}=c^\dagger_{i\sigma}(1-n_{i\bar\sigma})$ and $\tilde c_{i\sigma}=c_{i\sigma}(1-n_{i\bar\sigma})$ are projected creation and annihilation operators.
An additional nearest neighbor density-density term is not shown for simplicity. 

A minimal model to describe the coupling of a Mott insulator to phonons is  the Holstein-Hubbard model, which extends the Hubbard Hamiltonian (\ref{H_hubbard}) by a local coupling (strength $g$) of the density $n_{i}=n_{i\uparrow}+n_{i\downarrow}$ to dispersion-less phonons with frequency $\omega_0$ \cite{Holstein1959}, 
\begin{align}
H_\text{HH}=H_\text{Hub}+\sum_{i}\Big(\sqrt{2}g(n_{i}-1)X_i + \frac{\omega_0}{2} (X_i^2+P_i^2)\Big).
\label{H_hubbardholstein}
\end{align}
Here $X_i$ and $P_i$ are canonical displacement and momentum operators of the oscillator ($[X_j,P_j]=i$).
 The effective electron-phonon coupling strength is often characterized by the parameter $\lambda=2\frac{g^2}{\omega_0}$, which (up to the sign) corresponds to the phonon-induced static attraction. 

Another relevant extension of model (\ref{H_hubbard}) is the $U$-$V$ Hubbard model with local interaction $U$ and nearest-neighbor interaction $V$, 
\begin{align}
H_\text{UV}= H_\text{Hub}
+\frac{V}{2}\sum_{\langle i,j\rangle} n_{i}n_{j},
\label{H_uvhubbard}
\end{align}
which allows to describe the effects of nonlocal screening and the formation of charge order.  

Additional complexity is introduced by adding orbital degrees of freedom. The multiorbital extension of the Hubbard model combines a general hopping term $H_\text{hop}$ with the Kanamori interaction \cite{Kanamori1963},
\begin{align}
H=&\,H_\text{hop}+U\sum_{i,\alpha} n_{i\alpha\uparrow}n_{i\alpha\downarrow}
+U'\sum_{i,\alpha>\beta,\sigma} n_{i\alpha\sigma}n_{i\beta\bar\sigma}
\nonumber\\
&+(U'-J_\text{H})\sum_{i,\alpha>\beta,\sigma} n_{i\alpha\sigma}n_{i\beta\sigma}+H_\text{sf-ph},
\label{H_multiorbital}
\end{align}
where $\alpha$ labels the orbitals, $J_\text{H}$ is the Hund coupling, $U$ denotes the intra-orbital interaction, $U'$ the inter-orbital opposite-spin interaction, and $U'-J_\text{H}$ the inter-orbital same-spin interaction. An overline over the spin index $\sigma$ marks the opposite spin. For a two- or three-orbital Hamiltonian which is rotationally invariant in spin and orbital space, one has to choose $U'=U-2J_\text{H}$ and add the spin-flip and pair-hopping terms $H_\text{sf-ph}=-J_\text{H}\sum_{\alpha>\beta}(c^\dagger_{\alpha\downarrow}c^\dagger_{\beta\uparrow}c_{\beta\downarrow}c_{\alpha\uparrow} + c^\dagger_{\beta\uparrow}c^\dagger_{\beta\downarrow}c_{\alpha\uparrow}c_{\alpha\downarrow} + \text{h.c.})$. Multiorbital Hubbard models can realize Mott insulating phases with different integer 
fillings. The multiplet structure is reflected in substructures of the Hubbard bands, as well as $J_\text{H}$-related peaks in the metallic phase (Fig.~\ref{fig_mott}(a)).

\subsection{Coupling to electric fields}
\label{sec:coupling_to_light}

To describe the laser excitation, the lattice models introduced in Sec.~\ref{subsec:hamiltonians} need to incorporate the coupling to electromagnetic fields. There are different formulations of the light-matter coupling for general multi-orbital systems \cite{Li2020light-matter,Dmytruk2021,Schueler2021,Boykin2001,Tai2023,Golez2019}. A convenient setting for tight-binding models with localized Wannier orbitals is the dipolar representation, which describes the coupling of electrons of charge $q$ to the transverse vector potential ${\bf A}(t)$ and the transverse electric field ${\bf E}^T\!(t)=-\partial_t {\bf A}(t)$ in terms of dipolar matrix elements and a Peierls phase. The single-particle Hamiltonian becomes
\begin{align}
\label{eq:dipolar}
H_\text{hop}(t) = -\sum_{a,b} v_{ab}e^{i\chi_{ab}(t)}  c_{a}^\dagger c_b  - {\bf E}^T(t)\cdot {\bf P}(t),
\end{align}
omitting spin for simplicity. Here $c_a^\dagger$ creates an electron in a Wannier orbital $|a\rangle$ with center ${\bf R}_{a}$, and $v_{ab}$ is a hopping matrix element. The fields enter through the Peierls phase \cite{Peierls1933a,Luttinger1951a}
\begin{align}
\chi_{ab}(t) = \frac{q}{\hbar}({\bf R}_a-{\bf R}_b) \cdot {\bf A}(t)\label{eq:chi}
\end{align}
and the polarization operator
\begin{align}
{\bf P}(t) = \sum_{a,b}{\bf d}_{ab} e^{i\chi_{ab}(t)}\, c_{a}^\dagger c_{b},
\end{align}
with the dipolar matrix elements 
${\bf d}_{ab} = q\langle a | ({\bf r}-{\bf R}_a) |b \rangle = q\langle a | ({\bf r}-{\bf R}_b) |b \rangle  $. 
Equations~\eqref{eq:dipolar}  and \eqref{eq:chi} are valid 
for spatially homogeneous fields (dipole approximation). 
A straightforward extension to space-dependent fields is possible, as long as they vary slowly on the scale of the unit cell. 
In the dipolar gauge, a static electric field $E_0$ corresponds to $A(t) = -tE_0$, and the Hamiltonian becomes time-dependent in contrast to the corresponding Hamiltonian in the length gauge~\cite{Schueler2021}. 

More generally, various unitary transformations of the minimal coupling Hamiltonian lead to 
different hybrid light-matter basis sets, with a modified dependence of the Hamiltonian $H[{\bf A}]$ on ${\bf A}$ and the polarization density ${\bf P}$ 
\cite{LoudonBook}. The resulting theory relates to the macroscopic Maxwell equation ${\bf \nabla} \times {\bf B} = \mu_0({\bf J} + \partial_t {\bf D})$, with current operator ${\bf J} = -\delta H[{\bf A}] /\delta {\bf A}$ and displacement field ${\bf D}=\epsilon_0{\bf E}+{\bf P}$.  While the microscopic current density ${\bf j}$ always remains unique, its separation ${\bf j}={\bf J}+\partial_t {\bf P}$ into a polarization contribution $\partial_t {\bf P}$ and the current ${\bf J}$ is representation dependent.  
The different representations are formally equivalent, but they differ after projection to a few-band model, which matters in particular for an accurate description of the nonlinear response \cite{Li2020light-matter,Tai2023}. 
The dipolar Hamiltonian \eqref{eq:dipolar} is obtained by a field-dependent basis rotation of the Wannier orbitals, corresponding to a site-dependent Power-Zienau-Woolley transformation \cite{Li2020light-matter,Schueler2021,Luttinger1951a}. 

In a single-band model with an inversion center we have ${\bf P}=0$, and Eq.~\eqref{eq:dipolar} reduces to the well-known Peierls substitution  \cite{Peierls1933a}. The current, written in momentum space, is thus  
\begin{align}
{\bf J}(t) = q\sum_{\bf k} n_{\bf k} {\bf v}_{{\bf k}-\frac{q}{\hbar} {\bf A}(t)} \label{Eq:current}
\end{align}
with the velocity ${\bf v}_{\bf k}=\frac{1}{\hbar}\partial_{\bf k} \epsilon_{\bf k}$ determined by the lattice dispersion $\epsilon_{\bf k}$. Expressions for the current ${\bf J}$ in the multi-band case can be obtained by taking the derivative $\delta H[{\bf A}] /\delta {\bf A}$ of Eq.~\eqref{eq:dipolar}. The additional  contribution $\partial_t {\bf P}$ can be evaluated at the operator level, $\partial_t {\bf P}=i[H,{\bf P}]$  (see  \onlinecite{Golez2019} and \onlinecite{Schueler2021} for explicit expressions). However, because the commutator has in general also contributions from interaction terms, it may be easier in numerical studies to measure $\langle {\bf P}(t) \rangle $ 
and to calculate $\partial_t {\bf P}$ numerically. 

In general, after the transformation to the light-matter basis, also the interaction terms depend on the field. In the dipolar representation, nonlocal Coulomb scattering matrix elements such as pair hoppings get multiplied with Peierls phases \cite{Li2020light-matter}, and thus contribute to the current operator. Coulomb scattering contributions to the current are particularly relevant for nonlinear responses in flat band systems \cite{Tai2023}.

\subsection{Spectroscopic observables}
\label{sec:observables}

{\em Optical conductivity --- } 
The optical conductivity $\sigma(t,t')$ defines the linear response relation between the induced current ${\bf \delta j}$ and a weak probe electric field ${\bf E_{\rm probe}}(t)$,
\begin{align}
{\bf \delta j}(t)
=
\int^t_{-\infty}
 dt'\sigma(t,t') {\bf E_{\rm probe}}(t').
\end{align}
In general, ${\bf\delta j}$ has contributions from the current $\bf{J}$ and the polarization ${\bf P}$, as discussed in Sec.~\ref{sec:coupling_to_light}. 
 $\sigma(t,t')$ can be expressed in terms of response functions $\chi(t,t')$ between ${\bf J}$ and/or ${\bf P}$ using the nonequilibrium Kubo formula, see ~\onlinecite{Eckstein2008b,Filippis2012,Shao2016} for single-band cases.
In nonequilibrium simulations, one can evaluate $\sigma(t,t')$ either by calculating $\chi(t,t')$ or by explicitly simulating  a $\delta$-function probe field (step-like change of ${\bf A}(t)$) and  measuring the induced current ${\bf \delta j}(t)$ \cite{Shao2016,Werner2019}.  In Green's function based approaches, the latter strategy automatically takes into account the vertex corrections in the diagrammatic expression of $\chi(t,t')$.

In experiments, the nonequilibrium conductivity is measured using a pulse of nonzero duration ${\bf E_{\rm probe}}(t) = {{\bf E}_{t_p}}(t)$ centered at $t_p$. 
A possible working definition for the nonequilibrium conductivity is $\sigma_\text{meas}(\omega, t_p)\equiv{\bf \delta j}(\omega)/{\bf E}_{t_p}(\omega)$.
 In a quasi-equilibrium situation, $\sigma_\text{meas}(\omega, t_p)\simeq \int \sigma(t_p+s,t_p)e^{i(\omega+i\delta)s}ds$.
However, in general, the relation between $\sigma_\text{meas}(\omega, t_p)$ and $\sigma(t,t')$  depends on the shape of the probe \cite{Shao2016}.

For the single-band Hubbard model~\eqref{H_hubbard} under a pump field described by ${\bf A}_{\rm pump}(t)$, 
\eqq{
\sigma_{\alpha\beta}(t,t') =& \,\,\theta(t-t') \tau_{\alpha\beta}(t) + \int^t_{t'} d \bar{t} \chi_{J^\alpha\!,J^\beta}(t,\bar{t}). \label{sigma_Hubbard}
}
Here, $\tau_{\alpha\beta}(t) \equiv \frac{q^2}{\hbar^2}\sum_{\bf k} 
\big(\partial_{k_\beta} \partial_{k_\alpha} \epsilon_{{\bf k}-\frac{q}{\hbar}{\bf A}_{\rm pump}(t)}\big) \langle n_{\bf k}(t)\rangle  $  is the stress tensor, $\langle \cdots \rangle$  indicates the expectation value for the system with pump field only, and $\chi_{J^\alpha\!,J^\beta}(t,t')\equiv -\frac{i}{\hbar}\theta(t-t')\langle [J^\alpha(t), J^\beta(t')]\rangle$.
$J^\alpha(t)$ is the $\alpha$-component of the current operator in the Heisenberg representation with pump field.
The first term of Eq.~\eqref{sigma_Hubbard} captures the modification of the current operator by the probe field and is called the diamagnetic term. The second term is the paramagnetic contribution.

{\em Spectral functions --- } 
The time-dependent spectral function ($A$) and occupation function ($N$) provide useful information on the single-particle properties.
They are related to the retarded ($R$) and lesser ($<$) parts of the single-particle Green's function $G(t,t')\equiv -i\langle \mathcal{T}_\mathcal{C} c(t)c^\dagger(t')\rangle$, where $\mathcal{T}_\mathcal{C}$ is the contour-ordering operator~\cite{Aoki2014}. 
 For a mixed frequency-time representation, one can use the ``forward" Fourier transforms
\begin{align}
A(\omega,t')&=-\frac{1}{\pi}\text{Im}\int_{t'}^\infty dt e^{i\omega(t-t')}G^R(t,t'),\label{A}\\
N(\omega,t')&=\frac{1}{\pi}\text{Im}\int_{t'}^\infty dt e^{i\omega(t-t')}G^<(t,t') \label{N}.
\end{align}
or related ``backward" or symmetric (Wigner) transforms.
The occupation 
function is related to the time-resolved photo-emission spectrum $I(\omega,t_p)$ by convolutions 
with the envelope function $S(t)$ of the probe pulse centered at probe time $t_p$  \cite{Freericks2009pes,Eckstein2008c},
\begin{align}
I(\omega,t_p)=-i\int dt dt' S(t)S(t')e^{i\omega(t-t')}G^<(t+t_p,t'+t_p).
\label{I}
\end{align}
In contrast to $A(\omega,t')$ and $N(\omega,t')$, this spectrum is guaranteed to be positive. Variants of this expression, in particular for momentum-resolved spectra, or in the presence of fields acting also outside of the solid, have been discussed in several review articles \cite{Kemper2017,SchuelerArpes2021,EcksteinArpes2021}.

%%%%%%%%%%%%%%%%%%%%%%%%%%%%%%%%%%%%%%%%%%%%%%%%%%%%%%%%%%%%%%%%%%%%%%%%
\section{Simulation of nonequilibrium strongly correlated systems}
\label{sec:simulation}
%%%%%%%%%%%%%%%%%%%%%%%%%%%%%%%%%%%%%%%%%%%%%%%%%%%%%%%%%%%%%%%%%%%%%%%%

In this section, we explain relevant methods and concepts for the study of strongly correlated nonequilibrium systems. 
The numerical approaches can be classified into Green's function methods, such as %those based on 
various flavors of
dynamical mean-field theory (Sec.~\ref{sec:DMFT}),
wave-function approaches, such as exact diagonalization or matrix-product state methods (Sec.~\ref{sec:WF}), and density functional theories (Sec.~\ref{sec:DFT}).
We also discuss the concept of quasi steady-state descriptions of photo-doped large-gap Mott systems (Sec.~\ref{sec:quasiseq}).

\subsection{Methods based on dynamical mean field theory} \label{sec:DMFT}

The nonequilibrium dynamics of the Hubbard model and its extensions is generally not analytically solvable and we have to resort to numerical techniques to study a perturbed Mott state. 
In the limit of infinite dimensions or infinite coordination number \cite{Metzner1989}, an exact numerical solution is provided by dynamical mean field theory (DMFT)~\cite{Georges1996}. If applied to finite-dimensional systems, DMFT neglects nonlocal correlations but provides a good description of the local correlation effects, which are important in Mott systems. In the following subsections, we will briefly discuss nonequilibrium DMFT  
\cite{Aoki2014} and various extensions which have been used to study nonequilibrium states in Mott insulators.   

\subsubsection{Nonequilibrium DMFT}

\begin{figure}[t]
\begin{center}
\includegraphics[angle=0, width=\columnwidth]{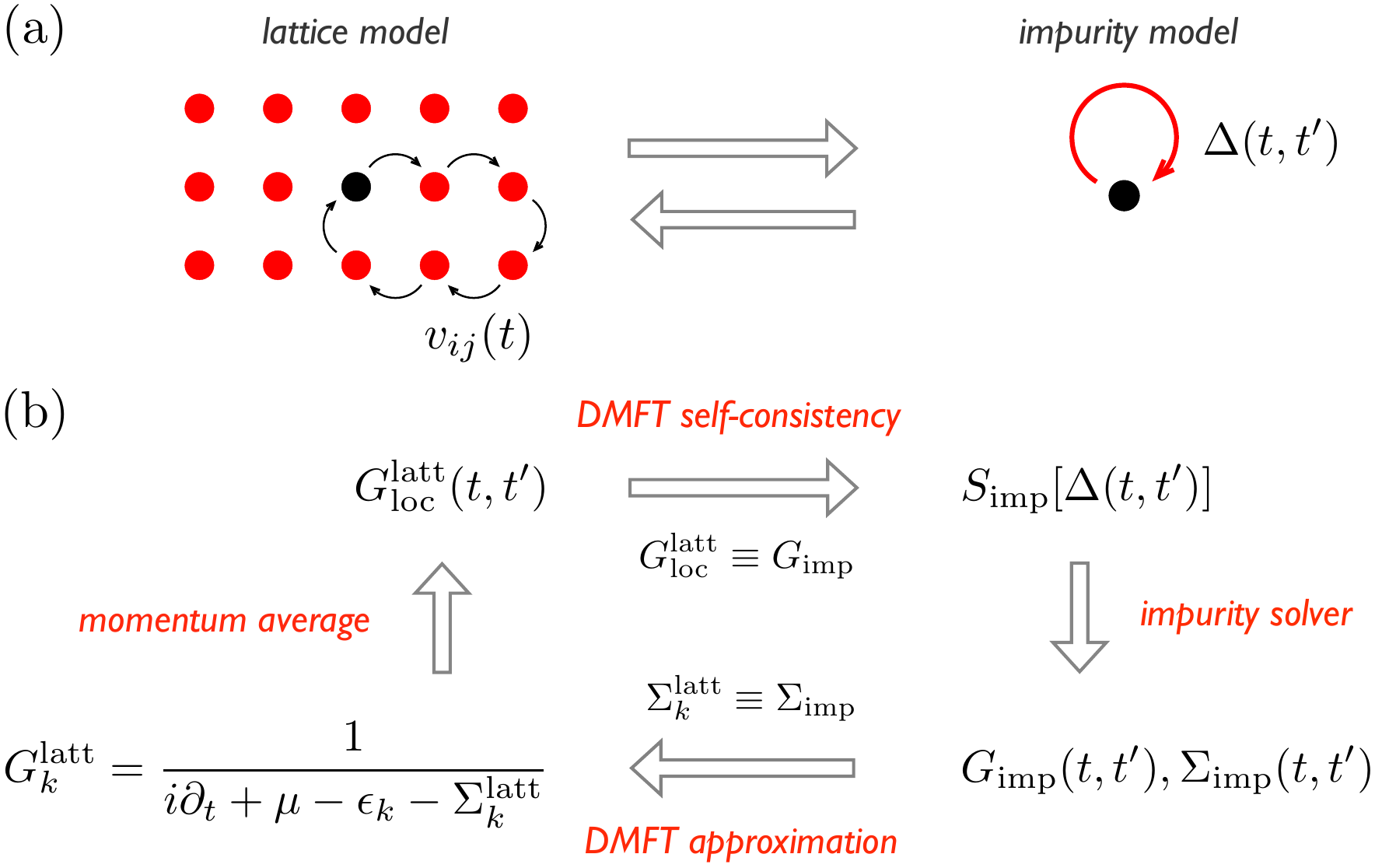}
\caption{(a) Self-consistent mapping of a correlated lattice model to an effective impurity problem with hybridization function $\Delta$. (b) Sketch of the DMFT self-consistency loop with the lattice~(impurity) Green's function $G_{\text{latt}}$ ($G_{\text{imp}}$), the lattice~(impurity) self-energy $\Sigma_{\text{latt}}~(\Sigma_{\text{imp}})$ and the impurity action $S_{\text{imp}}[\Delta]$.
}
\label{fig_dmft}
\end{center}
\end{figure}

DMFT maps a correlated lattice model onto a self-consistently determined single-site model, a so-called quantum impurity model \cite{Georges1992a}.  This approximate mapping identifies the momentum-dependent self-energy $\Sigma_{\bf k}^\text{latt}$ of the lattice model with the local (momentum-independent) impurity self-energy $\Sigma_\text{imp}$. The bath properties of the impurity model are fixed by a self-consistency condition, which demands that the local lattice Green's function is identical to the impurity Green's function, $G^\text{latt}_\text{loc}=\int (d{\bf k}) G_{\bf k}^\text{latt}\equiv G_\text{imp}$ (see Fig.~\ref{fig_dmft}). In the imaginary-time formalism, the bath is represented by a hybridization function $\Delta(\tau)$, which encodes how electrons hop in and out of the impurity site, and thus plays the role of a dynamical mean field. To generalize DMFT to systems out of equilibrium \cite{Schmidt2002,Freericks2006}, the equations are rewritten within the Keldysh formalism. This generalization is also possible for cluster \cite{Maier2005}, inhomogeneous~\cite{Potthoff1999} and diagrammatic \cite{Rohringer2018} extensions of DMFT.  
Below, we describe the formalism for the real-time evolution from an equilibrium initial state. Nonequilibrium steady state approaches are discussed in Sec.~\ref{eq:NESS_DMFT}.

%%%%%%%%%%%%%%%%%
%%%%%%%%%%%%%%%%%

In practice, there are two main challenges:
The first is the solution of the impurity problem (Sec.~\ref{impsolvers}). 
Second, in nonequilibrium DMFT, the functions $G(t,t')$, $\Sigma(t,t')$ and $\Delta(t,t')$ are defined on the Keldysh time contour $\mathcal{C}$ instead of the imaginary time axis. 
To evaluate $G$, 
one must solve the Dyson equation $G=G_0 + G_0*\Sigma*G$, where $*$ indicates the convolution on  $\mathcal{C}$, and $G_0$ is the free-electron Green's function.
When the system is in equilibrium or in a steady state, the Dyson equation can be solved by Fourier transformation with respect to the time-difference $t-t'$. 
However, generic cases require the explicit solution of the equation on $\mathcal{C}$.
A standard implementation using an equidistant time mesh can be found in the open-source library NESSi \cite{Nessi}.
With such a scheme, the required computer memory (cpu time) scales like $\mathcal{O}(N_t^2)$ ($\mathcal{O}(N_t^3)$) with the number of time points $N_t$, which makes simulations of the long-time behavior challenging. Several ideas have been recently put forward  to overcome this computational bottleneck in simulations of Mott systems, including quasi-steady-state approaches (Sec.~\ref{sec:quasiseq}) and compressed representations~\cite{Kaye2021,Shinaoka2022}. 
Systematic truncations of the memory 
kernel in the Dyson equation \cite{Schueler2018,Stahl2022} also allow to significantly extend the simulation time in certain parameter regimes \cite{Dasari2020,Picano2021b}.

DMFT provides direct access to single-particle observables and spectroscopic probes such as the time-resolved photoemission spectrum~(Sec.~\ref{sec:exp_probes}). The time-resolved optical conductivity can be  evaluated diagrammatically~\cite{Eckstein2008b}, or by simulating the probe process with a time-dependent electric field 
and computing the current \eqref{Eq:current}. The latter approach includes vertex corrections originating from the two-particle irreducible vertex (functional derivative of the DMFT 
self-energy with respect to the probe field)~\cite{stefanucci2013,Golez2019,murakami2016}. 
Moreover, DMFT is well suited to predict core-level X-ray photoemission spectroscopy (XPS) and X-ray absorption spectroscopy (XAS). These probes involve a localized core hole and the spectroscopic signal can be computed by considering an extended impurity model which contains the core level in addition to the valence orbitals \cite{Cornaglia2007,Haverkort2014,Hariki2018}. 
Within this extended impurity problem, the relevant interactions between the core hole and the valence electrons are treated exactly. 
Such a scheme has been used to calculate time-resolved XAS in a two-orbital Hubbard model \cite{Werner2022xas} and charge-transfer insulator~\cite{lojewski2023}. Similarly, DMFT can be used to predict the resonant inelastic X-ray scattering (RIXS) signal, although this approach only captures the dynamics of momentum-integrated excitations \cite{Eckstein2021,Werner2021rixs}. 

\subsubsection{Cluster DMFT} 

The DMFT approximation neglects spatial fluctuations, which can qualitatively alter the physics in 1D or 2D lattice models. Cluster extensions of DMFT have been developed to capture short-range correlations within small clusters, while longer-range correlations are treated at a mean-field level \cite{Maier2005}. The dynamical cluster approximation (DCA) \cite{Hettler1998} approximates the self-energy $\Sigma_{\bf k}^\text{latt}$ as constant within specific regions (``patches") of the Brillouin zone, and associates it with the corresponding momentum component of the self-energy in a cluster impurity model: $\Sigma_{\bf k}^\text{latt}(t,t')\approx \sum_{\bf K} \Sigma^\text{imp}_{\bf K}(t,t')\phi_{\bf K}({\bf k})$, where $\phi_{\bf K}({\bf k})=1$ for ${\bf k}$ inside momentum patch ${\bf K}$ and zero otherwise. The DCA self-consistency loop is analogous to Fig.~\ref{fig_dmft}, but with momentum averaging performed over the respective patch. The patch-averaged lattice Green's function $G^\text{latt,av}_{\bf K}$ is identified with the cluster impurity Green's function $G^\text{imp}_{\bf K}$ for momentum ${\bf K}$. For nonequilibrium simulations, DCA is preferred over real-space cluster DMFT (CDMFT) 
\cite{Lichtenstein2000}: Due to its diagonal representation in momentum space it avoids solving matrix-valued Dyson equations, and leads to a linear (rather than quadratic) scaling of the memory requirement with the number of cluster sites~\cite{Tsuji2014}. Nonequilibrium plaquette DCA has been used to investigate several phenomena in photo-doped Mott systems, such as the effects of spin-charge coupling \cite{Eckstein2016dca},  geometric frustration \cite{Bittner2020triangle} and Mott exciton formation \cite{Bittner2020exciton}. 
A limitation is that DCA lacks gauge invariance \cite{lin2009,Bittner2020triangle}, which  could possibly be overcome by a nonequilibrium extension of periodized CDMFT~\cite{kotliar2001,biroli2004}. 

\subsubsection{Extended DMFT and GW+EDMFT}
\label{edmftsec}

{\it Extended DMFT (EDMFT) ---}
The extended DMFT formalism \cite{Sengupta1995,Si1996,Sun2002,Ayral2013,Golez2015} enables the treatment of models with nonlocal interactions, like the $U$-$V$ Hubbard model (\ref{H_uvhubbard}), and incorporates dynamical screening effects on the local level \cite{WernerCasula2016}. 
EDMFT maps the lattice system to a single-site impurity model, but now includes 
both a fermionic dynamical mean field, corresponding to the hybridization function $\Delta(t,t')$, and a bosonic dynamical mean field, represented by a self-consistently computed retarded impurity interaction $\mathcal{U}(t,t')$. $\Delta$ is determined by the usual fermionic self-consistency loop (Fig.~\ref{fig_dmft}(b)), while $\mathcal{U}$ is calculated by an additional bosonic self-consistency loop which connects the screened interaction $W$ to the polarization $P$. The fully screened interaction in the lattice is determined by the bosonic lattice Dyson equation  $W_{\bf k}=V_{\bf k}+V_{\bf k}\ast P_{\bf k} \ast W_{\bf k}$, where $V_{\bf k}$ is the bare interaction and $P_{\bf k}$ the momentum-dependent polarization function. Within EDMFT, $P_{\bf k}$ is approximated by the local polarization function $P_{\rm imp}$ of the impurity model, which relates the fully screened impurity interaction  $W_\text{imp}$ and the ``bare impurity interaction'' $\mathcal{U}$ through $W_\text{imp}=\mathcal{U}+\mathcal{U}\ast P_{\rm imp} \ast W_\text{imp}$. The condition $\int (d{\bf k}) W_{\bf k}=W_\text{imp}$ closes the self-consistency. 

The impurity $\mathcal{U}$ and the fully screened interaction $W$ reflect the time-dependent modifications of the dynamically screened interactions out of equilibrium. The dominant peaks in the spectrum of $\mathcal{U}$, 
\begin{align}
-\frac{1}{\pi}\text{Im}\,\mathcal{U}^\mathrm{R}(\omega,t)&=-\frac{1}{\pi}\text{Im}\int_t^\infty dt' e^{i\omega(t'-t)}\mathcal{U}^\mathrm{R}(t',t),
\label{Uomega}
\end{align}
(and similarly for $W$) represent the screening modes associated with single-particle and collective charge fluctuations. By integrating over these modes, we can define an effective coupling strength between the electrons and charge fluctuations~\cite{Ayral2013}
\begin{align}
\lambda(t) = -\frac{2}{\pi}\int_0^\infty d\omega \frac{\text{Im}\,\mathcal{U}^\mathrm{R}(\omega,t)}{\omega}.
\label{lambda_screening}
\end{align}

The EDMFT framework is also applicable to models with nonlocal spin interactions and has been used to study the $t$-$J$ model (\ref{H_tj}) \cite{Haule2002,Bittner2018} and Heisenberg model (\ref{H_heisenberg}) \cite{Otsuki2013}. In this case, the nonlocal magnetic interactions determine a retarded impurity spin-spin interaction $\mathcal{J}(t,t')$,  whose modes describe the local magnetic excitations.

{\it GW+EDMFT ---}
To incorporate nonlocal correlation and screening effects into the single-site EDMFT formalism for the $U$-$V$ Hubbard model, we can add nonlocal $\Sigma$ and $P$ components taken from the GW approximation \cite{Hedin1965} and feed the resulting fermionic and bosonic self-energies back into the self-consistency loop~\cite{Biermann2003,Ayral2013,Golez2017}. In this so-called GW+DMFT formalism, we define 
\begin{align}
\Sigma^{\text{latt}}_{ij}&=\Sigma^\text{EDMFT}\delta_{ij}+\Sigma^{GW,\text{nonloc}}_{ij},\\
P^{\text{latt}}_{ij}&=P^\text{EDMFT}\delta_{ij}+P^{GW,\text{nonloc}}_{ij},
\end{align}
where $\Sigma^{GW,\text{nonloc}}_{ij}$ and $P^{GW,\text{nonloc}}_{ij}$ are the GW expressions for $\Sigma$ and $P$ with the local contributions subtracted.
This formalism allows to study nonlocal (charge) fluctuations in the Mott insulating phase, and the inclusion of the nonlocal polarization substantially improves the description of the fluctuation spectrum~\cite{Ayral2013}. GW+EDMFT can lead to non-causal behavior of the auxiliary fields $\Delta$ and $\mathcal{U}$ \cite{Nilsson2017}. Recently, an alternative self-consistency scheme which ensures causality has been proposed~\cite{backes2022,chen2022}, but the nonequilibrium implementation of this variant has not been realized yet. 

The GW+EDMFT formalism can  be extended to a multi-tier approach \cite{Boehnke2016,Nilsson2017} in which the weakly correlated bands are treated at the GW level, while the non-perturbative corrections  $\Sigma^\text{EDMFT}$ and $P^\text{EDMFT}$ are obtained for a subset of strongly correlated orbitals from a self-consistently embedded impurity model. The nonequilibrium implementation of this formalism \cite{Golez2019,Golez2019a} provides a self-consistent description of the dynamically screened interactions and Hartree shifts in photo-excited Mott systems (Sec.~\ref{sec:dynamical}).

\subsubsection{Steady-state DMFT and Floquet DMFT} \label{eq:NESS_DMFT}
If a system is connected to heat baths and is subjected to time-translation invariant perturbations, the balance between the energy injection and dissipation 
leads to a nonequilibrium steady state (NESS). If the external perturbations are time-periodic, the system reaches a time-periodic nonequilibrium state. Steady-state and  Floquet DMFT enable the description of such steady and time-periodic states, 
without explicitly simulating the initial transient evolution \cite{Schmidt2002,Tsuji2008,Joura08,Qin2017}. 
For the heat baths, one typically considers either a bath of noninteracting fermions with a finite bandwidth, or a boson bath. The self-energy produced by these baths is added to the impurity self-energy in the DMFT self-consistency loop. Both steady state and  Floquet formalisms are implemented on the two-branch Keldysh contour using  functions  $\hat G(t,t')$, $\hat \Sigma(t,t')$ and $\hat \Delta(t,t')$ which depend on the real physical time and have a 2$\times$2 Keldysh matrix structure \cite{Tsuji2008,Aoki2014}. The transformation to average time $t_\text{av}=(t+t')/2$ and relative time $t_\text{rel}=(t-t')$ allows to perform a Fourier transform with respect to $t_\text{rel}$, which introduces the frequency $\omega$. In a NESS,  two-time functions do not depend on $t_\text{av}$, and the Dyson equation becomes a $2\times2$ matrix equation in Keldysh space, $\hat G(\omega)^{-1}=\hat G_0(\omega)^{-1}-\hat \Sigma(\omega) $. In a time-periodic state, we assume that all two-time quantities are periodic with respect to $t_\text{av}$, with period $\mathcal{T}=\frac{2\pi}{\Omega}$. 
The  Floquet matrix representation is then defined by a continuous Fourier transform with respect to $t_{\text{rel}}$, and the discrete transform
\begin{align}
\hat {\underline G}_{nm}(\omega) = \frac{1}{\mathcal{T}}\int_0^\mathcal{T} \! dt_\text{av} \,
e^{i(m-n)\Omega t_\text{av}}
\hat G\Big(\omega+\frac{m+n}{2}\Omega,t_\text{av}\Big),
\end{align}
where the continuous frequency $\omega$ is restricted to the 
Floquet Brillouin zone $[-\frac{\Omega}{2}, \frac{\Omega}{2})$ and $\hat {\underline G}$ indicates a matrix with respect to the Floquet indices $n,m$ . 
With this matrix representation, the Dyson equation can be simply expressed as a matrix equation, $\underline {\hat G}(\omega)^{-1}=\underline {\hat G}_0(\omega)^{-1}-\underline {\hat \Sigma}(\omega)$.

\subsubsection{Impurity solvers}
\label{impsolvers}
The main challenge for the implementation of nonequilibrium DMFT and its extensions is the solution of the effective impurity model. In the case of the Hubbard model \eqref{H_hubbard} and single-site DMFT, this is the
Anderson impurity model with action
\begin{align}
S_\text{imp}=&\int_\mathcal{C} dt dt' \sum_\sigma c_\sigma^\dagger(t)\Delta(t,t')c_\sigma(t')
+\int_\mathcal{C} dt H_{\rm loc}(t).
\label{S_imp}
\end{align}
$H_{\rm loc}$ is the local Hamiltonian, which in particular contains the interaction $U$. 
In equilibrium, the Anderson model can be solved efficiently using a variety of methods, like quantum Monte Carlo (QMC) techniques \cite{Gull2011} or wave-function based techniques~\cite{Bulla08,Schollwoeck2005}. 
For nonequilibrium studies of Mott systems, there exists a number of approximate solvers, while exact solvers are currently limited to short-time simulations.  

{\it Self-consistent strong coupling expansion ---} 
The self-consistent strong-coupling (hybridization) expansion is a systematic approach which starts from 
the atomic limit. Its formulation in the nonequilibrium context has been discussed in detail in \onlinecite{Eckstein2010nca} and \onlinecite{Aoki2014}. In brief, by performing a power-series expansion in $\Delta(t,t')$, the partition function of the impurity problem (\ref{S_imp}), $Z=\text{Tr}[T_\mathcal{C} e^{-iS_\text{imp}}]$, can be expressed as an infinite sum over strong-coupling diagrams. Up to prefactors, the weight of a given diagram is given by the product of hybridization functions and atomic propagators $\tilde g_{\alpha,\beta}(t,t')$ with indices $\alpha,\beta$ in a basis of the local Hilbert space. 
Dressed propagators $\tilde G_{\alpha,\beta}(t,t')$, which include all hybridization events, then satisfy a Dyson equation with a self-energy $\tilde \Sigma[\tilde G,\Delta]$ expressed in terms of $\tilde G$ and $\Delta$. 
The impurity Green's function $G$ can be evaluated from the converged
$\tilde G$ and $\Delta$.
Most nonequilibrium DMFT simulations for Mott insulators to date have used the non-crossing approximation (NCA) \cite{Keiter1971} or one-crossing approximation (OCA) \cite{Pruschke1989}, where $\tilde \Sigma[\tilde G,\Delta]$  is calculated to first or second order in $\Delta$, respectively. This is a versatile approach, which works for arbitrary local Hamiltonians and generic hybridizations. 
The constraint on the crossing of hybridization functions however introduces unphysical correlations, which can become problematic especially in multi-orbital or doped systems.

An variant of the model \eqref{S_imp}, which plays an important role in (GW+)EDMFT, is the Anderson Holstein model with additional local bosonic degrees. After integrating out the bosons one obtains an additional retarded term in the action, $S=\int dtdt'n(t)\mathcal{B}(t,t') n(t')$, where $\mathcal{B}$ is the noninteracting boson propagator \cite{Aoki2014}. 
Three complementary NCA/OCA impurity solvers have been developed to handle this term: i) a combined strong/weak-coupling expansion approach \cite{Golez2015,Chen2016}, where the partition function is expanded in both the hybridization $\Delta$ and the retarded interaction $\mathcal{B}$ (for retarded spin-spin interactions, see \onlinecite{Bittner2018}), ii) a method based on a Lang-Firsov decoupling \cite{Lang1962} of the electron-boson interaction and a transformation to polaron operators \cite{Werner2013}, and iii) the direct inclusion of phonons into the local Hilbert space, which is then truncated~\cite{Grandi2021}.

{\it Monte Carlo based solvers ---}
The direct extension of continuous-time QMC algorithms from imaginary time to the Keldysh contour \cite{Muehlbacher2008,Werner2009}  leads to a severe dynamical sign problem, which restricts simulations to very short times. 
Numerically exact nonequilibrium DMFT results have been obtained with the real-time implementation of the weak-coupling continuous-time Monte Carlo method \cite{Werner2010}, but accessing the Mott regime with this solver is challenging \cite{Eckstein2010}. The inchworm algorithm is a more promising diagrammatic Monte Carlo algorithm, which uses the causal structure of the self-consistent  strong-coupling expansion while sampling the contributions to the propagation kernel in the Dyson equation \cite{Cohen2015,chen2017a,chen2017b}. Due to the high numerical cost, this algorithm has not yet been used within nonequilibrium DMFT to simulate photo-induced  dynamics. The more recent steady state inchworm algorithm  \cite{Erpenbeck2023} could be used as a solver for steady state techniques (Sec.~\ref{sec:quasiseq}). A possibly even more compact resummation of the diagrammatic series can be achieved with triangular vertices \cite{Kim2022,Kim2023}. 

{\it Hamiltonian representation and auxiliary master equation approach ---}  
The action \eqref{S_imp} is equivalent to the Anderson impurity Hamiltonian $H_\text{imp}=H_\text{loc}+H_\text{bath}$, where $H_{\rm bath}$ includes additional bath orbitals $p$ with fermion annihilation operators $a_{p,\sigma}$, and a time-dependent hybridization $V_{p}(t) a_{p,\sigma}^\dagger c_\sigma + h.c $ between the impurity site and the bath orbitals. $H_\text{imp}$ can be solved using exact diagonalization~\cite{Gramsch2013} or matrix product state techniques~\cite{Wolf2014}, but the main challenge is to represent the time evolving bath. The technique developed so far \cite{Gramsch2013} requires a number of bath sites which increases with time, leading to an exponential increase of the numerical cost. 

 Alternatively, one can represent the impurity action \eqref{S_imp} in terms of a Markovian open quantum system, which includes both explicit bath orbitals as in $H_{\rm bath}$, and Lindblad dissipators which add and remove particles from the bath \cite{Arrigoni2013}. This auxiliary Master equation approach has been formulated for NESS simulations~\cite{dorda2014,dorda2015,titvinidze2015} and also applied to Floquet DMFT~\cite{Sorantin2018,Gazzaneo2022,Mazzocchi2022,Werner2023,Mazzocchi2023}. 
 The Lindblad terms can be adapted to independently represent $\Delta^<(\omega)$  (occupied density of states of the bath) and $\Delta^>(\omega)$ (unoccupied density of states), which are not linked by the fluctuation dissipation relation in a NESS.

\subsection{Wave-function based methods} \label{sec:WF}

\subsubsection{Exact diagonalization} 

The time evolution  of small systems, such as Hubbard clusters up to $20$ sites, can be calculated using the time-dependent Lanczos method \cite{Park1986,Prelovsek_springer}. Expanding the time evolution of the wave function over one time step $\delta t$ into a power series, $e^{-iH(t)\delta t}|\psi(t)\rangle\approx \sum_{n=0}^{p-1}\frac{(-i\delta t H(t))^n}{n!}|\psi(t)\rangle$, yields $p$ independent vectors  $H(t)^n |\psi(t)\rangle$ which define the so-called Krylov space. The Lanczos algorithm \cite{Lanczos1950} recursively generates an orthonormal Krylov space basis $\{|\phi_l\rangle\}$, starting from $|\phi_0\rangle = |\psi(t)\rangle$, in which the Hamiltonian $H(t)$ becomes a tridiagonal matrix. The time evolution  is calculated in this small space 
(typically $p\sim 10$, depending on the step $\delta t$), and transformed back to yield the  state vector $|\psi(t+\delta t)\rangle$. 
The normalization of the wave function is thus preserved exactly. The main source of error are finite-size effects and artificial recurrences. Assuming that correlations spread with a characteristic light-cone velocity $v_\text{lc}$,  finite-size artifacts can be expected for times $\gtrsim L/(2v_\text{lc})$, where $L$ is the linear size of the system. Larger systems can in some cases be treated by  using the time-dependent Lanczos procedure within a suitable ``limited functional space", which is iteratively constructed from a simple initial product state~\cite{Edwards1987,Trugman1988,Inoue1990}. This approach has been used to study the dynamics of (one or two) photo-doped charge carriers forming spin~\cite{Golez2014,bonca2012} or lattice~\cite{Vidmar2011,golez2012} polarons.

\subsubsection{Matrix product state approach}

The time evolution  of large one-dimensional systems can be calculated using numerical methods based on a matrix product state (MPS) representation of the wave function, such as the density-matrix renormalization group (DMRG) \cite{White1992,White2004,Haegeman2011} or the time-evolving block decimation (TEBD)~\cite{Vidal2004,Daley2004,Vidal2007PRL}.
In these methods, the state-vector $|\psi\rangle$ is expressed as an MPS:
\begin{equation}
|\psi\rangle \approx \sum_{\sigma_1\ldots\sigma_L} A^{\sigma_1}\ldots A^{\sigma_L} |\sigma_1\ldots\sigma_L\rangle,
\label{eq_dmrg}
\end{equation}
where the $\sigma_i$ represent the local degrees of freedom at site $i$,  the $A^{\sigma_i}$ are $m\times m$ matrices, and $m$ is called the bond dimension. A truncation in $m$ 
is implemented by decomposing the system into subsystems $A$ and $B$, performing a Schmidt decomposition $|\psi\rangle=\sum_{\alpha} s_\alpha |\alpha\rangle_A|\alpha\rangle_B$ and keeping only the $m$ largest coefficients $s_\alpha$ \cite{Schollwoeck2011, Schollwoeck2005}. The efficiency of the MPS ansatz  depends on  the  scaling of the entanglement entropy $S_\text{vN}=-\sum_\alpha s_\alpha^2 \text{log}s_\alpha^2$ with system size. Ground states of gapped quantum systems satisfy an area law  \cite{Schollwoeck2011,Stoudemire2012}. In dimension $D=1$ the entanglement entropy of a large subsystem then becomes system-size independent, and the ground state wave function can be accurately described with a finite bond dimension. 

In TEBD, the ground state is obtained by an imaginary-time evolution, whose algorithm is essentially the same as for the real-time evolution ~\cite{Vidal2004}. A Suzuki-Trotter decomposition is used to split the time evolution operator $\exp(-i H \delta t)$ for a time step $\delta t$ into a product of operators $\exp(-i h_{i,i+1} \delta t)$, which act on just two neighboring $A$ matrices in Eq.~(\ref{eq_dmrg}). Then, one applies $\exp(-i h_{i,i+1} \delta t)$ to the MPS and performs a singular value decomposition at the bond $[i,i+1]$ to update the MPS. 
 In DMRG, the MPS for the ground state is evaluated (for a given $m$) using the variational principle for the energy. The real-time evolution  is either implemented with the TEBD algorithm~\cite{White2004} or using the time-dependent variational principle for MPSs~\cite{Haegeman2011,shinjo2021}.
After a global perturbation, such as a global quench or a perturbation with a uniform field, one usually finds that the entropy 
grows linearly with time, which in turn implies an exponential growth of the bond dimension $m$ needed to keep a given accuracy.
 
In analogy to exact diagonalization (ED), response functions can either be evaluated by the time-dependent Kubo formalism, or by simulating a small probe pulse. For example, the single-particle spectrum can be evaluated by introducing an auxiliary fermionic band and considering the excitations from the system to this auxiliary band~\cite{Bohrdt2018,Feiguin2019,Murakami2021PRB}. X-ray spectra can be calculated by adding core levels to the Hamiltonian \cite{Zawadzki2019}.

The thermodynamic limit can be accessed using a translationally invariant MPS ansatz~\cite{Vidal2007PRL,Pollmann2013PRB}. To treat nonequilibrium systems in dimensions $D>1$, related tensor-network ans\"atze, such as projected entangled pair states, have been developed~\cite{Phien2015,Hubig2020}.

\subsubsection{Time-dependent variational techniques}

{\it Time-dependent Gutzwiller approach ---}\label{sec_gutzwiller}
A widely used variational approach for correlated electrons employs the Gutzwiller wavefunction \cite{Gutzwiller1965} and the Gutzwiller approximation (GA), which becomes exact in the limit of infinite dimensions~\cite{Metzner1989}.
This ansatz has provided an early description of the Mott transition in the single-band Hubbard model \cite{Brinkman1970}, and was later generalized to multi-band systems \cite{Buenemann1998}
and time-dependent problems \cite{Schiro2010,Schiro11,Fabrizio_springer}. The idea is to use the ansatz
\begin{equation}
\label{Gansatz}
|\psi(t)\rangle = \prod_j \mathcal{P}_j(t)|\psi_\text{sd}(t)\rangle
\end{equation}
 for the time-evolving state-vector, where $|\psi_\text{sd}(t)\rangle$ is a time-dependent variational Slater determinant, and the $\mathcal{P}_j(t)$ are time-dependent variational operators which change the weights of the local many-body configurations at lattice site $j$ and depend on a set of variational parameters $\phi_{j}$. In the simplest case, the operator $\mathcal{P}_j$ can be parametrized in terms of projectors $\hat P_{j,\alpha}$ on the local Fock states at each site (e.g., $\alpha=0,\uparrow,\downarrow$, $2$ for the Hubbard model), 
$\mathcal{P}_j(t)=\sum_{\alpha}\frac{\phi_{j,\alpha}(t)}{(P_{j,\alpha}^{0})^{1/2}} \hat P_{j,\alpha}$, with $P_{j,n}^{0}=\langle \psi_\text{sd}|\hat P_{j,\alpha}|\psi_\text{sd}\rangle$. The dynamics is determined by requesting stationarity of the real-time action $\mathcal{S}(t)=\int_0^t dt' \langle \psi(t') | (i\partial_{t'}-H(t'))|\psi(t')\rangle$. When evaluating this action, one uses the GA, assuming the two constraints $\langle \psi_\text{sd}|\mathcal{P}_j^\dagger \mathcal{P}_j |\psi_\text{sd}\rangle=1$ and $\langle \psi_\text{sd}|\mathcal{P}_j^\dagger \mathcal{P}_jc_{j,a}^\dagger c_{j,b}|\psi_\text{sd}\rangle=\langle \psi_\text{sd}|c_{j,a}^\dagger c_{j,b}|\psi_\text{sd}\rangle$. With the GA, the expectation value of the kinetic energy in the state $|\psi(t)\rangle$ can be evaluated from the Slater determinant $|\psi_\text{sd}\rangle$ and a renormalized hopping Hamiltonian ${H_0}_*$ in which the operators $c_{j,a}$ are modified by a renormalization factor. The stationarity condition for the action translates into a set of coupled differential equations for the variational parameters $\phi_{j,\alpha}(t)$, while  $|\psi_\text{sd}(t)\rangle$  evolves according to a  single-particle Schr\"odinger equation with the renormalized single-particle Hamiltonian $H_{0*}$ \cite{Sandri2013,Fabrizio_springer}.

The time-dependent Gutzwiller approach yields qualitative agreement with nonequilibrium DMFT simulations for the short-time dynamics after an interaction quench in the Hubbard model \cite{Schiro2010}. It has also been applied to open quantum systems \cite{Lanata2012}, quenches in multi-orbital systems \cite{Behrmann2013,Oelsen2011}, the dynamics of antiferromagnetic \cite{Sandri2013} and superconducting order~\cite{Mazza2017}, and to photo-induced metal-insulator transitions (Sec.~\ref{sec:dynamical}).  
A main drawback is that the time-dependent Gutzwiller approach cannot describe thermalization. 

{\it Time-dependent variational quantum Monte Carlo ---}
While the Gutzwiller approach is semi-analytical,  more general variational wave functions have been used in numerical studies of strongly correlated nonequilibrium systems. To propagate a wave function $|\psi(\theta)\rangle $ depending on a vector ${\bf \theta}=\theta_1,\theta_2,...$ of time-dependent variational parameters, one can use the variational equation~\cite{carleo2012,Haegeman2013,Ido2017} 
\begin{align}
\Big\langle \delta \psi(\bar\theta)\Big| \Big(1-\frac{|\psi(\theta)\rangle\langle \psi(\bar\theta)|}{\langle \psi(\bar\theta) | \psi(\theta)\rangle }\Big) (H-i\partial_t) \Big|\psi(\theta)\Big\rangle =0,
\end{align}
where $\bar\theta$ is the complex conjugate of $\theta$.
This generates a norm-conserving time evolution  by projecting the full evolution onto the variational manifold. The projected time evolution  is equivalent to the coupled equations of motion $i\dot\theta = S(\theta)^{-1}g(\theta)$ for the parameters $\theta\equiv(\theta_j)$, where $S(\theta)_{ij}=\partial_{\bar\theta_i}\partial_{\theta_j} \ln \langle \psi(\bar\theta) | \psi(\theta) \rangle$ and $g_j=\partial_{\bar{\theta}_j} \frac{\langle  \psi(\bar\theta) | H|\psi(\theta)\rangle }{\langle \psi(\bar\theta) | \psi(\theta)\rangle}$. 

This formulation has been used in 
variational QMC simulations. For the wave function, one typically considers an ansatz similar to \eqref{Gansatz}, where $|\psi_\text{sd}\rangle$ is a pair-product wave function $\prod_{ij} e^{f_{ij}(t) c_i^\dagger c_j^\dagger} |0\rangle$  with variational parameters $f_{ij}$, and the $\mathcal{P}$ are operators of the Gutzwiller and Jastrow type. The differential equation for $\theta$ is then solved by Monte Carlo sampling the expectation values of $S$ and $g$.
Benchmarks of the short-time evolution  in small Hubbard clusters after a quench have shown good agreement with exact diagonalization and MPS calculations \cite{Ido2015}, and the method has been used to study the dynamics of the Hubbard model \cite{Ido2017,Orthodoxou2021} and Kondo lattice model \cite{Fauseweh2020}.

\subsection{Density-functional theory (DFT) based methods}  \label{sec:DFT}
 
Density functional theory (DFT) in combination with the local density approximation (LDA) \cite{Kohn1965} is a highly successful theory for weakly correlated materials, but fails to describe Mott insulators. 
The time-dependent version of DFT (TDDFT) represents the density $n(r,t)=\sum_j \rho_j | \langle r | \psi_j(t)\rangle |^2$ in a set of single-particle Kohn-Sham orbitals $|\psi_j\rangle$ (with occupations $\rho_j$), and solves the  time-dependent Kohn-Sham equations \cite{Runge1984}
\begin{align}
i\partial_t |\psi_j(t)\rangle &= H[n](t)|\psi_j(t)\rangle.\label{eq_LDA}
\end{align}
Besides the kinetic term, the Hamiltonian in Eq.~\eqref{eq_LDA} contains the external potential $V_\text{ext}(t)$ describing the effects of the nuclei and possible external fields, the Hartree potential $V_H[n(r,t)]$, and the exchange-correlation potential $V_\text{xc}[n(r,t)]$.

By adding a local Hubbard $U$ to the most strongly correlated orbitals (DFT+$U$) \cite{Liechtenstein1995,Anisimov1997}, a better description of strongly correlated materials can be obtained, although a frequency-dependent self-energy with an imaginary part would be needed for a proper modeling of paramagnetic Mott states. In conventional DFT+$U$ studies, the parameters $U$ and $J_\text{H}$ (Hund coupling) are either determined for a given ground state, or treated as empirical parameters, while out of equilibrium one can expect transient changes of the screening environment (Sec.~\ref{sec:dynamical}). A possible way to include this screening is the ACBN0 functional \cite{Agapito2015}, in which $U$ and $J_\text{H}$ are defined self-consistently  in terms of the density. The ``+$U$'' contribution has the standard form \cite{Dudarev1998} 
$E_U[\{\rho_{mm'}^\sigma\}] = \frac{U-J_\text{H}}{2}\sum_{m,m',\sigma} (\delta_{mm'}-\rho_{m'm}^\sigma) \rho_{mm'}^\sigma$, 
where the sum is over a set of localized orbitals $\{\phi_m^\sigma \}$, and $\rho^\sigma$ is the density matrix of this subspace. As in DFT+$U$, this correction is constructed to match the local Hartree-Fock energy (modulo double counting corrections). However, while in DFT+$U$ the latter is defined in terms of partially screened Coulomb matrix elements, 
the ACBN0 functional uses bare matrix elements and instead renormalizes the density matrix by a factor measuring the overlap of the occupied orbitals 
with the target space. This defines $U$ and $J_\text{H}$ as functionals of the density (for explicit expressions, see \onlinecite{Agapito2015}). By taking the derivative of $E_U$ with respect to $\rho^\sigma_{mm'}$ one obtains an extra potential $V_U^\sigma$, which is added to $H[n]$ in Eq.~(\ref{eq_LDA}).

The ACBN0 functional improves the standard DFT results for transition metal oxides \cite{Agapito2015}. When used  within TDDFT \cite{Tacogne2017} an additional approximation is the adiabatic approximation, but results for equilibrium optical properties show an improvement over TDDFT+$U$. This approach has also been used to estimate the effective on-site screening \cite{Tacogne2018}. 

An alternative strategy for incorporating strong correlation effects into (TD)DFT calculations is to construct an exchange-correlation functional $V_\text{xc}$ from equilibrium DMFT solutions of a lattice model. This functional and the adiabatic approximation can then be used  to compute the time evolution  of the model within TDDFT \cite{Verdozzi2008,Karlsson2011,Bostroem2019}.

\subsection{Nonequilibrium quasi steady state approaches}
\label{sec:quasiseq}

In large-gap Mott insulators, the recombination time of photo-generated charge carriers can be much longer than the time needed for intra-band thermalization.
Quasi steady state approaches 
approximate the resulting slowly-evolving state 
at a given instance of time by a true steady state. In band insulators, this steady state is often well described by 
a Gibbs state with separate chemical potentials for the holes in the valence band and the electrons in the conduction band~\cite{haug_quantum_1990,Keldysh1986review}. 
In large gap Mott insulators, one can expect a similar situation, but with holons and doublons instead of holes and electrons.   

More generally, we may assume that in addition to the energy $E_{\rm tot}$ and particle number $N$, there are macroscopic quantities $O$ (we write only one for simplicity) which are almost conserved. $O$ could represent, e.g., the conduction band occupation in a semiconductor, or the doublon number in a Mott insulator. 
If $O$ were exactly conserved, the steady state would be a universal function of $E_{\rm tot}$, $N$, and $\langle O\rangle$, usually taken to be a generalized Gibbs ensemble with a generalized chemical potential $\mu_{O}$ conjugate to $O$ \cite{Jaynes1957b,Rigol2007}. In situations where $\langle O(t)\rangle$ evolves sufficiently slowly, the dynamics can result in so-called prethermal states, 
which 
at a given time $t$ still have a universal form depending only on the conserved quantities  ($E_{\rm tot}$, $N$) and on the slow variable $\langle O(t)\rangle$ (see also Sec.~\ref{sec:general_comments_hidden}).
For the Hubbard model, the emergence of such a universal description is seen explicitly in long-time DMFT simulations \cite{Dasari2020}, which yield distribution functions that depend only on few parameters (Fermi functions with separate chemical potentials in the range of the upper and lower Hubbard band).

While the hypothesis of a universal time-evolving state is not new, the question is how to construct such a state for a strongly correlated quantum system. There are two complementary strategies, the quasi equilibrium approach and the quasi NESS approach, which we will briefly discuss with a focus on photo-doped Mott states in the Hubbard model.

{\it Quasi equilibrium approach---}
The quasi equilibrium approach approximates the quasi steady state as an equilibrium state of an approximate Hamiltonian $H_{\rm eff}$, which conserves $O$~\cite{Takahashi2002PRB,Rosch2008,Ishihara2011PRL,Murakami2022}. In this description $\langle O \rangle$ can be fixed by introducing a generalized chemical potential $\mu_O$. 
For a Mott-Hubbard system with hopping $v_0 \ll U$, the slow variables are the doublon and holon numbers, which correspond to 
$\hat{N}_{\rm doub} = \sum_i \hn_{i,d}$ and $\hat{N}_{\rm holon} = \sum_i \hn_{i,h}$, dressed by virtual charge fluctuations. 
Here $\hn_{i,d}=\hn_{i\uparrow}\hn_{i\downarrow}$ and $\hn_{i,h} =(1-\hn_{i\uparrow})(1-\hn_{i\downarrow})$.
The effective Hamiltonian $H_{\rm eff}$ in terms of these dressed doublons and holons can be obtained by the  unitary Schrieffer-Wolff (SW)  transformation $\hat{H}_{\rm SW} = e^{i\hat{S}}\hat{H}e^{-i\hat{S}}$ \cite{MacDonald1988PRB}, where the generator $\hat{S}$ is chosen  to eliminate from $\hat{H}_{\rm SW}$, up to a given order in $v_0/U$,  all recombination terms which change the doublon or holon number. 
For the $U$-$V$ Hubbard model \eqref{H_uvhubbard} with $U\gg v_0, V$, $H_{\rm eff}$ up to $\mathcal{O}(v^2_0/U)$ is given by \cite{Murakami2022}
\eqq{
 \hH_{\rm eff}  =& \hH_U + \hH_{\rm kin,holon} +  \hH_{\rm kin,doub} + \hH_V \nonumber\\
 &+  \hH_{\rm spin,ex} +  \hH_{\rm dh,ex} + \hH_{U,\rm{shift}} + \hH_{\rm 3-site}.
 \label{eq:Heff}
}
Here, $ \hH_U = U \sum_i \hn_{i\uparrow}\hn_{i\downarrow}$, $\hH_V = \frac{V}{2} \sum_{\langle i,j\rangle} \hn_i \hn_j$ and $ \hH_{\rm kin,holon}$ and $ \hH_{\rm kin,doub}$ describe the hopping of holons and doublons of $\mathcal{O}(v_0)$, respectively. The remaining terms are of $\mathcal{O}(v^2_0/U)$: $ \hH_{\rm spin,ex}=\frac{J_{\rm ex}}{2}\sum_{\langle i,j\rangle} {\bf S}_i\cdot{\bf S}_j$ is a Heisenberg spin exchange between singly occupied sites. 
$\hH_{\rm dh,ex}=-\frac{J_{\rm ex}}{2}\sum_{\langle i,j\rangle} {\boldsymbol \eta}_i\cdot{\boldsymbol \eta}_j$ is the analogous exchange for doublons and holons. Here, the $\eta$ spins are defined as 
\begin{equation}
\eta^+_i = (-)^i c^\dagger_{i\downarrow} c^\dagger_{i\uparrow}, \,\, \eta^-_i = (-)^i  c_{i\uparrow} c_{i\downarrow}, \,\, \eta^z_i = \frac{1}{2} (n_i-1).
\label{def_eta}
\end{equation}
$\hH_{U,\rm{shift}}$ describes a shift of the local interaction, and $\hH_{\rm 3-site}$ consists of three-site terms such as correlated doublon hoppings. Model~\eqref{eq:Heff} is a natural extension of the $t$-$J$ model~\eqref{H_tj}, which is obtained by restricting $H_\text{eff}$ to the sector with $\hat{N}_{\rm doub}=0$ (hole doping) or $\hat{N}_{\rm holon}=0$ (electron doping), and ignoring $\hH_{\rm 3-site}$. To explore the steady state, one may then study the equilibrium properties of Eq.~\eqref{eq:Heff} for a given density of doublons and holons. In the thermodynamic limit, this can be done by introducing separate chemical potentials for the doublons and holons. Some applications of this formalism can be found in Sec.~\ref{sec:hidden_phases}.

{\it Quasi NESS approach---}
An alternative approach is to approximate the quasi steady state by coupling suitable particle and energy reservoirs to the system, with some overall coupling strength $g$. 
The resulting state is a NESS, i.e., its properties are time-independent, while the spectral and distribution functions do not necessarily satisfy the universal fluctuation-dissipation relation. 
However, if $O$ is conserved, the limit $g\to 0$ should give the universal Gibbs state, where the effect of the bath is merely to fix the thermodynamic quantities $E_{\rm tot}$, $N$, and $\langle O\rangle$. 
For almost conserved $O$, already a weak bath coupling is sufficient to control $\langle O\rangle$ independently from the other thermodynamic variables. 

The strategy of controlling the values of almost conserved quantities was developed as a general framework for systems close to integrability~\cite{lange2017}. 
In the context of photo-exited Mott insulators, \onlinecite{Jiajun2021PRB} proposed to approximate the slowly evolving photo-doped state by coupling the system to particle reservoirs. If fully occupied (empty) electron baths are weakly coupled to the upper (lower) Hubbard band, a NESS with a given $E_{\rm tot}$, $N$, and density of doublons and holons is established. The nonthermal occupations in the system are  maintained by a small current from occupied to unoccupied states, while energy can be dissipated to the electron baths, or to additional bosonic heat baths. 
The NESS approach was applied to photo-excited Mott insulators,  
with reasonable agreement of the spectral and distribution functions  to  time-dependent simulations~\cite{Jiajun2021PRB}. 
In the phase diagram of the photo-doped Hubbard model, the NESS approach predicts the emergence of $\eta$-pairing~\cite{Li2020}, consistent with the quasi equilibrium approach (Sec.~\ref{sec:super_eta}). 

%%%%%%%%%%%%%%%%%%%%%%%%%%%%%%%%%%%%%%%%%%%%%%%%%%%%%%%%%%%%%%%%%%%%%%%%
\section{Mott insulators in static electric fields}
\label{sec:static_fields}
%%%%%%%%%%%%%%%%%%%%%%%%%%%%%%%%%%%%%%%%%%%%%%%%%%%%%%%%%%%%%%%%%%%%%%%%

In this section, we discuss the effects of static electric fields on Mott insulators. On the one hand, strong fields modify the kinematics of the charge carries (Sec.~\ref{eq:WS_local}), which also affects the low-energy spin physics (Sec.~\ref{sec:effective_jex}). On the other hand, they create charge carriers via quantum tunneling, which can result in a metallic state (Sec.~\ref{sec:breakdown}). We also discuss the mobility of charge carriers in strong fields, where the energy dissipation plays a crucial role (Sec.~\ref{mobility}).

\subsection{Field-induced localization} \label{eq:WS_local}

In the presence of a static (DC) electric field $E_0$, a charge carrier with charge $q$
(e.~g., a conventional band electron, or a doublon in a Mott insulator), which hops $n$ times the lattice spacing $a$ in the direction of the field, gains an energy $nqE_0a$. If this energy exceeds the width of the (Hubbard) band $W$, the energy absorbed from the field  cannot anymore be converted into kinetic energy. Hence, the charge carrier cannot move farther in the direction of the field, unless it can pass on energy to other degrees of freedom. 
For free electron systems, this results in Bloch oscillations \cite{Bloch1929,Zener1934,Kruchinin2018}, where carriers move back and forth over a distance $na\sim W/qE_0$, and thus get localized in the direction of the field. This explains the phenomenon of field-induced dimensional reduction \cite{Aron2012}. 

The energy of the localized state depends on its central position. 
Thus, the lattice system with static field exhibits an infinite series of localized states whose energy is shifted by $lqE_0a$ (the Wannier-Stark ladder), where $l$ corresponds to the site at which the state is localized. 
This  ladder structure is directly evident in the single-particle spectrum,
which can split into a series of $\delta$-peaks separated by $\Omega = qE_0a$ for noninteracting systems. Analytic expressions can be obtained, e.g., for the hypercubic lattice \cite{Tsuji2008}.

%%%%%%%%%%%%%%%%%
\begin{figure}[t]
\begin{center}
\includegraphics[angle=0, width=\columnwidth]{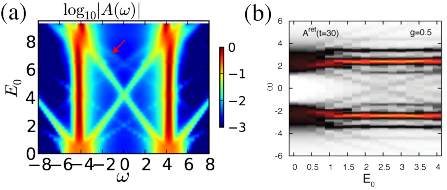}\hfill
\caption{
(a)~Spectral function of the Hubbard model with $U=8$ under a static electric field with amplitude $E_0$ (hypercubic lattice, $v_*=1$, $q=a=1$). The arrow marks the spectra weight at $\frac{3}{2}U - 2qaE_0$. (b)~Corresponding result for the Holstein-Hubbard model with $U=5$, $\omega_0=1$ and $g=0.5$. Both results were obtained by DMFT+NCA.~(From \onlinecite{Werner15epl,Murakami2018PRB}.) 
}
\label{fig:WS_DC}
\end{center}
\end{figure}   
%%%%%%%%%%%%%%%%%

In Mott insulators, DC fields result in a similar localization of charge carriers~\cite{Eckstein2013heating,Park2014,Murakami2018PRB,Udono2023}. 
Again, the localization effects can be observed in the spectral function. Figure~\ref{fig:WS_DC}(a) shows DMFT+NCA results for an initially Mott insulating Hubbard model on the infinite-dimensional hypercubic lattice 
(Gaussian noninteracting density of states $\rho(\epsilon)=\frac{1}{\sqrt{\pi}v_*}\exp(-\epsilon^2/v_*^2)$), 
obtained by applying the Floquet DMFT formalism to this steady-state problem in the dipolar gauge. 
Due to the localization, the Hubbard bands get narrowed compared to the field-free case and we notice the emergence of side-bands, which are split off from the original band center by $\pm lqaE_0$. The side peak at $\omega=U/2\pm lqaE_0$ is associated with adding an electron at site $i$ into a many-body Wannier-Stark state localized at site $i\pm l$. 
In this interacting system, 
the charge carriers can transfer excess potential energy to other particles or spin-excitations, and the localization is not perfect. This is reflected in the width of the Wannier-Stark bands. In addition, the many-body system can produce signals which are absent in free systems. For example, the faint spectral weight at $\frac{3}{2}U - 2qaE_0$, highlighted by the arrow in Fig.~\ref{fig:WS_DC}(a), represents a process where the insertion of one electron at a given site leads to the creation of a doublon and the simultaneous creation of another doublon-holon (D-H) pair. 

Recently, \onlinecite{Udono2023} studied the signature of the Wannier-Stark localization in the optical conductivity of the one-dimensional (1D) Hubbard model using iTEBD. They demonstrated the emergence of multiple peaks in the optical conductivity and the reduction of the optical gap under the dc electric field. The latter can be regarded as the Franz-Keldysh effect in this Mott system.

Furthermore, since the DC field quenches the kinetic energy, it can enhance local interaction effects. 
Figure~\ref{fig:WS_DC}(b) illustrates the three effects of a static electric field in the Holstein-Hubbard model: (i) the narrowing of the main Hubbard band for $qaE_0\gtrsim W$, (ii) the appearance of Wannier-Stark sidebands which are split off from the main Hubbard bands by an energy $\pm lqaE_0$, and (iii) the appearance of phonon sidebands with energy splitting $\pm n\omega_0$. In this simulation, as is representative for typical electron-phonon systems, the ratio $g/W$ between the phonon coupling and bandwidth is too small for the appearance of well-resolved phonon sidebands in equilibrium. 
However, if the field-induced localization narrows the bands, the relative coupling strength is enhanced and phonon sidebands can be clearly resolved in the nonequilibrium spectral function. The measurement of these sidebands would allow to extract the phonon energy $\omega_0$. Similarly, strong DC fields applied to multi-orbital systems produce well-defined multiplet peaks in the single-particle spectrum, from which the Hund coupling $J_\text{H}$ could be deduced~\cite{Dasari2020}.

\subsection{Effective spin exchange}
\label{sec:effective_jex}

\begin{figure}[t]
\begin{center}
\includegraphics[angle=0, width=0.9\columnwidth]{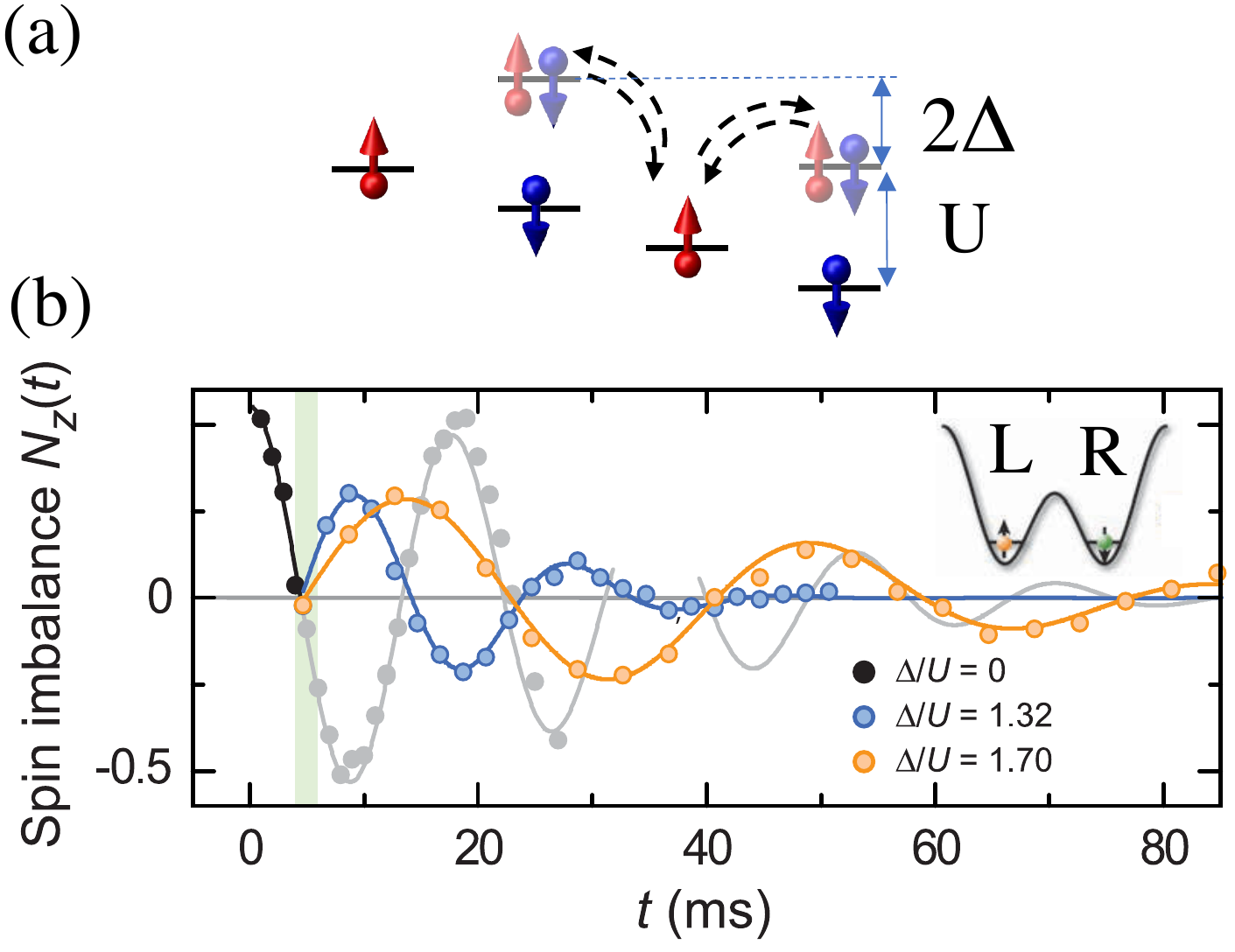}
\caption{(a) Real-space energy level structure of a Mott insulator with interaction strength $U$ in a DC field corresponding to the nearest neighbor energy difference $\Delta$. Dashed arrows indicate second-order virtual hopping processes.
(b) Time evolution of the spin imbalance $N_z(t)$ for different field strengths in a cold atom experiment on a two-site Bose Hubbard system.
The initial state is shown in the inset. The oscillation of $N_z(t)$ can be related to the value of the spin exchange coupling. The hopping-to-interaction ratio is $v_0/U=0.08.$
(From \onlinecite{Trotzky2008}.)
}
\label{fig:DC_spin}
\end{center}
\end{figure}

In the previous section, we showed that the energy levels of states with doublons and holons are strongly modified by a DC field.
Since such states are involved in the virtual excitation processes underlying the spin exchange mechanism, the DC electric field also modifies the spin exchange couplings~\cite{Trotzky2008,Takasan2019PRB}. 
The basic idea can be understood by considering the 1D Hubbard model in the length gauge,
\eqq{
H = - v_0 \sum_{\langle i,j\rangle\sigma} c^\dagger_{i\sigma}c_{j\sigma} + U\sum_i n_{i\uparrow}n_{i\downarrow} -q a E_0 \sum_i i n_i.
}
In this case, the energy level difference between neighboring sites becomes $\Delta\equiv qaE_0$. If we assume a Mott insulating phase and consider a Hilbert space consisting of states with only singlons, then second order perturbation theory (see Fig.~\ref{fig:DC_spin}(a)) yields the spin Hamiltonian
\eqq{
H_{\rm spin} =  \sum_{\langle i,j\rangle} \Bigl[\frac{v_0^2}{U-\Delta} +  \frac{v_0^2}{U+\Delta}\Bigl] {\bf S}_i \cdot {\bf S}_j.
\label{H_spin}
}
The first~(second) term is the contribution from the virtual excitation of an electron along~(opposite to) the field direction, see Fig.~\ref{fig:DC_spin}(a). This expression implies that for  $U>\Delta$, the spin exchange coupling is enhanced, while for $U<\Delta$ its sign flips. 
The spin Hamiltonian \eqref{H_spin} describes the physics of the spin sector if the excitation of long-lived D-H pairs by the field  (see Sec.~\ref{sec:breakdown}) can be neglected.   
In the dipolar gauge (Sec.~\ref{sec:coupling_to_light}), the vector potential for a DC field is $A(t)=-E_0 t$. Hence, the Hubbard model becomes time periodic with frequency $\Omega = qE_0a$, 
and Eq.~\eqref{H_spin} can be obtained as a Floquet effective Hamiltonian, see Sec.~\ref{sec:ac_fields}. 

%%%%%%%%%%%%%%%%%
%%%%%%%%%%%%%%%%%

Similar ideas for modifying the spin exchange coupling have been discussed both for bosonic systems~\cite{Trotzky2008,Dimitrova2020PRL,Sun2021} and fermionic systems \cite{Takasan2019PRB,Furuya2021PRR,Takasan2021arxiv}.
The bosonic case was considered in a cold atom context, where
the spin model can be derived from the two-component Bose Hubbard model in the strong coupling regime.
It becomes ferromagnetic at $\Delta=0$, unlike the fermionic Hubbard model, and it becomes antiferromagnetic for $\Delta > U$.
For a two-site system, such a sign change of the spin exchange coupling has been demonstrated experimentally~\cite{Trotzky2008}. 
The experiment measured the spin imbalance $N_z = (n_{\uparrow L}+n_{\downarrow R}-n_{\uparrow R}-n_{\downarrow L})$ between the left (L) and right (R) site, 
starting from an initial product state as shown in the inset of Fig.~\ref{fig:DC_spin}(b). The observed pseudospin precessional dynamics depends on the value of the exchange coupling.
For $\Delta>U$, 
$N_z$ exhibits an opposite dynamics, compared to the system with $\Delta = 0$, which demonstrates the sign change of the exchange coupling, see Fig.~\ref{fig:DC_spin}(b).
A modification of the exchange coupling based on this idea has also been experimentally realized in extended systems~\cite{Dimitrova2020PRL}.
For fermionic systems, the modification of magnetic superexchange couplings using DC fields has been discussed for the $d$-$p$ model \cite{Furuya2021PRR}.
Control over the exchange couplings is helpful for realizing exotic spin phases which are hard to access in equilibrium, 
such as the antiferromagnetic state in {\it bosonic} systems~\cite{Sun2021} or a chiral spin liquid state in a frustrated fermionic system~\cite{Schultz2023}.

\subsection{Dielectric breakdown} 
\label{sec:breakdown}

\begin{figure}[t]
\begin{center}
\includegraphics[angle=0, width=\columnwidth]{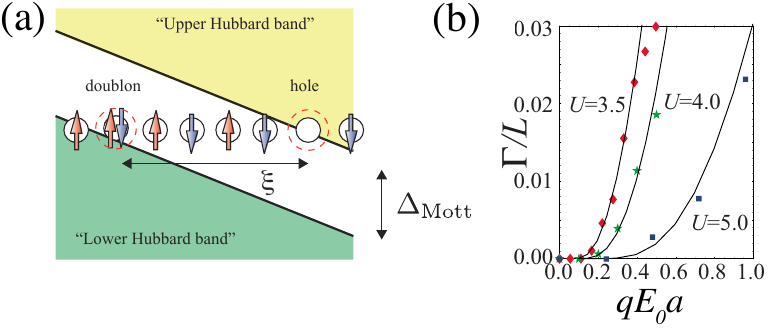}
\caption{
(a) Sketch of the dielectric breakdown of a Mott insulator in a static electric field $E_0$.
The Hubbard bands are tilted because of the potential $V(x) = -qE_0x$. 
$\Delta_\text{Mott}$ denotes the Mott gap and $\xi$ the characteristic size of a D-H pair.
  (b) Decay rate of the ground state $\Gamma$ 
  calculated with time-dependent DMRG (symbols) for a 1D system of length $L$. The solid lines indicate the analytical results from Eqs.~\eqref{dielec_break} and \eqref{eq_threshold}.
  (From \onlinecite{Oka2012} and \onlinecite{Oka2010}.) 
}
\label{fig_tunneling}
\end{center}
\end{figure}

The destruction of an insulating state by a static electric field is called dielectric breakdown.
A well-studied scenario in the context of Mott insulators is the creation of charge carriers via quantum tunneling. 
In a Mott insulating single-band Hubbard model, the tunneling induced by a static field $E_0$ produces D-H pairs.
This process is illustrated in the length gauge in Fig.~\ref{fig_tunneling}(a),  which 
sketches the lower and upper Hubbard bands that are tilted by the potential $V(x)=-qE_0x$ and separated in energy by the Mott gap $\Delta_\text{Mott}$. If a D-H pair is created on neighboring sites, and
then  separated by a distance $na$, 
the energy  $n q E_0a$ is absorbed from the field. If this energy exceeds the gap $\Delta_\text{Mott}$, D-H pairs can be freely generated, similar to the Schwinger mechanism for electron-positron production \cite{Schwinger1951}, or the Zener breakdown of band insulators \cite{Oka2005}. 

Numerical and analytical calculations \cite{Oka2003,Oka2005,Oka2010,HeidrichMeisner2010,Oka2012,Eckstein2010breakdown,Tanaka2011,Lenarcic2012,Eckstein2013heating,Takasan2019arxiv}
demonstrated that the D-H production rate $\Gamma$ follows a threshold behavior 
\begin{equation}
\Gamma\propto E_0 \exp\Big(-\frac{\pi E_{0,\text{th}}}{E_0}\Big)
\label{dielec_break}
\end{equation} 
with some threshold field $E_{0,\text{th}}$. It is generally expected that  the threshold field is related to the Mott gap $\Delta_\text{Mott}$ and the 
D-H correlation length $\xi$ in equilibrium as 
\begin{equation}
E_{0,\text{th}} \sim \frac{\Delta_\text{Mott}}{2q\xi}.
\label{eq_threshold}
\end{equation}
This expression was analytically derived for the 1D Hubbard model \cite{Oka2010,Oka2012} by applying Laundau-Dykhne tunneling theory for the D-H pair states, and using the Bethe ansatz solution for the non-Hermitian Hubbard model~\cite{Fukui1998}.  
Expressions~\eqref{dielec_break} and \eqref{eq_threshold} show that a field $qE_0 \xi \sim\Delta_\text{Mott}$ is required to turn the fluctuations of D-H pairs into real excitations. 
In connection with experiments, the dependence of  $E_{0,\text{th}}$ on the gap is important. For the 1D Hubbard model with small gap, $E_{0,\text{th}}\propto \Delta_{\rm Mott}^2$, analogous to the Landau-Zener result~\cite{Oka2003}. 
A different scaling $E_{0,\text{th}}\propto \Delta_{\rm Mott}^{3/2}$ has been reported for the fully spin polarized 1D Hubbard model~\cite{Lenarcic2012}.
For the infinite-dimensional case, a DMFT+NCA analysis showed $E_{0,\text{th}}\rightarrow 0$ when the system approaches the metal-insulator crossover regime, as expected from Eq.~(\ref{eq_threshold})~\cite{Eckstein2010breakdown}.
Figure~\ref{fig_tunneling}(b) shows the ground state decay rate, which is related to the D-H production rate ($|\langle \psi(t) | \psi(0)\rangle|^2\sim e^{-\Gamma t}$), from time-dependent DMRG simulations of a 1D Mott insulator in a constant electric field switched on at $t=0$ \cite{Oka2010}. The decay rate shows a threshold behavior and matches the analytical results from Laundau-Dykhne tunneling theory.

%%%%%%%%%%%%%%%%%
%%%%%%%%%%%%%%%%%

Strictly speaking, expression~\eqref{dielec_break} is expected to hold asymptotically for small $E_0$. 
For large $E_0$, one should consider the reconstruction of the spectral function and the emergence of in-gap states due to the development of the Wannier-Stark ladder, as discussed in Sec.~\ref{eq:WS_local}. In particular, for large $U$, the resonant enhancement of the pair creation at $q E_0 a=U/n$ ($n$ integer) is clearly observed in the DMFT simulations~\cite{Eckstein2013heating}. A related enhancement of the doublon/holon density has also been reported in the steady state of a Mott insulator under DC fields~\cite{Aron2012b,Park2014,Murakami2018PRB}.  

Experimentally, the dielectric breakdown of Mott insulators induced by DC fields has been extensively studied \cite{Taguchi2000,Janod2015}. In many experiments, dielectric breakdown is observed for much weaker electric fields than predicted by the quantum tunneling analysis. This low threshold has been attributed to other mechanisms, such as filament formation and avalanche effects \cite{Guilot2013}, efficient carrier generation from impurity levels~\cite{Diener2018,Kalcheim2020}, or a simultaneous structural phase transition~\cite{nakamura2013}. Theoretical scenarios for a transition induced by weak fields consider the possibility of a gap collapse in the metal-insulator coexistence region \cite{Mazza2016} or the inhomogeneous properties of resistor networks \cite{Stoliar2013}.  

%%%%%%%%%%%%%%%%%
\begin{figure}[t]
\begin{center}
\includegraphics[angle=0, width=\columnwidth]{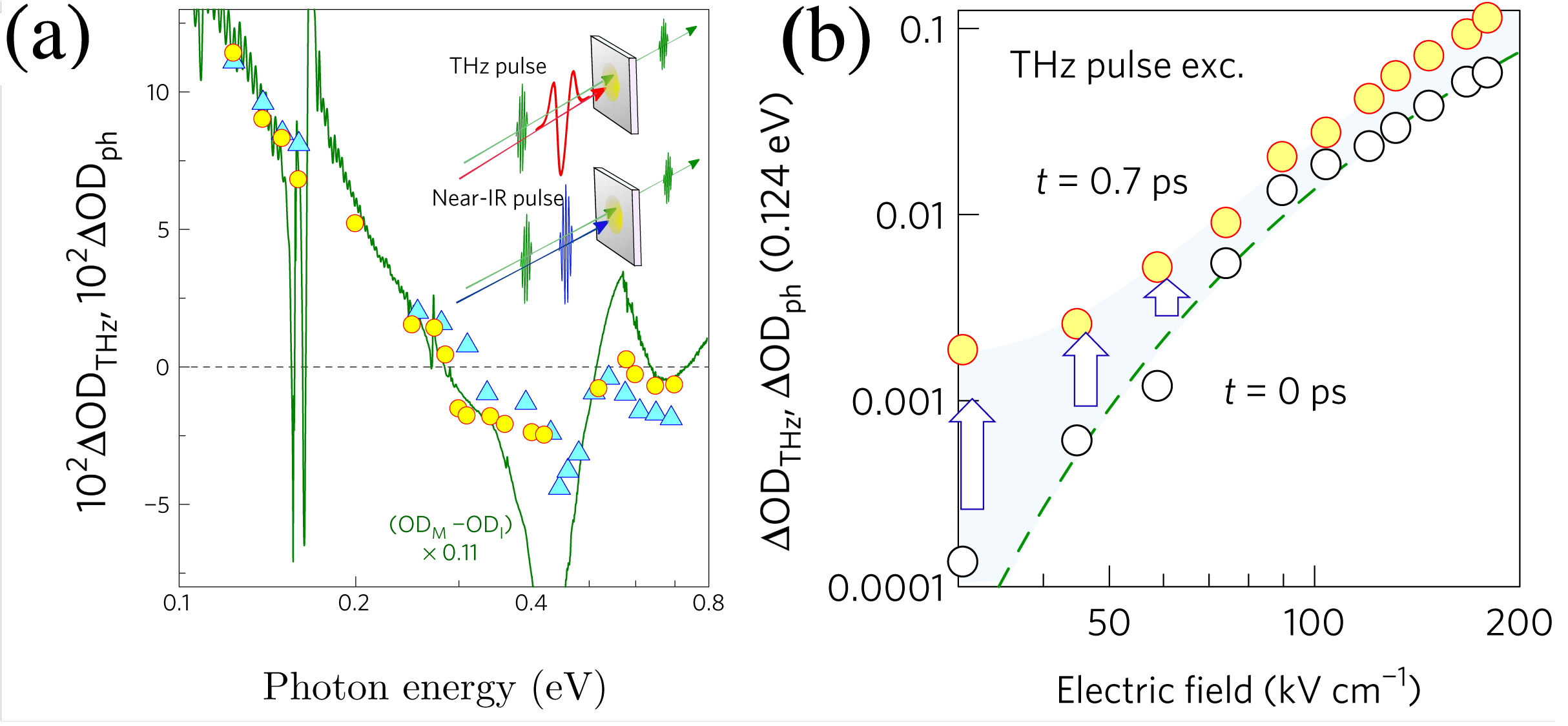}
\caption{(a) Transient absorption changes at $t=0.7$~ps induced by a THz pulse $\Delta$OD$_{\rm THz}$~(circles) and near-IR light pulse $\Delta$OD$_{\rm ph}$~(triangles) with frequency $0.93$~eV. 
The green solid line shows the difference of the absorption spectra of the Mott insulating and metallic phase in equilibrium.
(b) Transient absorption change $\Delta$OD$_{\rm THz}$ at delay $t=0$~ps~(empty) and $t=0.7$~ps~(yellow) at photon energy $0.124$~eV versus electric field strength $E_0\equiv E_{\rm THz}(t=0)$.
  The green broken line is a fit to Eq.~\eqref{dielec_break}.
(From \onlinecite{Yamakawa2017}.) 
}
\label{fig:THz_tunnel}
\end{center}
\end{figure}   
%%%%%%%%%%%%%%%%%

To see the physics induced by DC fields, one may practically use THz field, which vary slowly on electronic timescales. \onlinecite{Liu2012} studied the dielectric breakdown induced by strong THz fields in VO$_2$. The insulator-metal transition occurs on a timescale of several picoseconds and is associated with carrier generation from impurities and heating caused by carrier acceleration in the field. 
More recently, the dielectric breakdown of a Mott system has been investigated using THz-pump optical-probe spectroscopy~\cite{Yamakawa2017,Takamura2023PRB}. In \onlinecite{Yamakawa2017}, the 2D system $\kappa$-(ET)$_2$Cu[N(CN)$_2$]Br ($\kappa$-Br) on a diamond substrate, with a Mott gap of approximately 30~meV, was subjected to a strong pump pulse with a frequency of about 1~THz. By observing changes in the optical density (OD) in the mid-infrared range, it was possible to trace the insulator-metal transition, see Fig.~\ref{fig:THz_tunnel}(a). The field-strength dependence of the change in the OD at $t = 0$ (at the peak of the THz pulse) was found to be consistent with the tunneling formula \eqref{dielec_break} with $E_{0,{\rm th}}=64$~kVcm$^{-1}$, see Fig.~\ref{fig:THz_tunnel}(b). The corresponding correlation length $\xi=\Delta_{\rm Mott}/2qE_{0,{\rm th}}$ is 23~\AA, which is a reasonable value for this system. At $t=0.7$~ps, the change in the OD no longer follows the tunneling formula, which was explained by the collapse of the Mott gap.

\subsection{Mobility of charge carriers in strong fields}
\label{mobility}

In this section, we discuss how the charge carriers contribute to the current in strong DC fields. In closed systems, the DC field in Mott insulators leads to a broad energy distribution~(high temperature) of the generated doublons and holons~\cite{Oka2012,Eckstein2013heating}, 
see also Sec.~\ref{sec:single_multi}. Such ``hot" doublons and holons hardly contribute to the current, which results in a very small conductivity. This is in contrast to photo-doped systems, where photo-doped populations with a relatively narrow energy distribution and finite conductivity can be realized. Besides the temperature effect, recent studies emphasized that kinetic constraints due to the high fields can lead to a transient nonergodic evolution~\cite{scherg2021,desaules2021,kohlert2023} and reduced conductivity~\cite{Shinjo2022}. 
In the strong-field regime, the mobility of the carriers in the field direction is limited by their ability to dissipate the energy gained from the field, which leads to a negative differential resistivity~\cite{Mierzejewski2011,Vidmar2011}. Thus, the coupling between the charge carriers and dissipation channels, which leads to cooling and the breaking of kinetic constraints, is crucial for understanding the current induced by DC fields.  

%%%%%%%%%%%%%%%%% 
\begin{figure}[t]
\begin{center}
\includegraphics[angle=0, width=0.75\columnwidth]{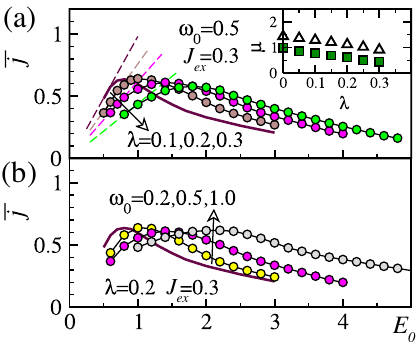}
\caption{
ED results for the quasi-steady current $\bar j$ of a single hole in the two-dimensional $t$-$J$-Holstein model with 
el-ph coupling  $\lambda=g^2/W\omega_0$ and exchange coupling $J=J_\text{ex}$.  
The inset in (a) shows how the mobility scales with $\lambda$. 
(Adapted from \onlinecite{Vidmar2011}.)
}
\label{fig_current_pulse}
\end{center}
\end{figure}
%%%%%%%%%%%%%%%%%

Possible dissipation channels in the single-orbital case are the interaction of doublons and holons with an antiferromagnetic spin background (see Sec.~\ref{sec:electron-spin}) and their coupling to phonons. 
The effects of the spin and phonon baths have been considered in the ED study by \onlinecite{Vidmar2011}, which simulated a single hole in the two-dimensional $t$-$J$ model with static electric field and additional coupling to Holstein phonons. Because of the energy dissipation, the current reaches a quasi-steady state value $\bar {\bf j}\equiv {\bf J}$(t$_{\text{max}}$) at the longest propagation time $t_{\text{max}}$ after an initial transient. Figure~\ref{fig_current_pulse} plots the value of $\bar j$ as a function of the field strength $E_0$. 
In the small-field regime, $\bar j$ increases linearly with $E_0$ and the slope
(conductivity) is plotted as a function of the electron phonon coupling  $\lambda$ in the inset. 
 The decrease in conductivity with increasing $\lambda$ is due to stronger phonon scattering, consistent with the linear-response result (triangles). 
 As the field strength is increased, the conventional positive differential resistivity gives way to a negative differential resistivity, i.e., the quasi-steady state current gets suppressed with increasing $E_0$. In this regime, the phonon coupling has the opposite effect: a stronger phonon coupling enables a more efficient energy dissipation and hence a higher mobility of the hole. For fixed $\lambda$, the current is furthermore enhanced for larger $\omega_0$, because the phonons can absorb larger energy quanta.

Comparing the effects of the phonons and the antiferromagnetic spin background on the dissipation, \onlinecite{Vidmar2011} concluded that the spin cooling is more effective for realistic parameters, as long as the antiferromagnetic correlations in the spin background remain significant.  
The energy transfer to the antiferromagnetic spin background is large, because in an antiferromagnetic state, each hopping breaks antiferromagnetic bonds and dissipates energy of the order of $J_\text{ex}$. In contrast, for small phonon coupling, there can be several hops before a phonon is emitted. 

%%%%%%%%%%%%%%%%%
\begin{figure}[t]
\begin{center}
\includegraphics[angle=0, width=0.9\columnwidth]{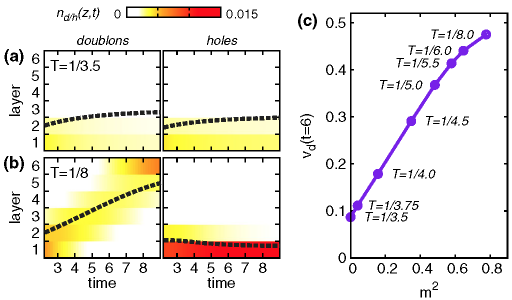}
\caption{
Layer-dependent doublon and holon densities in a 6-layer Mott insulating heterostructure within the high-temperature paramagnetic regime~(a) and the low-temperature antiferromagnetic regime~(b).
Panel (c) plots the drift velocity $v_d$ measured at time $t=6$ as a function of the staggered magnetization  squared. The results were obtained by DMFT+NCA.  
(From \onlinecite{Eckstein13layer}.) 
}
\label{fig_layer_driftvelocity}
\end{center}
\end{figure}   
%%%%%%%%%%%%%%%%%

The nonlinear strong-field transport behavior can also be relevant for strong {\it internal} fields. Such internal electric fields appear in polar heterostructures, and the separation of photo-doped doublons and holons in these fields has been discussed in connection with Mott insulating solar cells \cite{Assmann2013,Eckstein13layer,Petocchi2019}. Heterostructures can be simulated with the inhomogeneous extension of 
DMFT~\cite{Potthoff1999,Okamoto2004,Eckstein2013layer}.
 Also in this set-up, a fast diffusion and hence separation of the charge carriers is only possible in the presence of an efficient dissipation mechanism, since otherwise the strong field will localize the charge carriers \cite{Eckstein13layer}.  Mott insulating heterostructures with antiferromagnetic order have a built-in efficient dissipation channel (transfer of kinetic energy to the spin background) and are therefore particularly interesting for solar cell applications. The effect of the ordered spin background on the separation of doublons and holons is illustrated in Fig.~\ref{fig_layer_driftvelocity}, which plots the time evolution  of a 
 photo-induced doublon and holon population in a 6-layer heterostructure.  Panel (a) plots the evolution in a high-temperature paramagnetic structure. Here, the doublons and holons, which are produced by the pulse in layer 1, remain confined close to this surface layer, even though the heterostructure has a strong internal field that pushes the doublons in the direction of layer 6. 
 In the lower-temperature antiferromagnetic state (panel (b)), the mobility of the doublons is much higher, because they can now dissipate their kinetic energy to the spin background. The holons remain confined near layer 1, so that the doublons and holons are physically separated after just a few hopping times. The drift velocity of the doublons is plotted in panel (c) as a function of the squared staggered magnetization $m$. 
The quadratic scaling can be understood by considering that the spin energy is $\sim J_\text{ex}m^2$ and the energy dissipation rate is $\sim J_\text{ex}m(dm/dt)$ \cite{Eckstein13layer}.

%%%%%%%%%%%%%%%%%%%%%%%%%%%%%%%%%%%%%%%%%%%%%%%%%%%%%%%%%%%%%%%%%%%%%%%%
\section{Mott insulators in ac fields}
\label{sec:ac_fields}
%%%%%%%%%%%%%%%%%%%%%%%%%%%%%%%%%%%%%%%%%%%%%%%%%%%%%%%%%%%%%%%%%%%%%%%%

In this section, we discuss the AC driving of Mott insulators,
which includes the engineering of effective Floquet Hamiltonians under periodic driving, photo-carrier generation, nonlinear optical responses under strong driving fields (such as high-harmonic generation) and nonlinear phononics (for an overview of phenomena in periodically driven systems, see Sec.~\ref{sec:driven_overview}).

\subsection{Floquet Hamiltonian}\label{sec:Floquet}

The Floquet Hamiltonian provides useful insights into the effects of AC fields on Mott insulators. Let us assume that  the Hamiltonian is time periodic with frequency $\Omega$. Denoting the unitary time evolution operator from $t_0$ to $t$ by $\mathcal{U}(t,t_0)$, the Floquet Hamiltonian is defined  as the generator of the evolution $\mathcal{U}(t_0 + T_p,t_0)=e^{-i T_p \hH_{\rm Floq}[t_0]}$ over one period $T_p \equiv \frac{2\pi}{\Omega}$ (stroboscopic evolution). While this definition depends on the choice of $t_0$, the operators $\hH_{\rm Floq}[t_0]$ 
for different $t_0$ are unitarily equivalent  \cite{Bukov2015}. In general, it is hard to evaluate $\hH_{\rm Floq}$. However, when $\Omega$ is large compared to the typical energy scales of the system, the effective Hamiltonian can be approximated by a perturbative high-frequency expansion, which mainly comes in three flavors: i) the Floquet-Magnus expansion, where the Floquet Hamiltonian depends on the gauge freedom inherent in $\hH_{\rm Floq}[t_0]$~\cite{rahav2003}, ii) the van Vleck expansion, where the gauge freedom is removed by considering a non-stroboscopic Hamiltonian~\cite{Bukov2015}, and iii) the Brillouin-Wigner inverse frequency expansion~\cite{Mikami2016}. The latter is gauge invariant and often suitable for numerical studies, but it suffers from non-Hermitian high-order terms. Here, we proceed with the description of the van-Vleck formalism, where the effective Floquet Hamiltonian is given by 
\eqq{
\hH_{\rm eff} = \hH_0 + \sum_{l>0} \frac{[\hat{H}_l,\hat{H}_{-l}]}{l\Omega} + \mathcal{O}\Bigl(\frac{1}{\Omega^2}\Bigl). \label{eq:Magnus_exp}
}
$\hH_l$ is defined in terms of the Fourier expansion $\hH(t)=\sum_l \hH_l e^{il\Omega t}$. It describes the transition from $n$-photon dressed states to $(n+l)$-photon dressed states, so that $\hH_{\rm eff}$ takes  into account virtual fluctuations to photo-dressed states.

As a simple example, we consider the single-band Hubbard model \eqref{H_hubbard} in the Mott regime, with a hopping amplitude $v_{ij}(t)$ that oscillates with frequency $\Omega$. 
(This includes the driving with AC electric fields.) The high-frequency expansion can be used in two separate limits, leading to different effective Hamiltonians: 

{\it High-frequency regime ($\Omega\gg U,v_0$) ---}
When $\Omega$ is much larger than all system parameters, one can directly apply Eq.~\eqref{eq:Magnus_exp} to obtain the Floquet Hamiltonian. The lowest order Hamiltonian becomes
\eqq{
\hH_0 = -\sum_{\langle i,j\rangle,\sigma}  \mathcal{A}_{ij}^{(0)}\hc^\dagger_{i\sigma} \hc_{j\sigma}  + U \sum_j \hn_{j\uparrow} \hn_{j\downarrow} \label{eq:Heff_0},
}
where $\mathcal{A}_{ij}^{(l)}$ is defined by $v_{ij}(t) = \sum_l \mathcal{A}_{ij}^{(l)}e^{il\Omega t}$.  The higher order terms are obtained from Eq.~\eqref{eq:Magnus_exp} with
\eqq{
\hH_l = -\sum_{\langle i,j\rangle,\sigma}  \mathcal{A}_{ij}^{(l)}\hc^\dagger_{i\sigma} \hc_{j\sigma} \label{eq:Heff_l}.
}
At this level of approximation, the main effect of the field is to modify the hopping terms, and the effective Hamiltonian remains of the Mott-Hubbard type.

{\it High-frequency and large-$U$ regime ---} 
\label{sec:Floqeut_Ome_U}
When $\Omega, |U| \gg v_0$, but $U$ and $\Omega$ are of comparable magnitude, one can switch to a rotating frame to deal with $U$ and $\Omega$ on equal footing~\cite{Bukov2016}. For this, we introduce a common frequency $\Omega_0$ such that  $\Omega = k_0\Omega_0$  and $U\simeq l_0\Omega_0$ $(\equiv U_0)$, where $k_0$ and $l_0$ are co-prime integers. The condition $U\simeq l_0\Omega_0$ assumes that the detuning $\Delta U(\equiv U-U_0)$ is much smaller than $\Omega_0$. The rotating frame is defined by the unitary transformation $\hat{\mathcal{U}}(t) = \exp (-i U_0 t\sum_j \hn_{j\uparrow} \hn_{j\downarrow})$ so that the Hamiltonian in the rotating frame is given by 
$\hH^{\rm rot}(t) = -\sum_{\langle i,j\rangle,\sigma} \{ v_{ij}(t) \hat{g}_{ij\sigma} + \big[v_{ij}(t) e^{iU_0t} \hat{h}^\dagger_{ij\sigma} + h.c.\big]\} + \Delta U \sum_j \hn_{j\uparrow} \hn_{j\downarrow}$, where $\hat{g}_{ij\sigma} = (1-\hn_{i\bar{\sigma}}) \hc^\dagger_{i\sigma} \hc_{j\sigma} (1-\hn_{j\bar{\sigma}}) + \hn_{i\bar{\sigma}} \hc^\dagger_{i\sigma} \hc_{j\sigma} \hn_{j\bar{\sigma}}$ conserves the number of holons and doublons and $\hat{h}^\dagger_{ij\sigma} = \hn_{i\bar{\sigma}} \hc^\dagger_{i\sigma} \hc_{j\sigma} (1-\hn_{j\bar{\sigma}})$ generates a doublon. Then one applies the high frequency expansion with respect to $\Omega_0$ to $\hH^{\rm rot}$.
To lowest order, 
\eqq{
\hH^{\rm rot}_0 =& -\sum_{\langle ij\rangle \sigma} \left\{ \mathcal{\tilde{A}}^{(0)}_{ij} \hat{g}_{ij\sigma}  + [\mathcal{\tilde{B}}^{(0)}_{ij} \hat{h}^\dagger_{ij\sigma}  + h.c. ]\right\} \nonumber \\
& 
+ \Delta U \sum_j \hn_{j\uparrow} \hn_{j\downarrow},\label{eq:Heff_0_1}
}
where 
$\mathcal{\tilde{A}}$ and $ \mathcal{\tilde{B}}$ are defined as 
$v_{ij}(t) = \sum_l \mathcal{\tilde{A}}_{ij}^{(l)}e^{il\Omega_0 t}$ 
and $\mathcal{\tilde{B}}^{(l)}_{ij}=\mathcal{\tilde{A}}_{ij}^{(l-l_0)}$. 
Higher order terms are obtained with $\hH^{\rm rot}_l = -\!\!
\sum_{\langle ij\rangle \sigma} \!\!\left\{ \mathcal{\tilde{A}}^{(l)}_{ij} \hat{g}_{ij\sigma}  + [\mathcal{\tilde{B}}^{(l)}_{ij} \hat{h}^\dagger_{ij\sigma}  + \mathcal{\tilde{B}}_{ij}^{(-l)*}\hat{h}_{ij\sigma}] \right\}$ from Eq.~\eqref{eq:Magnus_exp}. The situation can be further categorized into the case $k_0=1$ ($U\simeq l_0\Omega$), corresponding to a driving in near resonance to single- or multi-photon absorption processes, and the nonresonant case ($k_0\neq1$). In the former case, the doublon-holon (D-H) creation/annihilation terms in the low-order effective Hamiltonian~(e.g. $\mathcal{\tilde{B}}^{(0)}_{ij}$ in \eqref{eq:Heff_0_1}) are nonzero, while they vanish in the non-resonant case. Non-resonant driving is relevant for  the control of the low energy physics, such as spin and orbital dynamics (Sec.~\ref{sec:effective_j_ac}).

\subsection{Band Renormalization}\label{sec_bandwidth_ren}

One of the most basic effects in a periodically driven lattice model is the renormalization of the hopping parameters, i.e. dynamical localization. This effect can be easily understood for noninteracting band electrons \cite{Dunlap1986,Holthaus1992,Eckardt2005,Tsuji11}, but it is equally relevant for the motion of doublons and holons in 
(photo)-doped Mott insulators. In the Floquet Hamiltonians~\eqref{eq:Heff_0} and \eqref{eq:Heff_0_1}, this renormalization of the tunneling appears in the coefficients of the kinetic terms. 
For an AC electric field with peak amplitude $E_0$ and frequency $\Omega$, the hopping along the direction of the field (and hence the corresponding ${\mathcal A}^{(0)}_{ij}$) is renormalized as
\eqq{
v_0 \rightarrow \mathcal{J}_0(A_0)v_0, \label{eq_bessel}
}
where $\mathcal{J}_0$ is the modified Bessel function of the first kind, $A_0=qaE_0/\Omega$, and $a$ is the bond length. When $\Omega\gg U$, there is no photon absorption in a single-band model and the main effect of the electric field is to drive the system into the strong coupling regime. In such a case, a reduction of the doublon density is expected~\cite{Sandholzer2019} and may tigger a metal-insulator transition~\cite{Ishikawa2014}.
In addition, the directional dependence of the hopping renormalization may be used to control the dimensionality of the system.

\begin{figure}[t]
\begin{center}
\includegraphics[angle=0, width=1\columnwidth]{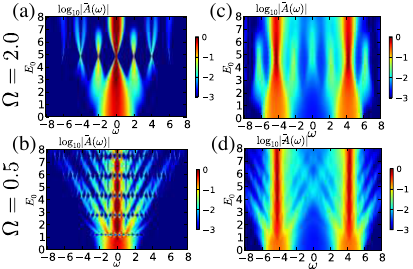}
\caption{Log scale plots of the time-averaged local (momentum-averaged) spectral function $A(\omega)$ of the Hubbard model on a hypercubic lattice driven by AC fields with  $\Omega=2.0$ or $\Omega=0.5$ as a function of the field amplitude $E_0$ (hypercubic lattice, $v_*=1$). Panels (a,b) are for the noninteracting system. Panels (c,d) show Floquet-DMFT+NCA results for Mott insulating systems with $U = 8$, coupled to a fermion bath to stabilize a time-periodic state. 
(From \onlinecite{Murakami2018PRB}.) 
}
\label{fig_floquet}
\end{center}
\end{figure}

The band renormalization effect is evident in the single-particle spectrum, as shown in Fig.~\ref{fig_floquet} for an infinite-dimensional hypercubic lattice 
with field along the body diagonal. In the noninteracting limit (panels (a,b)), the bandwidth shrinks to zero for the field strengths satisfying $\mathcal{J}_0(A_0)=0$, in accordance with Eq.~\eqref{eq_bessel}. A qualitatively similar evolution is seen in the spectral function of the Mott insulator with $U=8$ (panels (c,d)), where the width of the Hubbard bands shrinks and widens with increasing $E_0$.

%%%%%%%%%%%%%%%%%
%%%%%%%%%%%%%%%%%

\subsection{Floquet Sidebands}

Periodic driving does not only renormalize the (Hubbard) bands, but it also leads to new features in the spectral function, so-called Floquet side-bands. 
These bands correspond to photo-dressed states and are split off from the center of the original band by multiples of the driving frequency $\Omega$. 
Floquet bands have been observed experimentally in good metals \cite{Wang2013}, even for few cycle pulses \cite{Ito2023}.  However, the existence of Floquet sidebands depends on the electronic lifetime, as the electron-electron scattering prevents a coherent motion of Bloch electrons~\cite{Aeschlimann2021}. 
While doublons and holons in a Mott insulator move incoherently, theoretical simulations nevertheless show clear Floquet bands for sufficiently high frequency. 
Again, it is instructive to compare the spectra for the Mott insulator and noninteracting system in Fig.~\ref{fig_floquet}. 
 When the excitation frequency is high enough, one can identify clear Floquet side bands with an energy splitting $\Delta\omega=m\Omega$ from the 
 main band (panel (a)) or the Hubbard bands (panel (c)). In particular, when $\mathcal{J}_0(A_0)=0$, the Floquet bands in the noninteracting system (a) 
are clearly defined, with a position at $m\Omega$ regardless of the momentum. For the Mott insulator, the dynamical localization is not perfect, but nevertheless Floquet sidebands of the Hubbards bands are best defined at $\mathcal{J}_0(A_0)=0$. For small $\Omega$ (compared to the bandwidth $\simeq 2 v_*$) the sidebands overlap with each other and the spectrum shows structures resembling the Wannier-Stark ladders for a DC field, see panels (b,d) and Sec.~\ref{eq:WS_local}.

\subsection{Effective spin and orbital exchange}
\label{sec:effective_j_ac}

In Mott insulators, AC fields can be used to directly modify low-energy degrees of freedom related to magnetic and orbital orders. In particular the optical manipulation of magnetism is potentially interesting for ultrafast magnetic memories \cite{Kirilyuk2010}. On the one hand, Floquet spin Hamiltonians can be engineered based on the linear coupling of  spins to the magnetic field of light via the Zeeman term \cite{Takayoshi2014, Takayoshi2014b, Sato2016}. However, also the stronger electric field of the laser can be used, although it does not linearly couple to the spin. The underlying  principle is general and extends to basically all low-energy degrees of freedom:  Virtual transitions to photo-dressed states can modify the exchange couplings or even induce new terms in the Hamiltonian of spin and/or orbital degrees of freedom. 

A fundamental example is the half-filled single-band repulsive Hubbard model with nearest-neighbor hopping $v_0$, 
exposed to a linearly polarized AC field with frequency $\Omega$ and vector potential amplitude $A_0$. The corresponding effective Hamiltonian can be obtained following the discussion in Sec.~\ref{sec:Floqeut_Ome_U} for the case $\Omega,U\gg v_0$: For off-resonant driving ($U\neq l\Omega$), no doublons and holons are generated, and the leading term of the effective Hamiltonian becomes a spin Hamiltonian $\hH_{\rm spin} =\frac{J_{\rm ex}}{2} \sum_{\langle i,j\rangle} \mathbf{\hat S}_i \cdot \mathbf{\hat S}_j$ with renormalized exchange coupling 
\eqq{
J_{\rm ex} = \sum_l \frac{4 v_0^2}{U-l\Omega} |\mathcal{J}_{l}(A_0)|^2. \label{eq:J_ex_P}
}
This expression shows that the exchange coupling contains the contributions from virtual creations of D-H pairs dressed with $l$ photons (transition amplitudes 
${\mathcal A}^{(-l)}=v_0\mathcal{J}_{-l}(A_0)$). 
In addition to the Floquet Schrieffer-Wolff transformation \cite{Bukov2016}, the  Floquet spin Hamiltonian \eqref{eq:J_ex_P} has been obtained 
using a perturbation theory in $v_0/U$ in the extended Floquet Hilbert space \cite{Mentink2015}, a time-dependent Schrieffer-Wolff transformation~\cite{Kitamura2016PRB}, a variant of the high-frequency expansion~\cite{itin2015}, and a time-dependent perturbation theory~\cite{hejazi2019}.

%%%%%%%%%%%%%%%%%
\begin{figure}[t]
\begin{center}
\includegraphics[angle=0, width=1.0\columnwidth]{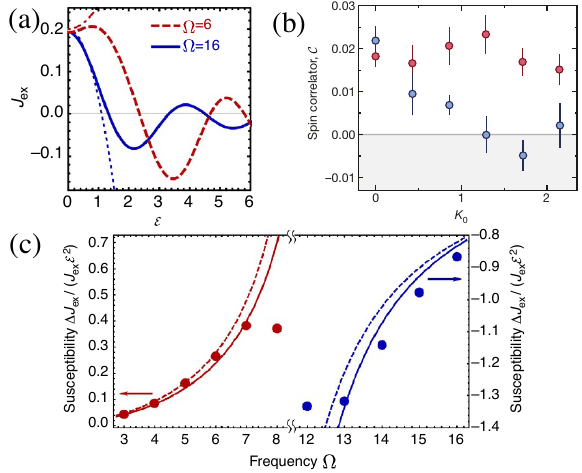}
\caption{(a) Superexchange for a two-site Hubbard model driven below ($\Omega<U$) and above ($\Omega>U$) the upper Hubbard band for $U/v_0=10.$ (b) Change in the nearest neighbor spin correlations in a cold atom setup versus the driving strength $K_0$ for the red~($\Omega<U$) and blue~($\Omega>U$) detuned periodic drive. (c) Simulated change in the superexchange $\Delta J_\text{ex}/(J_\text{ex} A_0^2)$ in the weak-driving regime $A_0\rightarrow0$ from a two-site Hubbard cluster~(full lines), perturbative expansion~(dashed lines) and time-dependent DMFT simulation~(dots) for driving $\Omega<U$~(red) and $\Omega>U$~(blue). 
(From \onlinecite{Mentink2015} and \onlinecite{Gorg2018}.)}
\label{fig:Floqeut_spin}
\end{center}
\end{figure}   
%%%%%%%%%%%%%%%%%

Equation \eqref{eq:J_ex_P} implies that, by tuning the field strength and excitation frequency, one can switch the sign of the  exchange coupling. In Fig.~\ref{fig:Floqeut_spin}(a), we show $J_{\rm ex}$ for a Hubbard model with $U=10$. In the small-field regime, $J_{\rm ex}$ increases with drive amplitude for $\Omega<U$, while it decreases for $\Omega>U$. In both cases, $J_{\rm ex}$ can be negative at certain field strengths. The perturbative prediction was compared with time-dependent DMFT simulations of the driven Hubbard model, where the modified exchange interaction was determined by analyzing the spin precession under a static magnetic field, see Fig.~\ref{fig:Floqeut_spin}(c). The agreement with Eq.~\eqref{eq:J_ex_P} is excellent away from the resonance $\Omega=U$. Close to the resonance, however, incoherent excitations dominate, and the perturbative treatment becomes unreliable~\cite{Mentink2015}.

The AC field induced modification of $J_\text{ex}$ has been experimentally confirmed in a cold atom setup~\cite{Gorg2018}, see also Sec.~\ref{sec:exp_probes}. Depending on the excitation frequency and the field strength, the spin correlations are either reduced or enhanced, and in some cases  
can become ferromagnetic (Fig.~\ref{fig:Floqeut_spin}(b)). 

Similar ideas can be applied to the attractive Hubbard model~\cite{Kitamura2016PRB,Fujiuchi2020,Murakami2023PRB}. 
The low-energy physics of the system in the strong-coupling regime is described by a Heisenberg-type 
model of the $\eta$-pseudo-spins (Eq.~\eqref{def_eta}), which represent doublons and holons. The AC electric field can effectively modify the anisotropy and/or the sign of the exchange couplings between the pseudo-spins, which allows to induce various phases involving  charge and $s$-wave superconducting orders. 
Other proposed applications involve the control of the exchange interaction in Kondo systems \cite{Nakagawa2015, Iwahori2016,eEckstein2017Kondo,MuellerShiba2023}.
Going beyond linearly polarized light, it was shown that circularly polarized light applied to frustrated lattices, like the triangular or honeycomb lattices, induces a scalar spin chiral term $\propto \mathbf{\hat S}_i\cdot (\mathbf{\hat S}_j\times \mathbf{\hat S}_k)$, which breaks time reversal symmetry while preserving the SU(2) symmetry~\cite{Claassen2017,Kitamura2017PRB}. The chirality originates from Floquet-induced virtual hoppings on 3-site clusters, and it could potentially lead to a chiral spin-liquid state. Suggested candidate materials include herbertsmithite ZnCu$_3$(OH)$_6$Cl$_2$, which is an antiferromagnet on a frustrated lattice, and PdCrO$_2$. 

In multi-orbital systems, the Floquet drive can be used to manipulate exchange couplings between spins and/or orbital degrees of freedom. Here, in general also dipolar light-matter coupling terms and higher order exchange paths via ligand orbitals may have to be considered \cite{chaudhary2019}.  \onlinecite{liu2018} and \onlinecite{hejazi2019} studied the partially-filled  three-band Hubbard model under time-periodic electric fields and discussed the conditions for modified or sign-inverted spin exchange couplings in ferromagnetic YTiO$_3$ and antiferromagnetic LaTiO$_3$. \onlinecite{Eckstein2017} demonstrated the electric-field control of the exchange couplings in the spin-orbital model (the Kugle-Khomskii model) obtained from the quarter-filled  two-orbital Hubbard model.

A further extension of the above studies is to consider multiband Mott-Hubbard~\cite{arakawa2021} or charge-transfer insulators~\cite{sriram2022} with strong spin-orbit coupling under circularly polarized light.  On a honeycomb lattice, besides the manipulation of the Heisenberg exchange, one can induce a Kitaev interaction $K S_{i}^{\gamma} S_{j}^{\gamma}$ with $\gamma=x,y,z$ representing the three inequivalent bonds. The competition between different magnetic instabilities leads to exotic phases within the Floquet prethermalized region, including a Kitaev spin liquid state~\cite{kitaev2006}. The most prominent candidate material is $\alpha$-RuCl$_3$, due to the honeycomb lattice and the strong spin-orbit coupling. 
Finally, Floquet engineering can be used to control fractionalized excitations in one-dimensional systems originating from spin-charge separation \cite{Gao2020} or spin-orbital separation \cite{MuellerKK2023}.

Many proposals for real materials rely on driving with frequencies not too far from the Mott gap. Even though the Mott gap can protect the system from rapid laser heating, for strong pulses, a substantial D-H production due to multi-photon processes is expected. An interesting proposal is therefore to observe the transient change of the interactions directly by motoring the spin or orbital dispersions during the Floquet pulse, which could be achieved using, for example, time-resolved RIXS \cite{Mitrano2020, MuellerKK2023}. In this context, it is important to note that an adiabatic Floquet Hamiltonian with a time-dependent amplitude parameter often provides an adequate description of the dynamics during a few cycle pulse 
(see,~e.g.,~\onlinecite{Eckstein2017}), while in the  limit of a single cycle, there is a crossover to the description in terms of a DC field modified exchange (Sec.~\ref{sec:effective_jex}). 

Another, more explorative proposal for Floquet engineering is to replace the laser drive by a strongly coupled mode of the quantum electromagnetic field, such as a confined cavity mode \cite{Schlawin0222cavity,BlochCavity2022}. At strong coupling, Floquet Hamiltonians may be realized as photon-dressed Hamiltonians with few photons \cite{Sentef2020cavity}. Various cavity analogs of the above mentioned Floquet Hamiltonians have been proposed \cite{Kiffner2019cavity,Li2020cavity,Bostroem2022}, although the experimental realization of such cavity quantum materials is a largely open problem. Further, driving the cavity with an external laser might open yet another avenue for Floquet engineering of long-range spin interactions \cite{Chiocchetta2021}.

\subsection{Floquet Prethermalization}
\label{sec:Floquet-prethermal}

%%%%%%%%%%%%%%%%%
\begin{figure}[t]
\begin{center}
\includegraphics[angle=0, width=\columnwidth]{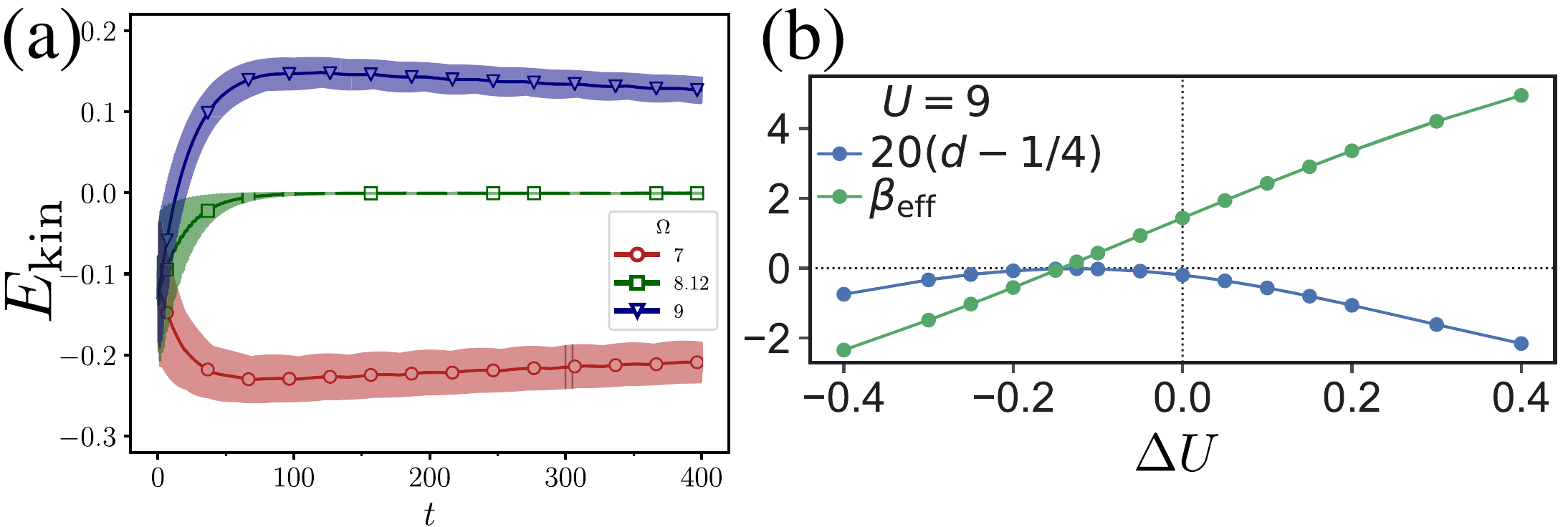}
\caption{
(a)~Kinetic energy $E_{\text{kin}}$ during a periodic interaction modulation with frequency $\Omega$ close to the resonance (Bethe lattice, $U=8$, $W=4$).~(b)~Doublon number ($d$) and inverse temperature ($\beta_{\rm eff}=1/T_\text{eff}$) of the AC field driven Hubbard model at the symmetric point (see text), with interaction $U+\Delta U$ and $\Omega=U/2$  (hypercubic lattice, $v_*=1$, $U=9$). (From \onlinecite{Herrmann2017} and \onlinecite{peronaci2018}.)}
\label{fig_prethermal}
\end{center}
\end{figure}   
%%%%%%%%%%%%%%%%%

A generic isolated system is expected to heat up to infinite temperature under periodic driving \cite{Alessio2014Heating,Lazarides2014}.
However, in some cases, such heating occurs only on relatively long timescales, while on shorter timescales, the system is trapped in a so-called Floquet prethermalized state 
\cite{Abanin2015, Mori2016, Dalessio2013,Citro2015,Canovi2016,Weidinger2017,ikeda2021,rubio2020,torre2021,Kuhlenkamp2020, Howell2019, Pizzi2021}.
For example, this behavior is observed in the high-frequency limit, where the prethermal state can last for exponentially long times in $\Omega$ and can be described by the effective Hamiltonian obtained from the high-frequency expansion. This partially justifies the analysis of equilibrium states of such effective Hamiltonians. The suppression of heating in the high-frequency limit is mathematically related to conventional prethermalization due to constraints arising from a large energy gap in closed quantum systems \cite{Abanin2017}. 
Another strategy for the suppression of heating was demonstrated in the integrable quantum Ising chain \cite{Das2010} and called dynamical freezing \cite{Bhattacharyya2012}.
It relies on approximately conserved quantities which emerge due to the strong drive, beyond  some threshold for the drive amplitude \cite{Haldar2018, Haldar2021}.

In the context of Mott insulators, Floquet prethermalization has been numerically observed in a wide parameter range, even close to resonant conditions. 
Prominent examples are time-dependent DMFT results for the single-band Hubbard model. Figure~\ref{fig_prethermal}(a) shows the evolution of the kinetic energy in a system with $U=8$, $W=4$ under periodic modulations of the Coulomb interaction~\cite{peronaci2018}. At infinite temperature, the kinetic energy should be zero. The system quickly approaches this state for the excitation frequency $\Omega=\Omega^*\equiv 8.12$, but even for slightly off-resonant conditions, the kinetic energy shows a nonthermal plateau. 
In the case of an electric field driven system, this prethermal state would correspond to a thermal state of $H^\text{rot}_{0}$ [Eq.~\eqref{eq:Heff_0_1}], up to corrections of order $1/\Omega$. This has been analyzed explicitly for the case where $E_0$ is tuned to the symmetric point $\mathcal{\tilde A}_{ij}^{(0)}=\mathcal{\tilde B}_{ij}^{(0)}$, such that $H^\text{rot}_{0}$ becomes the Hubbard model with renormalized hopping and interaction $\Delta U$ \cite{Herrmann2017}. This condition allows to directly compare the dynamics of the driven system and the equilibrium state of the effective Floquet Hamiltonian within DMFT. For a sudden switch-on of the driving, the effective temperature $T_\text{eff}$ of the Floquet state is obtained from the expectation value of $H^\text{rot}_{0}$, which is conserved under the  stroboscopic time evolution and can be evaluated from the initial Mott state. Based on this prediction, one expects a positive (negative) $T_\text{eff}$ for $\Delta U>0$ ($\Delta U<0$), which is confirmed by the simulation up to the deviations of $\mathcal{O}(1/\Omega)$ (Fig.~\ref{fig_prethermal}(b)).

\subsection{Single or Multi-Photon absorption}
\label{sec:single_multi}

\begin{figure}[t]
\begin{center}
\includegraphics[angle=0, width=1\columnwidth]{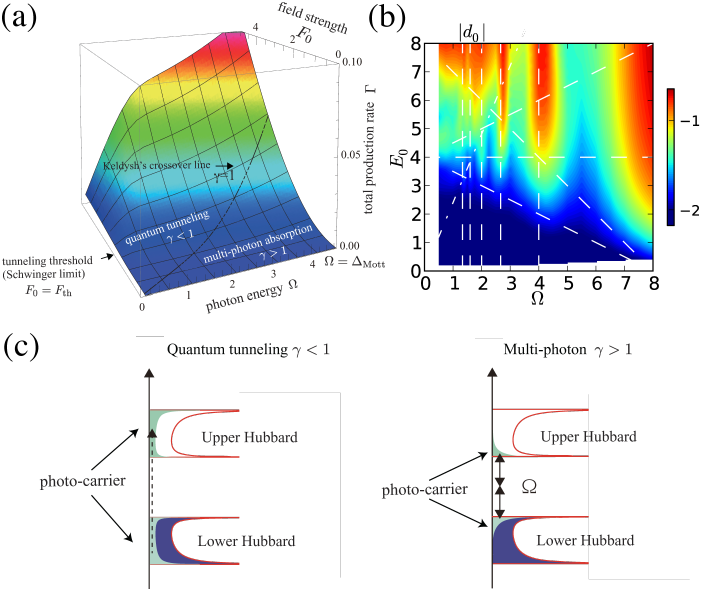}
\caption{(a) Production rate of doublon-holon pairs for the 1D Hubbard model with $U=8$, $W=4$ in an AC field obtained with 
the Landau-Dykhne method.~(b) Time-averaged double occupancy $d_{0}$ in the plane of field amplitude $E_0$ and frequency  $\Omega$, in a Mott insulating system with $U=8$, analyzed with Floquet DMFT and attached  fermion bath (hypercubic lattice, $v_*=1$). Vertical dashed lines at $\Omega=U/n$ mark the resonances associated with $n$-photon absorption.~(c) Schematic picture of the doublon-holon distribution for different excitations.
(From \onlinecite{Oka2012} and \onlinecite{Murakami2018PRB}.)}
\label{fig_multiphoton}
\end{center}
\end{figure}

Another important effect of the AC field is the creation of photo-carriers. In the weak-field (perturbative) regime, this can be described as single or multi-photon absorption. 
If $n\Omega\gtrsim\Delta_{\rm Mott}>(n-1)\Omega$, the probability for creating a D-H pair via $n$-photon absorption is proportional to $E_0^{2n}$. The effective  Floquet Hamiltonian introduced in Sec.~\ref{sec:Floqeut_Ome_U} captures the single or multi-phonon absorption in the D-H creation terms. The multi-photon absorption is complementary to the quantum tunneling mechanism discussed in Sec.~\ref{sec:breakdown}. For the 1D Hubbard model, the production rate of D-H pairs can be analytically obtained using a combination of the Landau-Dykhne method and the Bethe ansatz~\cite{Oka2012}. The crossover between the quantum tunneling and multi-photon absorption regime is controlled by the Keldysh parameter $\gamma\equiv \frac{\Omega}{qE_0\xi}$, where the length $\xi$ characterizes the spatial extent of the D-H pairs. The multi-photon absorption (tunneling) regime corresponds to $\gamma \gg 1$ ($\gamma \ll 1$). The value $\gamma=1$ defines the Keldysh line ~\cite{Keldysh1965}, see Fig.~\ref{fig_multiphoton}(a). 

In the multiphoton absorption regime, the creation probability of the D-H pairs is given by 
\eqq{
\mathcal{P}_\text{D-H} \simeq  \Bigl( \frac{q E_0 \xi}{2\pi\Omega}\Bigl)^{2 \frac{\Delta_{\rm Mott}}{\Omega}},
} 
which should be compared with Eq.~\eqref{dielec_break} for the tunneling case. These two D-H creation mechanisms lead to a different energy distribution of the charge carriers, as sketched in Fig.~\ref{fig_multiphoton}(c). In the case of tunneling, the D-H distribution becomes momentum-independent, i.e., the created doublons and holons are effectively very hot. Using multi-photon processes, the doublons and holons can instead be created near the edges of the Mott gap. 

The crossover from the multiphoton~(power-law) to the tunnelling~(exponential) behavior was experimentally observed in Ca$_2$RuO$_4$, an antiferromagnetic Mott insulator with a $0.6$~eV gap \cite{li2022}. The nonthermal distribution of charge carriers was investigated by analyzing the photo-induced reflectivity change $\Delta R(\omega)$
for a subgap excitation. With increasing excitation strength one observes a  shift of the characteristic frequency defined by $\Delta R(\omega)=0$, which was linked to the change in the D-H distribution associated with the Keldysh crossover.

%%%%%%%%%%%%%%%%%
%%%%%%%%%%%%%%%%%

A systematic numerical analysis of the generation of D-H pairs has been conducted using Floquet DMFT+NCA in a hyper-cubic system~\cite{Murakami2018PRB}. Figure~\ref{fig_multiphoton}(b) shows the number of  photo-generated doublons as a function of $\Omega$ and $E_0$. One can identify the resonant excitations around $\Omega=U/n$ ($n=1\sim 6$), which are associated with $n$-photon absorption. In the perturbative regime, the $n$-photon absorption probability is proportional to $|E_0|^{2n}$, consistent with the fact that higher order peaks become more prominent with larger $E_0$ for $E_0\lesssim U/2$. By increasing the field strength beyond $E_0 \approx U/2$, some higher-order resonances disappear or emerge at off-resonant conditions, which reflects nonperturbative effects.

\subsection{Nonlinear optical responses} 
\label{sec:nonlinear}

Nonlinear optical responses reveal excited states beyond the reach of linear responses, and thus provide a means to extract detailed information on the system's excitation spectrum. Strongly correlated systems can host excited states whose properties are significantly different from those of conventional semiconductors, and thus can show characteristic nonlinear optical signals.

\subsubsection{Third order responses}
\label{sec:TOR}

\begin{figure}[t]
\begin{center}
\includegraphics[angle=0, width=1\columnwidth]{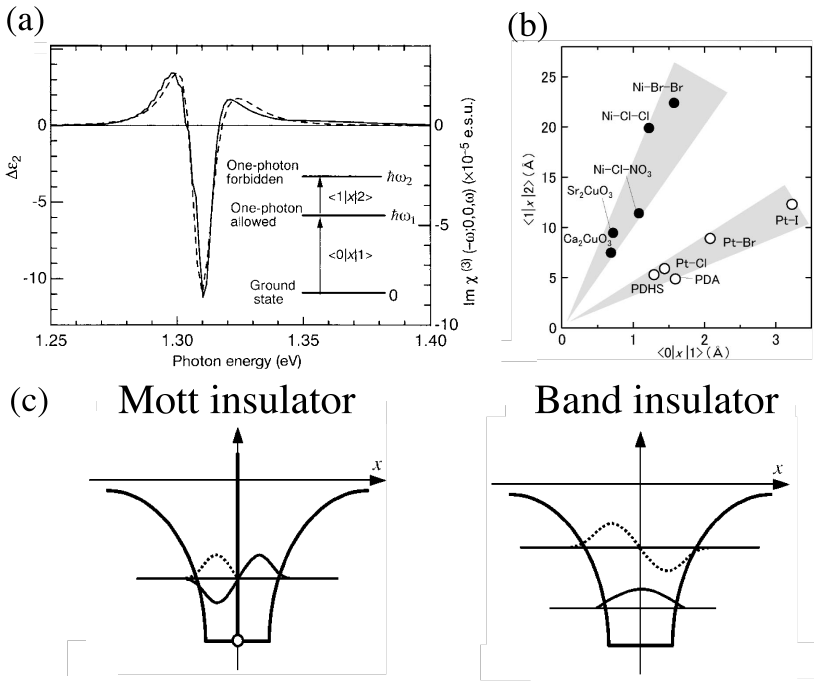}
\caption{
(a) Third-order nonlinear susceptibility spectra Im$\chi^{(3)}(-\omega;0,0,\omega)$ of [Ni(chxn)$_2$Br]Br$_2$ at $77$K. The solid line is the experimental result and the dashed line shows the results of the 3-level model illustrated in the inset. (b)~Relation between the transition dipole moments for various 1D materials, where the filled circles and open circles correspond to 1D Mott insulators and 1D semiconductors, respectively. (c) Schematic picture of the wave function of odd- and even-parity excitons in a 1D Mott insulator and 1D band insulator.
 ~(From \onlinecite{Kishida2000Nature} and \onlinecite{Ono2004PRB}.)
}\label{Fig:HTOR_Mott_exp}
\end{center}
\end{figure}

One-dimensional (1D) Mott insulators are known to show large third-order nonlinear optical (TNLO) responses due to the characteristic features of excited states originating from spin-charge separation~\cite{Kishida2000Nature,Kishida2001PRL,Ono2004PRB,Ono2005PRL}.~In general, the TNLO response is characterized by the 
susceptibility $\chi^{(3)}$ defined as $P(\omega_\sigma)\propto \epsilon_0\chi^{(3)}(-\omega_\sigma;\omega_a,\omega_b,\omega_c)E(\omega_a)E(\omega_b)E(\omega_c)$, where $P(\omega_\sigma)$ is the nonlinear polarization, $E(\omega_{i})$ the electric field, $\epsilon_0$ is the vacuum permittivity, and energy conservation implies $\omega_\sigma = \omega_a + \omega_b + \omega_c$. The nonlinear susceptibility ${\rm Im}\chi^{(3)}(-\omega;0,0,\omega)$ was measured by electroreflectance~(ER) experiments for 1D charge transfer insulators, like the  Ni-halogen and CuO chains, see Fig.~\ref{Fig:HTOR_Mott_exp}(a). In these materials, $\chi^{(3)}(-\omega;0,0,\omega)$ reaches $10^{-8}-10^{-5}$ e.s.u.~\cite{Kishida2000Nature,Ono2004PRB}, which is much larger than the values in semiconductors such as silicon polymers and Peierls insulators ($10^{-12}-10^{-7}$ e.s.u.).
The third order response $\chi^{(3)}(-\omega;0,0,\omega)$ can also be measured using strong THz fields \cite{Yada2013}. In addition, a large third harmonic generation, which corresponds to $\chi^{(3)}(-3\omega;\omega,\omega,\omega)$, was reported for 1D Mott insulators~\cite{Kishida2001PRL,Ono2005PRL}.

%%%%%%%%%%%%%%%%%
%%%%%%%%%%%%%%%%%

The observed spectral features of the nonlinear susceptibilities can be described by a 3-level model~\cite{Kishida2000Nature,Ono2004PRB}, which considers the ground state ($|0\rangle$), an optically active parity-odd state ($|1\rangle$) and an optically forbidden parity-even state ($|2\rangle$). $|1\rangle$ and $|2\rangle$ are associated with Mott excitons (Sec.~\ref{sec:MottExciton}) 
or with low-lying states in the D-H continuum. For example, within the 3-level model, the main term in 
$\chi^{(3)}(-\omega;0,0,\omega)$ becomes  
\eqq{
\chi^{(3)}(-\omega;0,0,\omega)
 \!\propto\!
 \frac{|\langle 0 | x | 1 \rangle|^2 |\langle 1 | x |2 \rangle|^2} {(E_1-\omega-i\gamma_1)^2(E_2-\omega-i\gamma_2)}.
}
Fitting the experimental results to this model (Fig.~\ref{Fig:HTOR_Mott_exp}(a)) provides information on the dipole moments $\langle i | x |j \rangle$ and 
the energies $E_i$ of the excited states $|i\rangle$. The main origin of the large TNLO responses in 1D Mott insulators turns out to be a large dipole moment between $|1\rangle$ and $|2\rangle$, see Fig.~\ref{Fig:HTOR_Mott_exp}(b), and the energy differences between the excited states, which is small compared to conventional semiconductors. 

The peculiar properties of the excited states in the 1D Mott insulator were clarified by the ``holon-doublon" model, which is based on spin-charge separation and focuses only on the charge degrees of freedom (doublons and holons). In this model, doublons and holons are treated as hard-core particles. 
Due to the combination of the hard-core nature and the 1D lattice structure, the odd-party and even-parity states are 
degenerate and have the same form of the wave function (Fig.~\ref{Fig:HTOR_Mott_exp}(c)), which explains the large transition dipole moment \cite{Mizuno2000PRB}. 
In conventional semiconductors, the lack of the hard-core nature makes the energy difference between the odd-parity and even-parity states large and the dipole moment between them small~\cite{Ono2004PRB}. 

In higher dimensions, due to the lack of spin-charge separation, the excitation spectrum is different. TONL responses become weaker in ladders or 2D systems, compared to 1D~\cite{Mizuno2000PRB}. Still, TONL signals provide a useful spectroscopy of Mott excitons. For example, the contribution of the spin-charge coupling to the formation of Mott excitons has been detected using TONL responses, see Sec.~\ref{sec:MottExciton}.

\subsubsection{High-harmonic generation}

Optical excitations with even stronger fields result in high-harmonic generation (HHG), i.e., the emission of higher harmonics of the injected light from matter.  HHG was initially 
studied in gases~\cite{Ferray_1988}, but it has also been observed in solids, in particular semiconductors~\cite{Ghimire2011,Schubert2014}. 
In a semiconductor, the light emission is associated with the independent dynamics of electrons and holes under the AC field. Both intraband and interband currents contribute to HHG 
\cite{Vampa2014}. The latter contribution can be analyzed by the three-step model~\cite{Corkum1993PRL,Lewenstein1994,Vampa2015}, which describes i) the creation of an electron-hole pair, ii) the electron/hole dynamics in the bands, and iii) the recombination via light emission. Since both the intra- and interband contributions depend on the band structure, 
HHG spectroscopy of semiconductors~\cite{Vampa2015PRL,Luu2015} allows to reveal band information.

%%%%%%%%%%%%%%%%%
\begin{figure}[t]
\begin{center}
\includegraphics[angle=0, width=1\columnwidth]{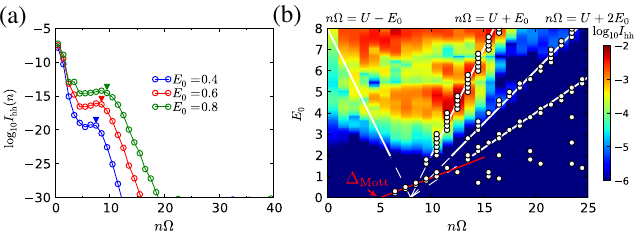}
\caption{
(a) HHG spectrum $I_{\rm HHG}(\omega)$ of a paramagnetic Mott insulating Hubbard model for different field amplitudes $E_0$. Arrows indicate the cutoff energies of the plateaus.  (b) $I_{\rm HHG}(\omega)$ as a function of $E_0$ and harmonic energy ($n\Omega$). White circle markers show cutoff energies. Red and white lines indicate the linear scaling of the cutoff energy with $E_0$. Results obtained with Floquet DMFT+NCA on a hypercubic lattice with $v_*=1$, for $U = 8$ ($\Delta_{\rm Mott}\simeq 5$), $\beta = 2$, 
$\Omega = 0.5$. (From \onlinecite{Murakami2018PRL}.)
}
\label{fig_HHG}
\end{center}
\end{figure}
%%%%%%%%%%%%%%%%%

More recently, the scope of HHG research has been extended to strongly correlated solids~\cite{Silva2018NatPhoton,Murakami2018PRL,Tancogne-Dejean2018}, where the 
excitations cannot be described by independent electrons and holes. The relevant questions are the physical origin of HHG, its relation to many body excitations, and potential application as a spectroscopic tool. Basic features of HHG in Mott insulators have been revealed by analyzing the Hubbard model with Floquet DMFT~\cite{Murakami2018PRL}. Figures~\ref{fig_HHG}(a,b) show the HHG spectrum $I_{\rm HHG}(\omega)$, which is evaluated from the current $j(\omega)$ as $I_{\rm HHG}(\omega)=|\omega j(\omega)|^2$, for a paramagnetic Mott insulator exposed to a strong subgap excitation. The HHG features change qualitatively depending on the field strength $E_0$. When $E_0$ is relatively weak, 
the HHG spectrum is characterized by  a  single plateau, as in atomic gases (Fig.~\ref{fig_HHG}(a)). On the other hand, for stronger fields, the HHG spectrum shows multiple plateaus. The cutoff frequencies ($\omega_{\rm cut}$) of the plateaus scale linearly with $E_0$, i.e., $\omega_{\rm cut}=\Delta+\alpha E_0$. In particular, in the strong field regime, the cutoff follows $\omega_{\rm cut}=U+mE_0$ with integer $m$, see Fig.~\ref{fig_HHG}(b). In both cases, the HHG signal is dominated by the recombination of D-H pairs, and the three-step picture applies to these D-H pairs. The qualitative change in the HHG spectra reflects the switch from the itinerant nature of the D-H pairs in the weaker-field regime to a localized nature of the pairs in the stronger-field regime (Sec.~\ref{eq:WS_local}). This study also revealed that, in contrast to semiconductors, the HHG spectrum of Mott insulators is not directly related to the dispersion reflected in the single-particle spectrum. 

%%%%%%%%%%%%%%%%%
\begin{figure}[t]
\begin{center}
\includegraphics[angle=0, width=1\columnwidth]{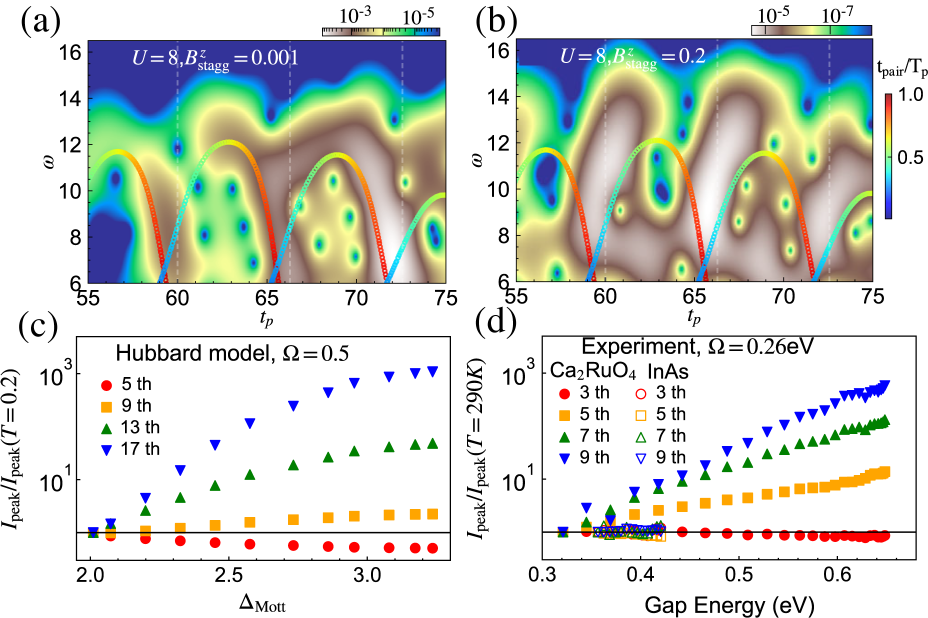}
\caption{(a,b)~Subcycle analysis of $I_{\rm HHG}(\omega,t_p)$ for the 1D Hubbard model with staggered magnetic field $B_{\rm stagg}^z$  analyzed with iTEBD for the pump parameters $\Omega=0.5,E_0=0.8$. The colored markers show the prediction from the three-step model combined with the D-H dispersion, with the color indicating the time interval $t_{\rm pair}$  between pair creation and recombination in units of $T_p=\frac{2\pi}{\Omega}$. The  larger staggered field $B_{\rm stagg}^z$ in (b) mimics the spin-charge coupling. 
The vertical dashed lines indicate the times when $A(t)=0$. (c)~Intensity of the HHG peaks as a function of the Mott gap $\Delta_{\rm Mott}$ for the single-band Hubbard model within DMFT+NCA. (d)~Experimental HHG intensity 
as a function of the optical gap for Ca$_2$RuO$_4$ (Mott insulator) and InAs (semiconductor).  
(From \onlinecite{Murakami2022hhg} and \onlinecite{Uchida2021}.)
}
\label{fig_HHG_2}
\end{center}
\end{figure}
%%%%%%%%%%%%%%%%%

Further insights into the relation between HHG and the excitation spectrum of Mott insulators have been obtained by studying the 1D Hubbard model with iTEBD~\cite{Murakami2021PRB}. This system shows spin-charge separation, and at half-filling hosts many-body elemental excitations called doublons, holons, and spinons, whose dispersion relations are known from the Bethe ansatz. It was shown that the HHG process is well explained by the semi-classical three-step model combined with {\em the D-H dispersion}. Figure~\ref{fig_HHG_2}(a) plots the subcycle signal $I_{\rm HHG}(\omega,t_p)\equiv|\omega j(\omega,t_p)|^2$ (frequency profile of the light emitted at time $t_p$), where $j(\omega,t_p) = \int dt e^{i\omega t}  F_{\rm window}(t-t_p)j(t)$ is a windowed Fourier transform of the current with  $F_{\rm window}(t)$ a window function centered at $t=0$. This subcycle signal is well explained by the light emission from the recombination of the D-H pair predicted by the three-step model combined with the D-H dispersion, see the colored line. This is in stark contrast to semiconductors, where HHG directly reflects the dispersions of the single-particle bands. These results suggest that HHG could be used to detect the dispersion of many-body elemental excitations~\cite{Imai2022}. Beyond this, an interesting proposal is to measure subcycle resolved nonlinear processes in correlated electron systems using ideas from multi-dimensional spectroscopy \cite{Valmispild2023}.

In higher dimensions, the dynamics of doublons and holons disturbs the spin background (Sec.\ref{electron-spin}), resulting in strong spin-charge coupling with profound effects on HHG. \onlinecite{Murakami2022hhg} showed that spin-charge coupling efficiently reduces the coherence time of the D-H pairs (Fig.~\ref{fig_HHG_2}(b)), which leads to a strong temperature dependence of the HHG signal. Although the band gap increases with decreasing temperature, which suppresses the D-H production, the change in the coherence time dominates this effect and leads to a drastic enhancement of HHG with increasing gap size (Fig.~\ref{fig_HHG_2}(c)). 
A similar behavior has been experimentally reported in the Mott insulator Ca$_2$RuO$_4$~\cite{Uchida2021}, where the HHG intensity is exponentially enhanced with increasing gap size (Fig.~\ref{fig_HHG_2}(d)) following a peculiar phenomenological scaling. This behavior was not observed in the conventional semiconductor InAs.

Doping and edge effects on HHG in the single-band Hubbard model have been studied with ED~\cite{Hansen2022,Hansen2022PRA}.
Going beyond the single-band Hubbard model, strongly correlated systems can host various types of elemental excitations, such as solitons in dimer systems~\cite{Ishihara2020}, string states in multi-orbital systems~\cite{Markus2020}, or Mott excitons~\cite{Udono2022,Yamakawa2023} and magnons in magnetic insulators~\cite{Takayoshi2019PRB,Ikeda2019}. HHG from the dynamics of these elemental excitation has also been investigated. Another direction for HHG research in correlated materials is to use HHG as a spectroscopic tool for phase transitions. The analysis of the Hubbard model revealed significant changes in the HHG signal during a photo-induced insulator-metal transition~\cite{Silva2018NatPhoton,Orthodoxou2021}. Experimental work on VO$_2$ tested this idea~\cite{Bionta2021PRR}, where 
the photo-induced transition to two types of metallic phases was analyzed by monitoring the time evolution  of the HHG intensity.  
In addition, the detection of equilibrium phase transitions using HHG has been theoretically studied in the Hubbard model~\cite{Shao2021} and experimentally reported in YBa$_2$Cu$_3$O$_{7-d}$~\cite{Alcala2022}.

\subsection{Nonlinear Phononics}
\label{sec:nonlinear_phononics}

THz laser fields allow to selectively excite certain infrared active (IR) phonon modes. The coherent oscillation of the IR mode can significantly affect the electronic properties 
through its nonlinear coupling to other lattice modes and to the electrons (nonlinear phononics).  
For example, due to a third-order nonlinear coupling between the IR mode and a Raman mode, the excitation of the IR mode distorts the lattice along the coordinate of the Raman mode, which changes the electronic properties~\cite{forst2011,subedi2014}. Nonlinear phononics has been used to induce insulator-metal transitions~\cite{Rini2007a}, 
manipulate ferroelectric order~\cite{nova2019,mankowsky2017}, magnetic states \cite{Nova2017,Disa2023}, and superconducting-like states ~\cite{Fausti11,mankowsky2014,Mitrano2015,Buzzi2020}, as reviewed by \onlinecite{subedi2021,Cavalleri2018}. 
Here, we focus on the modification of the electron-electron interactions by nonlinear phononics, which is also relevant for Mott systems. 
A modulation of the Hubbard interaction due to a coherent phonon and nonlinear electron-lattice couplings was experimentally reported in 
 the 1D Mott insulator ET-F$_2$TCNQ~\cite{kaiser2014b,singla2015}.
After a vibrational excitation, one observes a redshift of the optical  charge-transfer gap, and the appearance of mid-gap peaks with a separation of twice the phonon frequency. 
The red shift was connected with a strong asymmetry in the coupling of the lattice modes to holons and doublons. These experiments clearly show the importance of nonlinear electron-lattice couplings in photo-excited systems. Still, their relevance for photo-induced phase transitions in systems like  K$_3$C$_{60}$~\cite{Mitrano2015} or $\kappa$-(ET)$_2$Cu[N(CN)$_2$]Br~\cite{Buzzi2020} remain to be understood.

On the theory side, \onlinecite{kennes2017} proposed that in systems with quadratic electron-phonon coupling, which can be dominant due to the crystal symmetry, intense external driving leads to an attractive contribution to the electronic interaction, which could overcome the Hubbard repulsion and induce either a metal to superconductor or insulator to metal transition. The induction of the attractive interaction was confirmed by ED studies on small clusters~\cite{sentef2017}, and in the dilute limit of an extended system ~\cite{kovac2023}. 
For half-filled systems, an iTEBD study starting from a direct product state of a metal and coherent phonons showed that the final state is rather dynamically localized~\cite{sous2021}. 
A transition between a Mott insulator and a coherent metal was demonstrated in a model with nonlinear electron-phonon coupling using time-dependent DMFT \cite{Grandi2021}. 

Another proposed mechanism for inducing or enhancing an attractive (static) electron-electron interaction is the
parametric excitation of a phonon linearly coupled to electrons \cite{knap2016,babadi2017}. The effect 
is most prominent at the resonance, where heating is also strong, so that the pairing susceptibility 
may be reduced in the driven state in spite of an enhanced attraction~\cite{murakami2017}. In general, it is numerically challenging 
to  take into account both the effects of heating and modified interactions, but such results are important to understand the conditions under which nonthermal phase transitions can be induced.

%%%%%%%%%%%%%%%%%%%%%%%%%%%%%%%%%%%%%%%%%%%%%%%%%%%%%%%%%%%%%%%%%%%%%%%%
\section{Photo-doped  Mott insulators}
\label{sec:photo_doped}
%%%%%%%%%%%%%%%%%%%%%%%%%%%%%%%%%%%%%%%%%%%%%%%%%%%%%%%%%%%%%%%%%%%%%%%%

In this section, we explore processes which control the evolution of photo-doped charge carriers in Mott insulators. After an electronic excitation across the Mott gap, depending on the type of system, we can expect two fundamentally different scenarios, see Fig.~\ref{fig_mott}. In small-gap Mott insulators  (or correlated metals), the electrons thermalize quickly.  Conversely, in large-gap Mott insulators, the dynamics can be described in terms of nonthermal populations of electron- and hole-like carriers (doublons and holons in a single-band Mott insulator). The central questions are how these carriers interact with other degrees of freedom, how they develop into dressed quasiparticles or bind into excitons, how the presence of these carriers modifies the band structure and interactions, whether the nonthermal carrier populations can stabilize new types of electronic orders, and how the photo-doped carriers ultimately recombine to restore the original Mott phase. In the following subsections, we will discuss the different stages in the evolution of a photo-doped Mott system, and the underlying physical phenomena. 

\subsection{Excitation across the Mott gap}
\label{sec:photocarrier_generation}

{\em Intraband acceleration ---}  The easiest way to electronically excite a Mott insulator is to apply a light pulse with a frequency comparable to the Mott gap. What distinguishes Mott insulators from conventional band insulators is the possibility of optical transitions between the lower and upper Hubbard bands, which originate from a single orbital. These transitions result from the intra-band acceleration of the electrons which is induced by the Peierls phases of the hopping terms  (Sec.~\ref{sec:coupling_to_light}). For low amplitudes, the absorbed energy is proportional to the pump intensity ($\propto E_0^2$, linear optical absorption), while in the strong-field regime, a significant increase in double occupancy can be generated within just a few cycles. 

%%%%%%%%%%%%%%%%%
\begin{figure}[t]
\begin{center}
\includegraphics[angle=0, width=0.9\columnwidth]{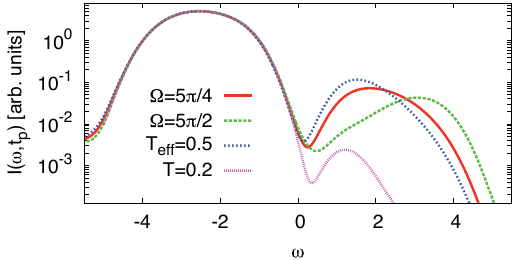}
\caption{
Time-resolved photoemission spectrum I$(\omega,t_p)$ after a resonant pump excitation in a single-band Mott insulator for two different pump energies $\Omega$. The pink (blue) dotted lines indicate the results for the initial (thermalized) system. The pulse amplitudes $E_0$ for the two driving frequencies $\Omega$ were chosen to be compatible with $T_\text{eff}$. Calculations based on DMFT+NCA on the hypercubic lattice ($v_*=1$) with interaction $U=5$. (Adapted from \onlinecite{Eckstein2011thermalization}.)} 
\label{fig_few_cycle}
\end{center}
\end{figure}   
%%%%%%%%%%%%%%%%%

{\em Interband dipole transitions ---} In generic multi-band systems, the light-matter interaction contains both dipolar transition matrix elements between orbitals of different symmetry and Peierls phases (Sec.~\ref{sec:coupling_to_light}). Hence, both intraband and interband transitions need to be considered. For sufficiently high frequency, excitations may occur also outside the valence band manifold, and in the absence of particle-hole asymmetry, transiently change the particle number in the valence band. This effect can be important for the initial stage of photo-induced insulator-metal transitions \cite{Stojchevska2014}, and exploited in tailored excitation protocols (Sec.~\ref{sec_entropy}).

{\em Nonlinear mechanisms ---}  Strong laser pulses enable non-resonant excitations across a Mott or charge transfer gap. The underlying mechanisms are discussed in Sec.~\ref{sec:single_multi} (multi-photon absorption) and Sec.~\ref{sec:breakdown} (Zener-type tunneling). The tunneling scenario applies to a situation where the pulse frequency is much lower than the Mott gap, as in the case of a THz pulse applied to a system with an eV gap.

In view of these different mechanisms, a targeted pump field excitation can create specific states. For instance, in the single-band simulation result shown in Fig.~\ref{fig_few_cycle}, the distribution function of electrons after the excitation differs for pulses with frequency $\Omega > U$ and $\Omega < U$, even if these pulses inject the same amount of energy. Similarly, multi-photon and tunneling processes lead to distinct doublon populations (Sec.~\ref{sec:single_multi} and Fig.~\ref{fig_multiphoton}). However, intra-band relaxation processes can quickly erase the information on the excitation pathway. For example, in a study of La$_2$CuO$_4$ based on the $d$-$p$ model for charge transfer insulators \cite{Golez2019a,Golez2019}, excitations producing holes in the lower Hubbard band or holes with mixed $d$-$p$ character (Zhang-Rice singlets \cite{Zhang1988}) were explored. It was shown that the rapid hole transfer between the Zhang-Rice band and the lower Hubbard band erases the information on the initial excitation on the femtosecond timescale, if electron-electron interactions and nonlocal charge fluctuations are taken into account.

\subsection{Thermalization of correlated electrons}
\label{sec:general_thermalization}

An isolated interacting many-particle quantum system is expected to eventually thermalize. The final state will then be indistinguishable from the Gibbs density matrix $\rho_{\rm th} \sim e^{-(H-\mu N)/T_{\rm eff}}$, with parameters $\mu$ and $T_{\rm eff}$ that depend only on the total energy and particle number. Also, correlation functions will satisfy a universal fluctuation-dissipation relation with the effective temperature $T_{\rm eff}$. Understanding the timescale on which this ergodic behavior is established is a fundamental problem of nonequilibrium quantum physics~\cite{polkovnikov2011,Alessio2016}, and crucial for determining whether laser excitations can lead to nonthermal electronic phases.

In the absence of symmetry-breaking, correlated electrons can exist in three regimes: conventional Fermi liquid states, strongly correlated unconventional metals, and robust Mott or charge transfer insulators. In ideal Fermi liquids, the phase space for quasiparticle scattering is restricted, causing thermalization to slow down as the temperature approaches zero. The relaxation proceeds in two stages, where the rapid formation of a prethermal phase~\cite{Berges2004,Moeckel2008a,Hackl2009a,Eckstein2009,epjst2010a} is followed by a slower thermalization described by a Boltzmann equation~\cite{Haug2008a,Kollar2011a}. Nevertheless, even in good metals, this second stage is often fast enough that the electronic subsystem thermalizes on femtosecond timescales, and thus before a significant amount of energy is transferred to the lattice. This assumption underlies widely used phenomenological few-temperature models \cite{Allen1987}, which describe photo-excited electrons as quasi-thermal hot electrons.

%%%%%%%%%%%%%%%%%
\begin{figure}[t]
\begin{center}
\includegraphics[angle=0, width=\columnwidth]{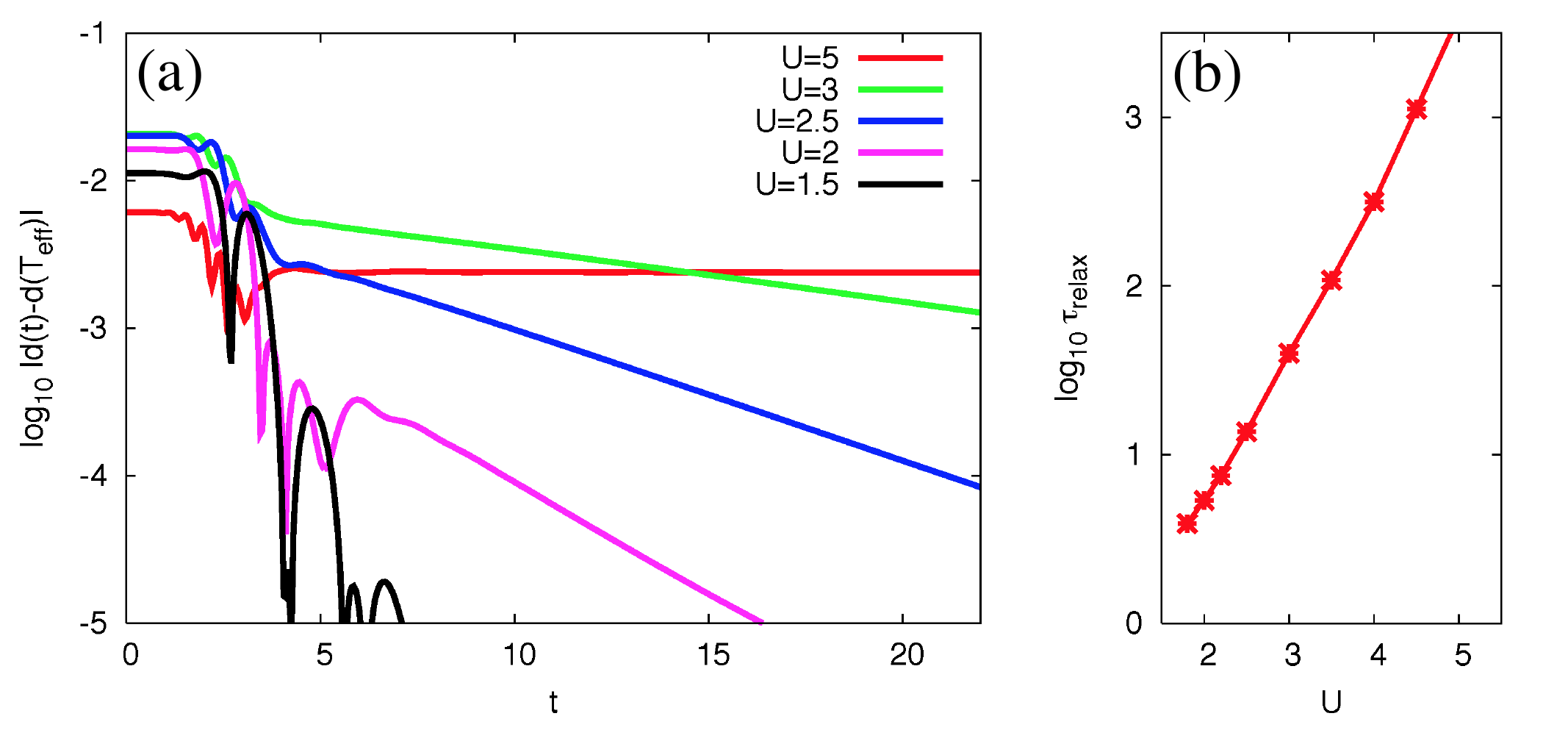}
\caption{
(a) Thermalization of the double occupancy in a photo-doped Hubbard model with indicated values of $U$ in the metal-insulator crossover region. 
(b) Thermalization time as a function of interaction strength. Same setting as in Fig.~\ref{fig_few_cycle}, with a resonant excitation $\Omega=U\pi/2$.  The pulse amplitudes $E_0$ are adjusted such that $T_\text{eff}=0.5$ for all values of $U$. (Adapted from \onlinecite{Eckstein2011thermalization}.)}
\label{fig_thermal}
\end{center}
\end{figure}  
%%%%%%%%%%%%%%%%%

Stronger interactions do not necessarily lead to faster thermalization. In particular, a large-gap Mott insulator exhibits a thermalization bottleneck, as observed in interaction quenches of the Hubbard model in both one and infinite dimensions \cite{Kollath2007,Eckstein2009}. Further simulations for the isolated single-band Hubbard model \cite{Eckstein2011thermalization} revealed a crossover between rapid thermalization at intermediate interactions (in the insulator-to-metal crossover regime) and an exponential slowdown in the Mott state. Figure~\ref{fig_thermal}(a) illustrates this behavior by plotting the absolute value of the difference between the double occupancy $d(t)$ and the thermalized double occupancy $d(T_{\rm eff})$ after a resonant photo-excitation. Here, $d(T_{\rm eff})$ is obtained from a Gibbs state with the final state temperature $T_{\rm eff}$ determined by the absorbed energy. The exponential decay of $|d(t)-d(T_{\rm eff})|$ provides an estimate for the thermalization time $\tau$, which is plotted in Fig.~\ref{fig_thermal}(b). Deep in the Mott insulator ($U=5$), the relaxation time is hundreds of inverse hoppings, while it becomes faster as we approach the insulator-metal crossover region~($U_{\text{crossover}}/v_*\approx 2.6$ 
for the present setting \cite{Werner2014}). In the strongly correlated metal ($U=1.5$), the system thermalizes within a few inverse hopping times. 

In the following, we separately discuss these two cases of fast thermalization and exponential slowdown.

{\em Fast thermalization in the correlated metal ---} 
At intermediate interactions, the thermalized state can become a bad metal with a temperature $T_{\rm eff}$ above the Fermi-liquid coherence temperature. In this state, quasiparticles are not well-defined~\cite{deng2013}, and rapid thermalization, as observed in Fig.~\ref{fig_thermal}, cannot be understood within conventional Boltzmann scattering theory. Sachdev-Ye-Kitaev models \cite{syk_rmp}, which can be solved with equations that are closely related to DMFT, capture rapid thermalization in a bad metallic non-Fermi liquid regime \cite{Eberlein2017}. Here, one finds a universal thermalization rate determined solely by the temperature $T$. However, a comprehensive understanding of thermalization in strongly correlated unconventional metals is still lacking. The discussion above pertains to single-particle quantities, while two-particle quantities like spin and charge correlation functions may thermalize on different timescales~\cite{Matveev2019,Sayyad2019,Simard2022}. Multi-orbital systems, which already show unconventional slow behavior such as spin freezing in equilibrium \cite{Werner2008}, have not yet been systematically explored. In disordered metals, thermalization can be delayed as shown in the analyses of the Falicov Kimball model \cite{Eckstein2008a,Eckstein2008b}. The interplay of strong disorder and interactions in a nonequilibrium context, and the possible relations to many-body localization \cite{Abanin2019}, remain largely unexplored.

The rapid electronic thermalization in the small-gap regime may appear as an obstacle to the realization of nonthermal electronic phases. However, it  provides a unique opportunity to study electronic states and their properties at effective temperatures which are much higher than those achievable in equilibrium, while the lattice still remains frozen. An early example of such an exploration is a time-resolved photoemission experiment on the correlated insulator $1T$-TaS$_2$  \cite{Perfetti2006}, see Sec.~\ref{sec:exp_probes}. Another more recent example is a tr-ARPES study on SrIrO$_4$ \cite{choi2023}. Here, the authors argued that the specific behavior of the spectral function in the high-temperature regime, which can be accessed by ultrafast laser excitations, allows to identify antiferromagnetic correlations rather than Mott physics as the primary origin of the insulating state. 

%%%%%%%%%%%%%%%%%
\begin{figure}[t]
\begin{center}
\includegraphics[angle=0, width=0.8\columnwidth]{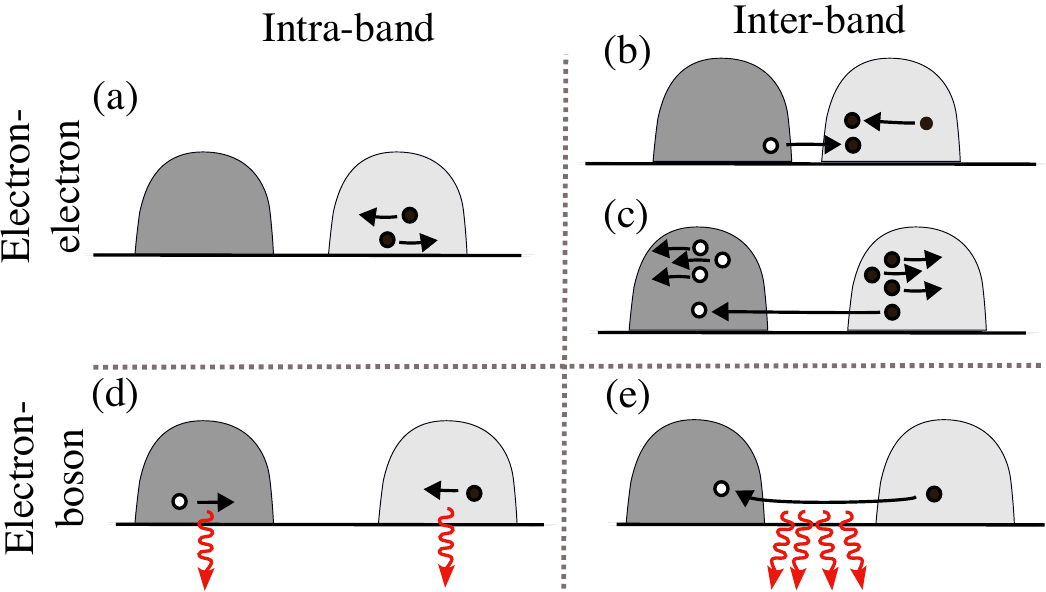}
\caption{
Illustration of different scattering processes in Mott insulators, as discussed in the text. Depending on the initial state, these processes can go in both directions.
}
\label{fig_relaxation_processes}
\end{center}
\end{figure} 
%%%%%%%%%%%%%%%%%  

{\em Thermalization bottleneck in the Mott regime---} 
In large-gap Mott insulators, the relaxation and thermalization dynamics can be understood by considering the scattering of carriers (doublons and holons in the single-band case) among each other and with bosonic excitations, such as spin or charge excitations and phonons. Figure~\ref{fig_relaxation_processes} schematically illustrates different processes: One may distinguish electron-electron scattering processes which can thermalize the electrons (upper panels) from electron-boson scattering processes where the energy is transferred to other degrees of freedom (lower panels). On the other hand, there are processes which change the occupation in the upper band (recombination/excitation, right panels), and intra-band processes which keep this number constant~(left panels). The intra-band relaxation modifies the distribution of photo-excited doublons and holons by either electron-electron scattering~(Fig.~\ref{fig_relaxation_processes}(a)) or electron-boson scattering~(Fig.~\ref{fig_relaxation_processes}(d)). The inter-band relaxation by single-particle scattering, i.e., impact ionization and its inverse (Fig.~\ref{fig_relaxation_processes}(b)), is possible only in small-gap insulators, see Sec.~\ref{sec:photo_doped}C. As the Mott gap $\Delta_{\rm Mott}$ (or the local interaction $U$) becomes large compared to other energy scales, the remaining high-order inter-band processes (Fig.~\ref{fig_relaxation_processes}(c) and (e))  become exponentially suppressed with increasing gap size \cite{Rosch2008,Strohmaier2010,Sensarma2010a}. This is because the recombination (creation) of charge carries releases (requires) an energy of order $U$, which must be transferred simultaneously to (from) many low-energy excitations with a smaller energy $\epsilon_0 \ll U$. Here, $\epsilon_0$ can be the kinetic energy of the carriers ($v_0$, hopping), as in Fig.~\ref{fig_relaxation_processes}(c), or the energy of the bosonic excitation, as in Fig.~\ref{fig_relaxation_processes}(e). Based on this, an asymptotic estimate how the rate $\Gamma$  for these processes depends on the gap $U$ can be obtained from a high-order perturbation theory~\cite{Sensarma2010a},
\begin{equation}
\Gamma\sim  \exp[-\alpha(U/\epsilon_0)\log(U/v_0)].
\label{eq_thermalization}
\end{equation}
The coefficient $\alpha$ in this expression is a phase space factor which depends on the number of available scattering partners, such as the density of 
doublons and holons in the final thermal state. The $U$-dependence of the thermalization time of the DMFT results in Fig.~\ref{fig_thermal}(b) can be well fitted by Eq.~\eqref{eq_thermalization} with $\alpha$ close to one.  

The slowdown of inter-band processes explains the electronic thermalization bottleneck in isolated Mott systems, because thermalization of the electronic single-particle properties requires that the change in the number of doublons is compensated by a change in the kinetic energy, as illustrated in Figs.~\ref{fig_relaxation_processes}(b,c,e). In the following sections, we discuss in detail the processes depicted in Fig.~\ref{fig_relaxation_processes} for the thermalization of Mott insulators. In particular, in open systems with a robust Mott gap,  the photo-carrier generation is followed by intra-band relaxation of the carriers (Sec.~\ref{sec:intraband_relax}), and subsequently by the recombination of carriers (Sec.~\ref{sec_recombination}), as shown in Fig.~\ref{fig_mott}(c). We will also discuss phenomena that are directly related to the presence of nonthermal charge carriers, such as Mott exciton formation (Sec.~\ref{sec:MottExciton}) and the renormalization of the interactions and band structure by the photo-carriers (Sec.~\ref{sec:dynamical}).

\subsection{Impact ionization}

Impact ionization is a process where a high-energy charge carrier loses kinetic energy by creating other charge carriers (Fig.~\ref{fig_relaxation_processes}(b)). This process is well understood in semiconductors \cite{bhattacharya1997} and finds applications in avalanche photodiodes. It has also been discussed as a potential mechanism to overcome the Shockley-Queisser limit for the efficiency of semiconductor solar cells \cite{werner1994,spirkl1995}. Analogous processes play a role in Mott insulators, if the Mott gap is smaller than the width of the Hubbard bands. Here, the injection of doublons and holons with high kinetic energy can lead to a multiplication of charge carriers \cite{Manousakis2010,Werner14impact}. In impact ionization, a doublon near the upper edge of the upper Hubbard band scatters to the lower edge, exciting an additional doublon-holon~(D-H) pair with low kinetic energy: $\text{doublon}_\text{high} \rightarrow \text{doublon}_\text{low} + \text{doublon}_\text{low} + \text{holon}_\text{low}$, see illustration in Fig.~\ref{fig_relaxation_processes}(b), where full and empty dots represent doublons and holons, respectively. A similar process occurs when a holon with high kinetic energy scatters within the lower Hubbard band. In a particle-hole symmetric situation, the net effect of these impact ionization processes is that each high-energy D-H pair produces three low-energy D-H pairs.

%%%%%%%%%%%%%%%%%
\begin{figure}[t]
\begin{center}
\includegraphics[angle=0, width=\columnwidth]{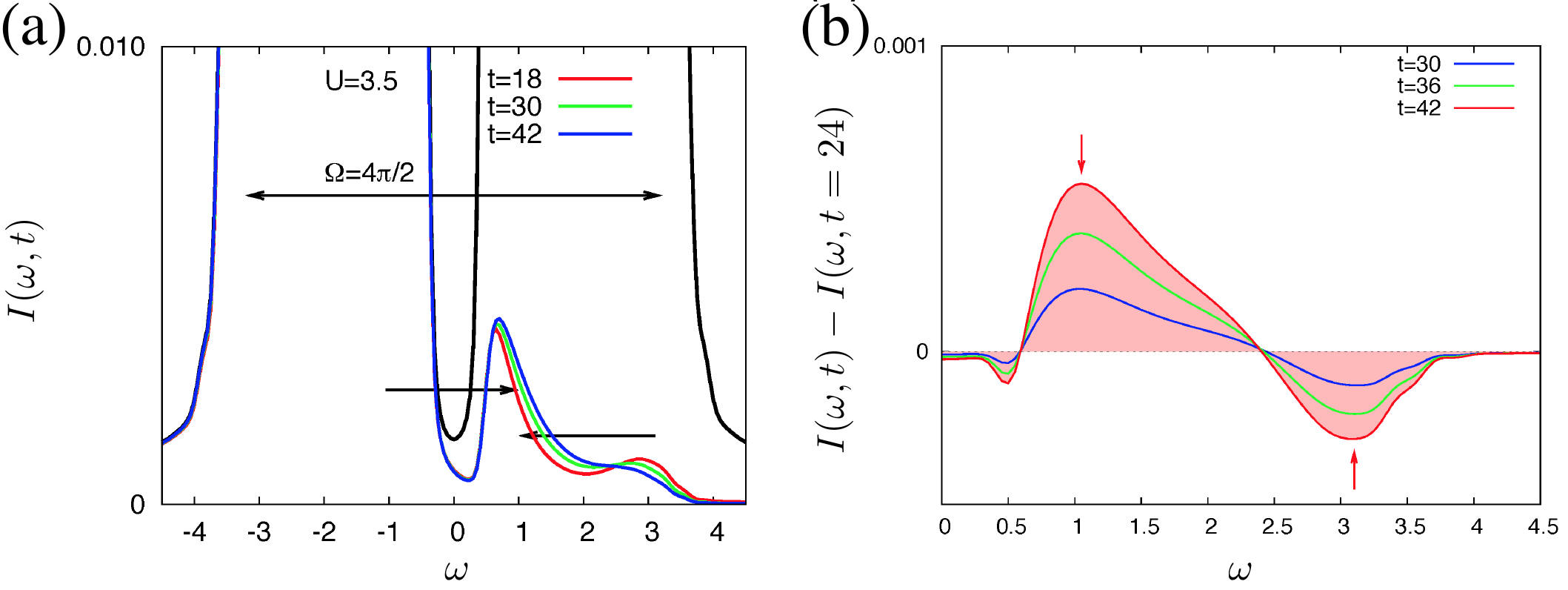}
\caption{
(a) Time evolution  of the photo-emission spectrum $I(\omega,t)$ after a pulse excitation with energy $\Omega=4\pi/2$ and amplitude $E_0=0.5$. (b) Time-dependent change of the photo-emission spectrum relative to time $t=24$. DMFT+NCA results for a single-band Hubbard model on a hypercubic lattice with $U=3.5$ and $v_*=1$. (Adapted from \onlinecite{Werner14impact}.)
}
\label{fig_impact}
\end{center}
\end{figure}
%%%%%%%%%%%%%%%%%

The multiplication of charge carriers and the associated changes in their energy distribution are evident in the short-time dynamics of the photoemission spectrum, as shown in Fig.~\ref{fig_impact} for a single-band Hubbard model. Panel (a) plots $I(\omega,t)$ at three times after a photo-doping pulse which inserted doublons (holons) near the upper (lower) edge of the upper (lower) Hubbard band. It is evident that the spectral weight associated with high-energy doublons decreases over time, while the spectral weight near the lower band edge increases. To analyze the redistribution of spectral weight, we display in panel (b) the change in the photoemission spectrum relative to the result at time $t=24$. This clearly reveals how spectral weight disappears at an energy $\omega\approx 3$, while there is an approximately 2.3 times larger increase in spectral weight near $\omega\approx 1$. If impact ionization were the only process responsible for this reshuffling of spectral weight, we would expect a $3$ times larger increase in low-energy spectral weight. The observed number is somewhat smaller due to intra-band scattering processes (Fig.~\ref{fig_relaxation_processes}(a)). Similar results for impact ionization were obtained using a quantum Boltzmann description without momentum conservation \cite{Wais2018}. Impact ionization in Mott insulators has also been studied using a steady-state Floquet set-up \cite{Sorantin2018}.

\begin{figure}[t]
\begin{center}
\includegraphics[angle=0, width=\columnwidth]{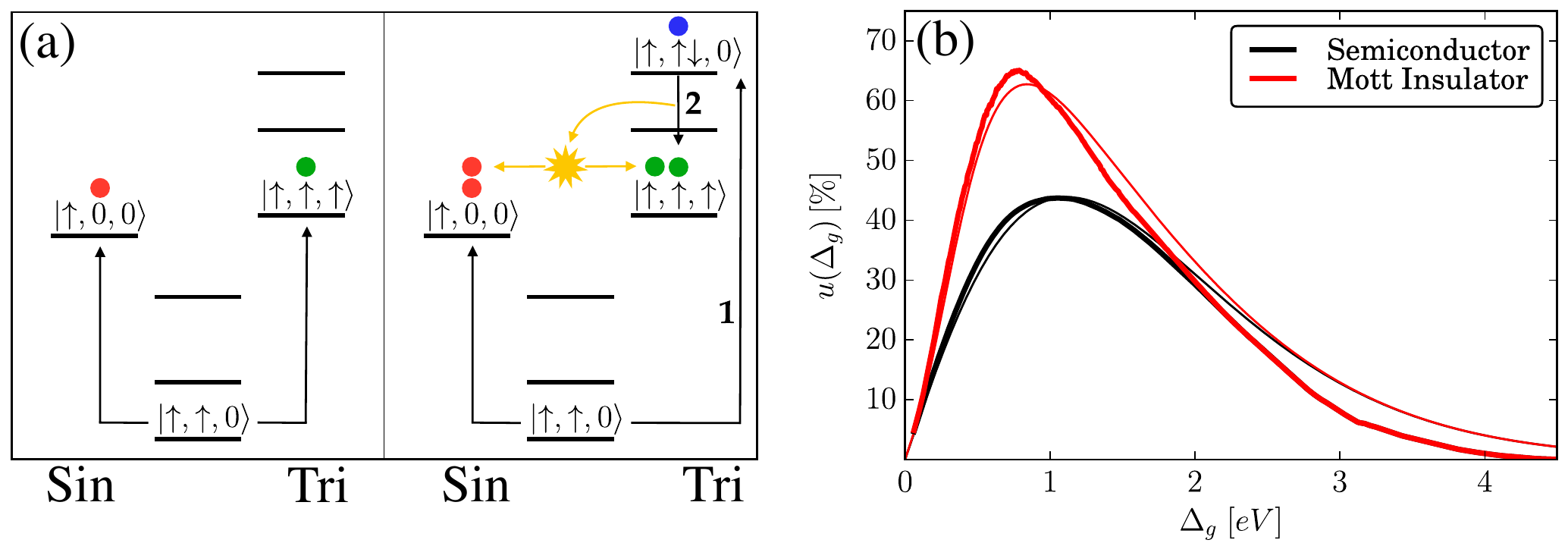}
\caption{
(a)~Excitation of a low-energy singlon~(Sin) and triplon~(Tri) pair in a three-orbital Mott insulator with average occupation $n=2$ by a low-energy field pulse (left subpanel), and excitation of a high-energy triplon (1), with subsequent Hund impact ionization (2). (b)~Ultimate quantum efficiency $u(\Delta_g)$ of the Mott insulating system compared to that of a semiconductor solar cell. Here the thin (thick) lines show the results for a black body (solar) spectrum. Inhomogenous DMFT+NCA results for the three-orbital Slater-Kanamori Hamiltonian.
(From \onlinecite{Petocchi2019}.) 
}
\label{fig_hund_impact}
\end{center}
\end{figure}

Following initial proposals in~\cite{Manousakis2010,Assmann2013,Werner2014}, several studies exploited the carrier multiplication due to impact ionization in Mott systems to design efficient photovoltaic devices. A strategy based on combining heterostructures and magnetic materials was presented in Sec.~\ref{mobility}. Further enhancements may be realized in multi-orbital systems, where the Hund coupling leads to subbands in the Hubbard bands, corresponding to local spin excitations (Hund excitations). Their energy separation is $U$, $U-2J_\text{H}$ and $U-3J_\text{H}$ in the case of model (\ref{H_multiorbital}) with density-density interactions. If the splitting between these Hubbard subbands is comparable to the gap, transitions between Hund excitations can produce additional charge carriers \cite{Petocchi2019}. For example, a three-band Mott insulator with two electrons per site and positive $J_\text{H}$ is dominated by high-spin doublon states of the type $|\!\uparrow,\uparrow,0\rangle$.  A photo-doping pulse with sufficiently high energy can produce a singlon-triplon pair, with the triplon in the low-spin configuration $|\!\!\uparrow,\uparrow\downarrow,0\rangle$ (corresponding to the highest-energy subband in the upper Hubbard band), see step 1 in the right subpanel of Fig.~\ref{fig_hund_impact}(a). Such a low-spin triplon can be converted into a high-energy triplon plus an additional low-energy singlon-triplon pair (step 2) if the gap size is comparable to or smaller than $3J_\text{H}$,  which in typical transition metal compounds can be as large as 2-3 eV. Impact ionization by Hund excitations is conceptually related to singlet-fission in molecular systems or semi-conductor quantum dots \cite{Nozik2010}.

Figure \ref{fig_hund_impact}(b) compares the ultimate quantum efficiency of a Mott insulator with impact ionization to that of a semiconductor. The Shockley-Queisser \cite{Shockley1961} estimate of the ultimate quantum efficiency assumes that charge excitations across a gap $\Delta_g$ in semiconductors contribute an energy $\Delta_g$. For the Mott insulator estimate, it was assumed that photons with frequency $\Omega>2\Delta_g$ ($\Omega>3\Delta_g$) produce twice (three times) as many charge carriers than photons with frequency $\Delta_g<\Omega<2\Delta_g$. This comparison suggests that the efficiency of a Mott system which fully exploits impact ionization can be larger than 60\%, and that the optimal gap size of a Mott solar cell is smaller than that of a semi-conductor solar cell.  

%%%%%%%%%%%%%%%%%
%%%%%%%%%%%%%%%%%

\subsection{Intra-band relaxation and cooling of charge carriers}
\label{sec:intraband_relax}

If the Mott gap is large, intra-band relaxation processes (Fig.~\ref{fig_relaxation_processes}(d)) can precede electron thermalization and carrier recombination. 
These processes  lead to a reduced kinetic energy and will be called carrier cooling. Cooling before recombination is a typical prerequisite for the realization of novel nonthermal phases. Moreover, the processes behind the carrier cooling describe the formation of dressed quasiparticles, such as spin and lattice polarons, and are therefore important for understanding the  properties of the long-lived photo-doped state. 

\subsubsection{Weak-coupling scenario}

Generic aspects of carrier cooling can be understood within a weak-coupling scenario. In Green's function methods, electron-boson scattering can be described by a self-energy contribution $\Sigma_B$ given by the Migdal diagram $\Sigma_k(t,t')=\sum_{q} i|g_q|^2G_{k-q}(t,t')D_{q}(t,t')$, where $G(t,t')$ is the electron Green's function, $D_q(t,t')$ the equilibrium bosonic propagator, and different types of bosons (phonons, magnons, plasmons) are distinguished only by their couplings $g_q$ and dispersions. This formalism is closely related to a kinetic description of population dynamics, where relaxation rates are determined by scattering matrix elements $\propto g_q$ and the available phase space for the bosons and electrons. This theory can be used to numerically simulate the carrier cooling rate (see below), and a reverse argument can be applied to extract information on the bosons (spectrum and couplings $g_q$) from the relaxation times \cite{Giannetti2016, DalConte2012,Novelli2014,golez2022}. For example, gapped bosonic modes relevant for the relaxation can be identified by a bottleneck effect, since electrons with energy $\epsilon-E_F<\omega_0$ cannot relax further via the emission of a bosons of energy $\omega_0$ at sufficiently low temperature \cite{Sentef2013,rameau2016, Murakami2015,Kemper2017}. 

The weak-coupling approach has been used to simulate the intra-band relaxation of photo-excited carriers in Mott insulators \cite{Eckstein2013}. In large-gap systems, the kinetic energy of the photo-doped doublons and holons is transferred to the boson bath, while the density of doublons and holons is approximately conserved. 
This cooling effect manifests itself in the accumulation of spectral weight of the doublons (holons) at the lower (upper) edge of the respective Hubbard band. Recent long-time DMFT simulations, up to $t\approx 2000$ inverse hoppings (corresponding to ps in realistic materials), explicitly demonstrated the formation of a quasi-stationary state in which the distribution function $N(\omega,t)/A(\omega,t)$ within the upper and lower Hubbard band is well approximated by Fermi functions $f(\omega-\mu,T_{\rm eff})$ with separate chemical potentials for the doublons and holons \cite{Dasari2020}. For sufficiently low temperature $T_{\rm eff}$ in the quasi-steady state, one observes the emergence of narrow bands close to the band edges, which shows that both the doublons and holons behave as strongly renormalized quasiparticles in the cold photo-doped state. 
Moreover, the comparison of different excitation protocols (such as laser excitations and ``temperature quenches'', in which the initial carrier distribution is thermally excited) revealed that the quasi-steady spectral function does not depend on details of the excitation protocol, but only on the effective temperature $T_{\rm eff}$ and the two effective chemical potentials.  This numerical observation justifies the quasi-steady description of photo-doped states with a few generalized thermodynamic variables, as explained in Sec.~\ref{sec:quasiseq}. Still, a detailed understanding of the formation time of this universal state is presently lacking. DMFT simulations indicate that the transfer of energy to the bath becomes less efficient close to the onset of the quasiparticle formation \cite{sayyad2016,Dasari2020}, so that for a remarkably long time after the excitation, a photo-doped metal state can still be less coherent than a doped Mott insulator with a comparable density of carriers and the temperature of the bosonic bath \cite{Eckstein2013}. 

\subsubsection{Electron-phonon interactions}
\label{sec:electron-phonon}
\label{sec_boson}

The coupling to phonons is a natural mechanism for carrier cooling, and qualitatively captured by the weak-coupling description of the previous subsection. The simplest relevant model is the Holstein-Hubbard model (\ref{H_hubbardholstein}), which describes the linear coupling of the electronic density to the displacement of an Einstein phonon mode.  The dynamics for a wide range of couplings has been studied by ED for a single Holstein polaron in \onlinecite{Golez2012b}, and using MPS simulations~\cite{Dorfner2015}. At weak couplings, the relaxation rate of photo-excited charge carriers is consistent with the rate expected from Fermi's golden rule, $1/\tau(\omega)\propto g^2 \rho(\omega-\omega_0),$ where $\rho$ is the electronic density of states, $\omega_0$ the boson frequency, and $g$ the coupling constant. At stronger couplings, the relaxation timescales deviate from this simple scaling and eventually the feedback of the excited lattice modes to the electronic dynamics is strong enough that the system oscillates in a self-formed potential trap. 

%%%%%%%%%%%%%%%%%
\begin{figure}[t]
\begin{center}
\includegraphics[angle=0, width=\columnwidth]{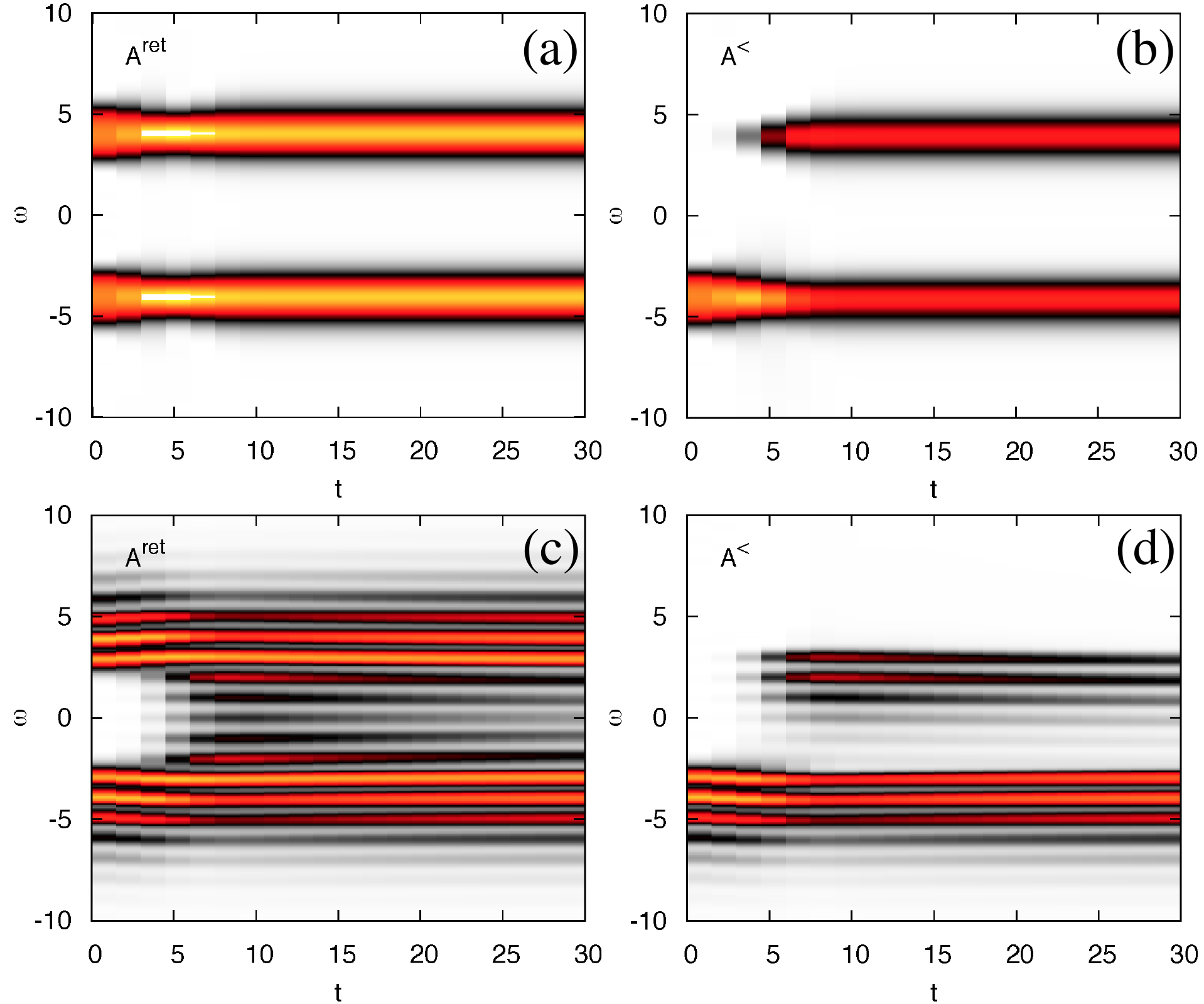}
\caption{
Spectral function $A(\omega,t)$ ($A^\text{ret}$) and occupied states $N(\omega,t)$ ($A^<$) of the photo-doped Mott insulating Hubbard model (a,b) and Holstein-Hubbard model (c,d). 
D-H pairs are produced by a few-cycle pulse with frequency $\Omega=U=8$, centered around $t=6$. DMFT+NCA calculations for a hypercubic lattice with $v_*=1$, $\omega_0=1$ and $g=1$. (Adapted from \onlinecite{Werner15epl}.) 
}
\label{fig_a_hubbard_hubbardholstein}
\end{center}
\end{figure}   
%%%%%%%%%%%%%%%%%

The properties of the photo-doped Mott insulator described by the Holstein-Hubbard model have been systematically studied using DMFT combined with a variant of NCA based on  the Lang-Firsov decoupling of the electron-phonon term \cite{Werner15epl}. The cooling rate $1/\tau$ of the kinetic energy, as determined from an exponential fit to the time-dependent kinetic energy, was found to be proportional to $g^2/\omega_0$ in the weak-coupling limit, in contrast to the scaling for a single polaron or in a metallic system. In particular, the decay time of the kinetic energy in metallic states is expected to be proportional to $g^2\omega_0$~\cite{Inayoshi}. For stronger couplings, the DMFT simulations help to clarify the effects of polaron formation on the electronic properties. Figure~\ref{fig_a_hubbard_hubbardholstein} compares the spectral functions (panels (a,c)) and occupation functions (panels (b,d)) of the Hubbard and Holstein-Hubbard models. In the Hubbard model (panels (a,b)), the photo-doping pulse centered near $t=6$ has little effect on the spectral function, and  there are no in-gap states generated by the photo-doping. The result is very different in the Holstein-Hubbard case (panels (c,d)). The effect is demonstrated with an artificially large phonon coupling $g=1$, which splits the Hubbard bands into phonon side-bands separated by the energy $\omega_0=1$. Already during the pulse, additional phonon sidebands emerge in the gap, corresponding to electron insertion or removal processes with simultaneous emission or absorption of phonons, c.f. Fig.~\ref{fig_mott}. They appear in the spectrum because in a dynamical phonon calculation, photo-doped doublons and holons get dressed by lattice excitations. The occupied part of the spectrum ($A^<$) shows that in the Holstein-Hubbard case, the photo-doped doublons and holons mainly occupy these induced in-gap states, which significantly affects the D-H recombination rate and thermalization process (Sec.~\ref{sec_recombination}).

\subsubsection{Electron-spin interactions}
\label{sec:electron-spin}
\label{electron-spin}

\begin{figure}[t]
\begin{center}
\includegraphics[angle=0, width=0.9\columnwidth]{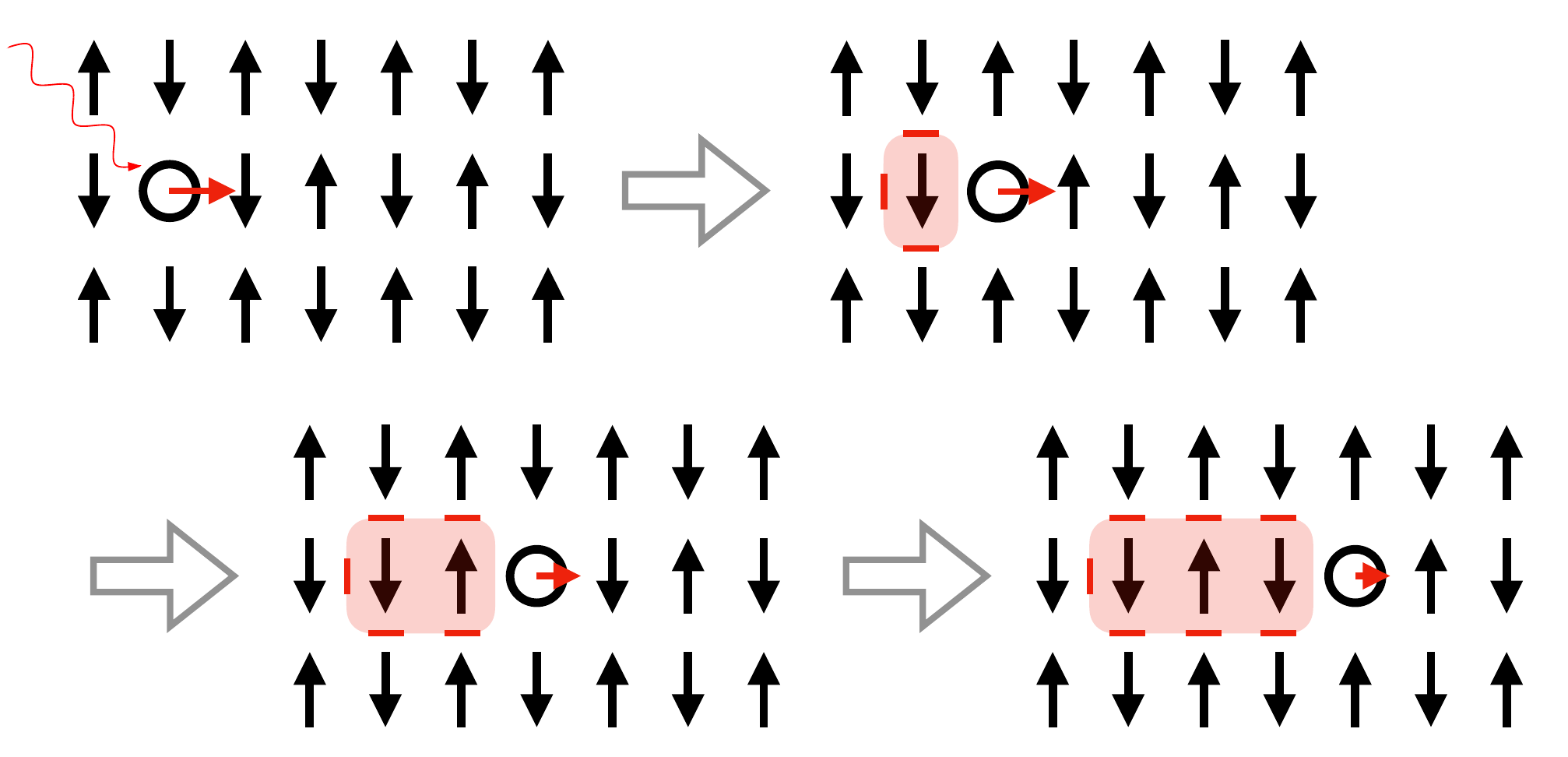}
\caption{
Formation of a string state by a holon moving in an antiferromagnetic spin background. The red bars indicate broken antiferromagnetic bonds, which cost an energy $J_\text{ex}$, and the red arrows the holon's motion~(kinetic energy). 
}
\label{fig_string}
\end{center}
\end{figure}

An efficient cooling mechanism for photo-doped doublons and holons is the interaction with antiferromagnetically ordered spins \cite{takahashi2002,Werner2012afm,Iyoda2014,Golez2014,DalConte2015,Eckstein2016dca,bohrdt2020}. In a half-filled Mott insulator with N\'eel order and dimension $D\ge 2$, each hopping of a doublon or holon flips a spin against the antiferromagnetically ordered surrounding, which costs energy of the order of the exchange coupling $J_\text{ex}$ (Fig.~\ref{fig_string}). The moving doublon or holon thus leaves behind a string of flipped spins, and exhausts its kinetic energy after just a few hoppings.  This mechanism is also relevant for systems with only short-range antiferromagnetic correlations, such as high-temperature superconductors in the pseudo-gap phase \cite{lee2006}. 

%%%%%%%%%%%%%%%%%
%%%%%%%%%%%%%%%%%

\begin{figure}[t]
\begin{center}
\includegraphics[angle=0, width=1.0\columnwidth]{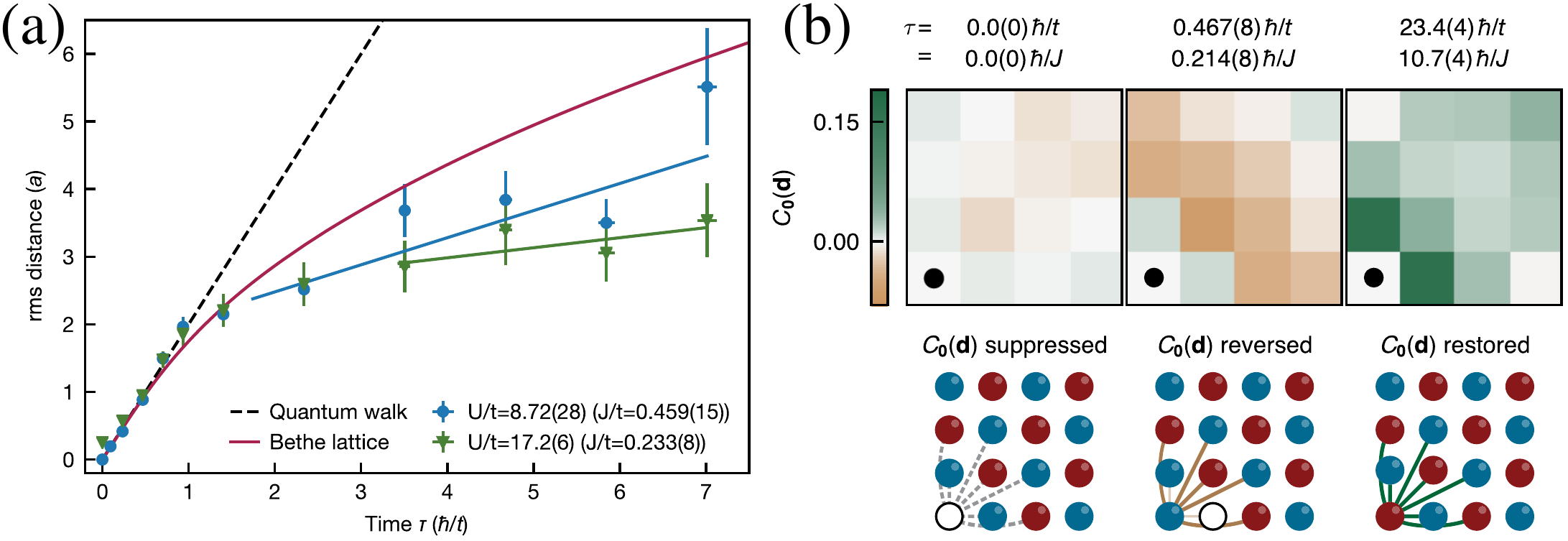}
\caption{(a)~Time evolution  of the root-mean-square distance for a hole suddenly inserted into the 2D square lattice Hubbard model, where time is measured in inverse hopping $\hbar/v_0\equiv \hbar/t.$ Results obtained with a cold-atom simulator~(dots) for two values on the Coulomb repulsion $U/v_0=8.72$ and $U/v_0=17.2$ are compared with a quantum walk model and the Bethe lattice evolution (see text). (b)~Time evolution  of the spin correlation function $C_0(d)$ with brown~(green) colors marking ferromagnetic~(antiferromagnetic) correlations with respect to the N\'eel background.~(Adapted from~\onlinecite{ji2021}.)}
\label{fig_polaron_spin}
\end{center}
\end{figure}

Detailed investigations of the coupled charge-spin dynamics have been conducted in the small-doping regime by either exciting a single holon in the AFM background or by inserting a holon into the $t$-$J$ model using ED \cite{takahashi2002,Golez2014,DalConte2015,hahn2022}, real-time QMC at high temperatures~\cite{kanasz2017} and MPS approaches~\cite{Hubig2020,bohrdt2020}. These studies revealed a two-stage dynamics, which was recently confirmed with a cold atom simulator~\cite{ji2021}, see Fig.~\ref{fig_polaron_spin}. At short times, the initially localized holon spreads ballistically with the maximal group velocity. The root-mean-square of the position in units of lattice spacing grows as $d_\text{rms}=\frac{2 v_0}{\hbar} t$, where $v_0$ is the hopping parameter, as in a non-interacting quantum walk, see dashed line in Fig.~\ref{fig_polaron_spin}(a). 
The subsequent stage exhibits a slowdown and diffusive-like spreading ($d_\text{rms}\propto \sqrt{t}$) with a sensitive temperature dependence. The slowdown can be analytically understood by considering a free-particle walk on the Bethe lattice in a spinful environment~\cite{Golez2014,kanasz2017}. In the low-temperature regime ($T<J_{\text{ex}}$), the spreading scales with the super-exchange $J_{\text{ex}}$~\cite{ji2021,kogoj2014}. At early times, holes leave behind a string  of antiferromagnetic correlations with the opposite sign, similar to the sketch in Fig.~\ref{fig_string}, although at later times spin fluctuations restore the antiferromagnetic correlations, see Fig.~\ref{fig_polaron_spin}(b). The high-temperature regime was modeled by assuming no quantum interference between paths leading to the same holon position, while different paths leave behind a different spin background~\cite{kanasz2017}. A recent $t$-$J_z$ model study showed that near the N\'eel temperature there exists a temperature-driven confinement-deconfiment transition between the two regimes~\cite{hahn2022}.

%%%%%%%%%%%%%%%%%
%%%%%%%%%%%%%%%%%

\begin{figure}[t]
\begin{center}
\includegraphics[angle=0, width=0.99\columnwidth]{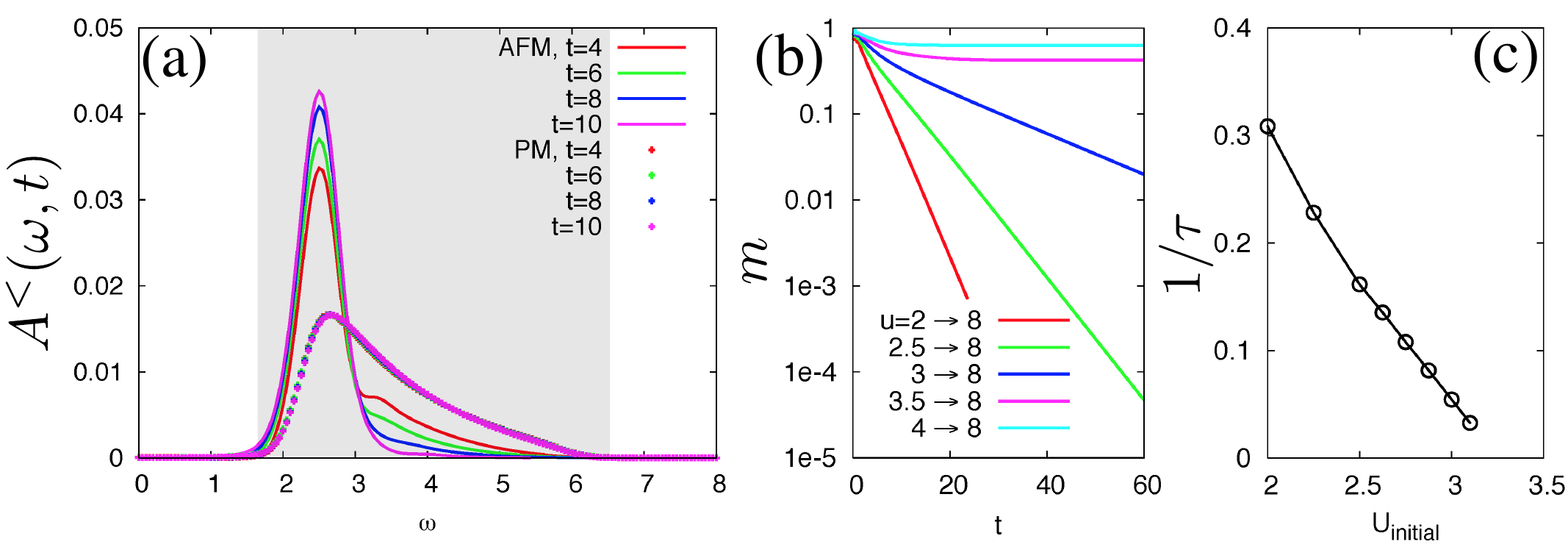}
\caption{(a) Occupied part of the spectral function within the upper Hubbard band~(shaded area) for an antiferromagnetically ordered Hubbard model after an interaction quench from $U=4\rightarrow 12$. Cross symbols show the paramagnetic results. (b) Time evolution  of the staggered magnetization $m$ after an interaction quench from the indicated values of the initial $U$ to $U_\text{final}=8$. (c) Inverse relaxation time as a function of the initial $U$. DMFT+NCA calculations on the Bethe lattice ($W=4$) for initial temperature $T=0.1$.~(Adapted from \onlinecite{Werner2012afm}.)
}
\label{fig_afm_semi}
\end{center}
\end{figure}

DMFT-based approaches were used to discuss the effects of larger photo-doping densities. Single-site DMFT allows to study photo-doping in a Hubbard model with long-range antiferromagnetic order, such as N\'eel order on a bipartite lattice \cite{Werner2012afm}. Figure~\ref{fig_afm_semi}(a) plots the occupied density of states $N(\omega,t)\equiv A^<(\omega,t)$ for a half-filled antiferromagnetically ordered Mott insulator (upper Hubbard band). The ``photo-doping" is mimicked here by a quench from a moderately correlated initial state $U=4\rightarrow12$, but the excitation with a short electric field pulse yields similar results. In the antiferromagnetically ordered system (lines) the spectral weight rapidly accumulates near the bottom of the Hubbard band (around $\omega\approx 2$) due to the energy dissipation to the spin background. In a paramagnetic calculation (cross symbols), this dissipation mechanism is absent, and there is no significant change in the energy distribution on the time scale of the calculation. The carrier cooling is accompanied by a reduction of the antiferromagnetic order (Fig.~\ref{fig_afm_semi}(b)). At low photo-doping concentrations, the order decreases only slightly and the system is trapped in a long-lived prethermal antiferromagnetic state. As the photodoping concentration increases, the system undergoes a dynamical phase transition with a critical slowdown (Fig.~\ref{fig_afm_semi}(c)). An analogous slowdown has been reported in Slater antiferromagnets with weaker interaction strengths~($U<W$)~\cite{tsuji2013}, although with a different lifetime of the prethermal phase~\cite{picano2021}.

A related intra-band carrier relaxation is observed in systems without long-range order, but strong short-range antiferromagnetic correlations, which can be realized either by heating above the N\'eel temperature~\cite{Eckstein2016dca,Gillmeister2020} or by chemical doping~\cite{Bittner2018}. DMFT simulations fail to capture these correlations, resulting in inefficient carrier cooling in the paramagnetic phase as shown in Fig.~\ref{fig_afm_semi}(a). Fast cooling in the paramagnetic phase has been demonstrated in the Mott-insulating half-filled square lattice Hubbard model using the 4-site DCA description \cite{Eckstein2016dca}. Here, spin correlations within a 4-site cluster are considered, while longer-range correlations are neglected. The cooling rate was determined from the decrease of the photoemission spectral weight near the upper edge of the upper Hubbard band, and was shown to scale quadratically with the nearest-neighbor spin correlations $S_\text{NN}$. This scaling can be rationalized by a simple rate equation argument. Since the antiferromagnetic energy $E_\text{AFM}$ is quadratic in the short-range order $m$, and the rate of generation of spin flips is proportional to $m$, one finds $dE_\text{AFM}/dt\propto m(dm/dt)\propto m^2$. With increasing temperature, $S_\text{NN}$ decreases, leading to a slower dissipation of kinetic energy to the spin background. Similarly, a saturation of the cooling rate occurs at high photo-doping density, where $S_\text{NN}$ is reduced by the effect of the pulse. The fragility of the cooling via electron-spin interactions should be contrasted with the cooling via phonons, which is more robust against the excitation intensity~\cite{Eckstein2016dca,Kogoj2016}. In non-bipartite lattices, the intra-band relaxation due to scattering with the spin background can be reduced as a result of magnetic frustration \cite{Bittner2020triangle}.

%%%%%%%%%%%%%%%%%
%%%%%%%%%%%%%%%%%  

The effect of chemical doping on the photo-induced dynamics was studied in the $t$-$J$ model using EDMFT \cite{Bittner2018}. The fastest cooling occurs at low temperatures in the weakly-doped regime, where antiferromagnetic spin correlations are strong, while the cooling rate decreases with increasing temperature or doping. This study also revealed oscillations in the time evolution  of the kinetic energy and the antiferromagnetic spin correlations, but  only in the regime where pseudogap signatures are present in equilibrium. Similar short-lived oscillations~(several femtoseconds) were also detected in cuprates \cite{miyamoto2018}, where the authors connected their presence with the two-magnon excitation signal in Raman scattering.

In multiorbital systems with Hund coupling, an additional highly efficient cooling mechanism based on local spin excitations exists even in the paramagnetic state \cite{Strand2017,rincon2018}. The Hund coupling $J_\text{H}$ favors high-spin configurations if the filling is $n>1$ electrons per site. As photo-doped charge carriers (e.g., singlons or triplons in a half-filled two-orbital system) move through this background of predominantly high-spin states, they induce local spin excitations into lower-spin states at an energy cost of $O(J_\text{H})$. Since $J_\text{H}$ is large (of the order of eV) in transition metal compounds, this allows the photo-doped carriers to release kinetic energy and to cool down to an effective temperature of the order of $J_\text{H}$ within a few hopping times. This cooling mechanism is expected to dominate over antiferromagnetic spin-flip scattering in the first stage of the relaxation process. 

\begin{figure}[t]
\begin{center}
\includegraphics[angle=0, width=1.0\columnwidth]{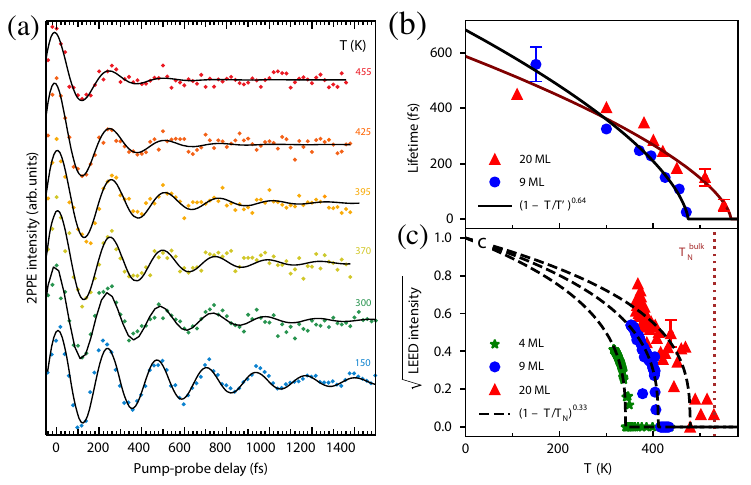}
\caption{
(a) Time evolution  of the two-photon photo-emission (2PPE) signal for a photo-induced in-gap state in NiO at various temperatures. (b) Lifetime of the oscillations as a function of temperature for a 9 and 20 mono-layer thin film. The solid line indicates the critical behavior with an exponent $\nu$ = 0.64.~(c)~The antiferromagnetic scattering amplitude extracted from superstructure spots in low-energy electron diffraction (LEED) as a function of temperature. The dashed lines indicate the critical behavior with an exponent $\beta$ = 0.33. ~(From \onlinecite{Gillmeister2020}.)
}
\label{fig_widdra}
\end{center}
\end{figure}

An experimental two-photon photo-emission (2PPE) study of the multiorbital Mott insulator NiO with long-range antiferromagnetic order revealed how the above effects cooperate \cite{Gillmeister2020}. The fast initial relaxation of the photo-doped spectral weight~(on a timescale $< 80$ fs) and the generation of photo-induced in-gap states was assigned to local spin~(Hund) excitations. The subsequent dynamics is governed by long-lived oscillations of the in-gap state intensity, whose lifetime exhibits a critical damping close to the N\'eel temperature~(extracted from low-energy electron diffraction (LEED)), which was interpreted as evidence for the strong coupling between the charge carriers and the magnetic background, see Fig.~\ref{fig_widdra}.

%%%%%%%%%%%%%%%%%
%%%%%%%%%%%%%%%%%

\subsubsection{Entropy cooling}
\label{sec_entropy}

Carrier cooling before recombination through interactions with bosonic degrees of freedom is a natural mechanism for generating effectively cold photo-doped states. Alternatively, one can try to design targeted excitation protocols which allow to directly generate photo-doped states without excess kinetic energy. In this section we discuss proposals along these lines, which suggest that photo-excitation can lead to a reduced effective temperature in the valence band.

\begin{figure}[t]
\begin{center}
\includegraphics[angle=0, width=\columnwidth]{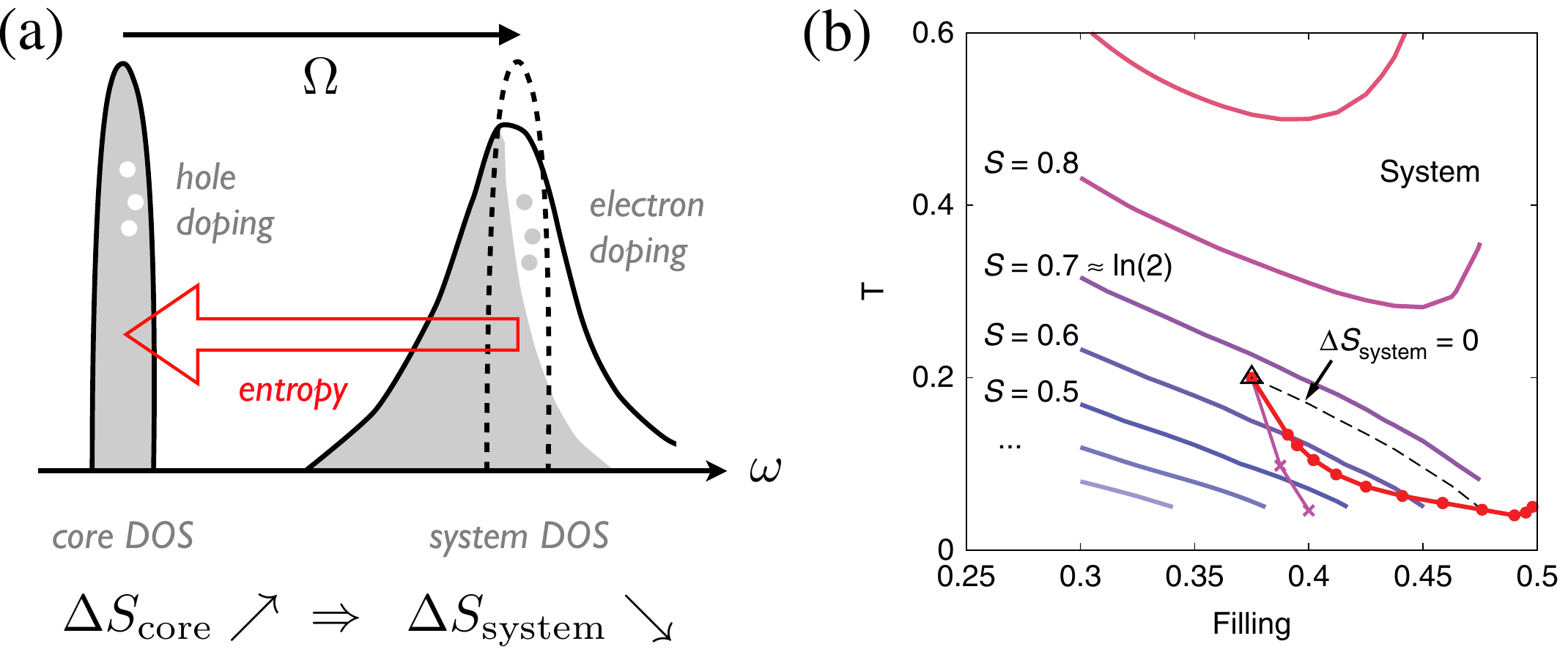}
\caption{
(a) Basic principle of the entropy reshuffling from a partially filled valence or Hubbard band (``system") to an initially filled band (``core"). 
(b) Isentropy lines of a single-band Hubbard model with bandwidth $W=4$ and $U=6$, where $S$ denotes the entropy per spin. The violet line with crosses shows the temperatures and fillings realized by an isentropic charge transfer from an initially full and narrow band. The red dots indicate the temperatures and fillings which can be obtained with realistic short (chirped) pulses. (Adapted from \onlinecite{Werner2019}.) 
}
\label{fig_entropy}
\end{center}
\end{figure}

A general strategy for cooling down a system, denoted by ``entropy cooling'' in the following, involves exchanging particles or energy with another subsystem, which is initially prepared in a state of low entropy.  This concept is utilized in various contexts, such as evaporative cooling, adiabatic demagnetization \cite{Pecharsky1999}, and protocols in cold-atom setups where the two subsystems are prepared by an inhomogeneous potential~\cite{bernier2009,Chiu2018}. Inter-band photo-excitation can be viewed as a similar process: Two bands are transiently coupled by light, and the transfer of electrons between different bands also implies a flow of entropy (Fig.~\ref{fig_entropy}(a)). To understand the possible cooling effects in this setup it is useful to consider the ideal situation where the charge transfer is adiabatic (isentropic with respect to the whole system). Within DMFT, the Mott insulating Hubbard model has a large entropy of $S_\text{Mott} = \ln 2$ per site down to the lowest temperatures, while the entropy of a Fermi-liquid metal is  $S_\text{FL}=\gamma T$ with $\gamma=\lim_{T\rightarrow 0} C/T$ and $C$ the specific heat. In a doped Mott insulator, the $\gamma$-factor diverges as $1/|\tfrac12 -n_\sigma|$ near half-filling \cite{Werner2007}, but also the Fermi liquid coherence temperature drops. Figure~\ref{fig_entropy}(b) shows that the isentropy lines of the doped Mott insulator for $S\lesssim \ln 2$ decrease to zero as $|n_\sigma|\rightarrow \tfrac12$, but increase for $S\gtrsim \ln 2$. If we assume that the isentropy lines  of a photo-doped state as a function of the total number of carriers (doublons and holons) have a similar form, the isentropic generation of doublons and holons always leads to a high temperature (corresponding to $S\gtrsim \ln 2$ at nonzero doping). This is consistent with the observation that the photo-doped metal states obtained by exciting charge carriers across the Mott gap typically have a broad energy distribution (see, e.g., the PM solution in Fig.~\ref{fig_afm_semi}(a)) and characteristic features of low-temperature strongly-correlated metal states, such as narrow quasi-particle peaks, are not observed in simulations of isolated single-band systems.

%%%%%%%%%%%%%%%%%
%%%%%%%%%%%%%%%%%

In a multi-band context, entropy considerations suggest that photo-doping can substantially cool the electronic state \cite{Werner2019}. Consider a two-band setup with an initially completely filled, narrow low-energy band (such as a core level or ligand band in a charge transfer insulator), and a partially filled, wider band near the Fermi level, as sketched in Fig.~\ref{fig_entropy}(a). For the cooling effect, it does not make a qualitative difference if this partially filled band is weakly \cite{Glazman1981} or strongly \cite{Werner2019} interacting, but in the following, we discuss the case of a Mott insulator with a partially filled lower Hubbard band. For simplicity, we assume that the initially filled narrow band is noninteracting, so that the entropy of this band, above a temperature scale determined by its bandwidth, is given by the infinite-$T$ (or atomic limit) result $S_\infty=2n_\sigma^\text{core} \ln n_\sigma^\text{core} + 2 (1-n_\sigma^\text{core}) \ln (1-n_\sigma^\text{core})$. In the initial state, $n_\sigma^\text{core}=1$ and $S_\infty=0$, while the entropy grows rapidly with hole doping. An isentropic photo-doping process, which conserves the total entropy of the system while electrons are transferred between the core and valence bands, therefore results in a reduced entropy and temperature of the valence band electrons.

The violet curve with crosses in Fig.~\ref{fig_entropy}(b) shows the effective temperature achieved by such an isentropic photo-doping process when starting from the initial filling and temperature marked by the black triangle. Even a small charge transfer (a few percent photo-doping) can produce a significant cooling. In reality, the photo-doping by a short and strong pulse may not be isentropic. Still, using suitably optimized chirped pulses \cite{Werner2019}, it is possible to achieve a net cooling down to a temperature which is essentially limited by the width of the low-energy band, as shown by the red dots in Fig.~\ref{fig_entropy}(b). 

Similar cooling effects can be produced by exciting electrons from the system into an initially empty narrow high-energy state. Such a process was proposed for K$_3$C$_{60}$, where the high-energy localized spin-triplet exciton visible in the mid-infrared absorption spectrum can be populated by light~\cite{nava2018}. Motivated by this, the cooling dynamics in a minimal exactly-solvable model consisting of two infinitely connected transverse-field Ising models was studied \cite{fabrizio2018}. It was shown that one of them can be cooled down from the disordered into the ordered phase by switching on a transient coupling between the two subsystems. In systems coupled to an environment, it was shown that the entropy-cooled state can lead to a very long-lived metastable state~\cite{nava2022}. Moreover, a generalized photo-doping set-up involving both full and empty narrow bands can be used to create an effectively cold strongly photo-doped Hubbard model and to realize the $\eta$-pairing state~\cite{Werner2019b}. These minimal models are useful for understanding the underlying mechanism, which should generically play a role in systems involving photo-excitations from or into narrow bands. More realistic modeling will be needed to understand competing relaxation channels that may destabilize the photo-doped state and result in heating, to estimate whether (and on which timescales) cooling by photo-doping can be realized in experiments.

\subsection{Mott excitons}\label{sec:MottExciton}

Excitons in photo-doped Mott insulators are bound pairs of photo-generated multiplets (holons and doublons in the single-band case). Similar to excitons in semiconductors, these states can be detected using linear or nonlinear optical responses, and the photo-generated carriers can relax into such exciton states. In this section we focus on excitons which are bound by the nonlocal Coulomb interaction and/or the exchange of spin and orbital fluctuations. 

{\em Excitons from nonlocal Coulomb interactions ---} 
In any spatial dimension, strong nonlocal Coulomb interactions can naturally lead to bound excitons. Their properties are well understood in one-dimensional~(1D) systems, thanks to the study of effective models~\cite{Gallagher1997,Koch1997} and numerical DMRG investigations of the extended Hubbard model (Eq.~\eqref{H_uvhubbard})~\cite{Essler2005,jeckelmann2003,matsueda2004}.  For sufficiently strong nonlocal interaction one finds excitons of odd and even parity, where the former~(latter) corresponds to a sub-gap optically active~(forbidden) mode. In 1D, the energy levels of odd and even excitons are close, and the transition dipole matrix element is large as a result of spin-charge separation~\cite{Mizuno2000PRB}. 
Linear optical measurements can directly access the optically allowed excitons, while the nonlinear optical response allows to detect the dark excitons, as explained in Sec.~\ref{sec:TOR}. Related to this is a modification of the optical response by strong excitations, e.g. the emergence of in-gap signals and/or the renormalization of the exciton energy \cite{lu2015,rincon2021,novelli2012}. In addition, similar to the observation of excitons in two-dimensional materials \cite{schmitt2022,dong2021,madeo2020}, Mott excitons should be observable in pump-probe  photo-emission spectroscopy (PES). This has been demonstrated  in a nonequilibrium DCA study of excitons in a two-dimensional extended Hubbard model, in which small excitons are bound by a sufficiently strong nonlocal Coulomb interaction~\cite{Bittner2020exciton}. If the binding energy decreases, the excitons merge into the particle-hole continuum and become short lived resonances. A similar excitonic signal in PES has been observed in the photo-doped 1D Mott insulator simulated with iTEBD~\cite{Sugimoto2023}. Experimentally, long-lived Mott excitons with a lifetime of more than several hundreds picoseconds have been detected using trARPES in $\alpha$-RuCl$_3$ ~\cite{Nevola2021}. 

%%%%%%%%%%%%%%%%%
\begin{figure}[t]
\begin{center}
\includegraphics[angle=0, width=1.0\columnwidth]{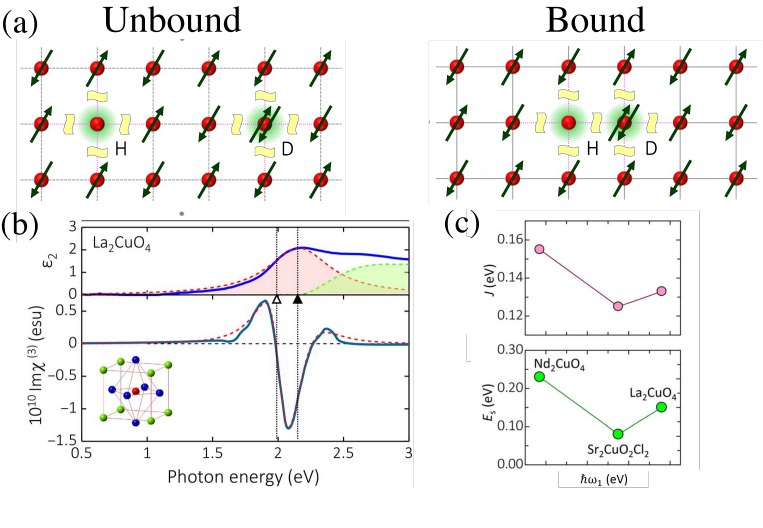}
\caption{(a) Sketch of an exciton pair stabilized by the shared disturbance of the antiferromagnetic background, with broken bonds marked by wiggly lines.
(b) Imaginary part of the dielectric function $\epsilon_2$~(blue solid) and nonlinear optical response $\chi^{(3)}$~(black solid) for La$_2$CuO$_4$. The red~(green) regions correspond to the optical transitions to the odd-parity exciton~(holon-doublon continuum). The left (right) vertical line indicates the energy of the odd-parity (even-parity) exciton. (c) Exchange interaction $J_{\text{ex}}\equiv J$ and splitting of the odd-parity and even-parity excitons $E_s$ as a function of the odd-parity exciton energy $\hbar\omega_1$. (From \onlinecite{Okamoto2019SciAdv}.) }
\label{fig_spinexciton}
\end{center}
\end{figure}   
%%%%%%%%%%%%%%%%%

{\em Exchange coupling assisted excitons ---} 
An alternative binding force of charge carriers originates from the energy gain due to the shared distortion of the background order, as illustrated for the case of a spin system in Fig.~\ref{fig_spinexciton}(a).  This mechanism is characteristic of strongly correlated systems in dimensions $D\ge 2$
and acts both on chemically doped and photodoped charge carriers. Spin-mediated excitons have been extensively studied due to their relevance for high-temperature superconductors.  Theoretical studies concluded that, while the wave-function of a pair of two holons has a $d$-wave symmetry, the wave function of a bound holon/doublon pair is $s$ ($p$) wave for even (odd) parity excitons~\cite{tohyama2006,Zala2013PRL,takahashi2002_TOR}. Furthermore, it has been pointed out that the even exciton can have lower energy than the odd exciton due to the spin contribution, unlike in the 1D case~\cite{Mizuno2000PRB,takahashi2002_TOR}. 

To relate the Mott exciton binding with the spin-charge coupling, the exciton energies were extracted by measuring the 3rd order response using a THz pump and an optical probe to detect the electric field induced changes in the optical response of  cuprates~\cite{Okamoto2019SciAdv}, see Fig.~\ref{fig_spinexciton}(b) and Sec.~\ref{sec:TOR}. 
The characteristic plus-minus-plus structure originates from a transition between the ground state and an  even-parity  and odd-parity  exciton. The exciton energies were extracted by fitting the nonlinear spectrum.  It was found that the even-parity exciton is below the odd-parity exciton and their energy difference scales with the superexchange $J_\text{ex}$. Theoretical calculations based on the $t$-$J$ model using ED~\cite{Okamoto2019SciAdv}, DMRG~\cite{shinjo2021} and the self-consistent Born approximation~\cite{huang2023} confirmed this scaling and showed that the presence of the Coulomb interaction further enhances the binding energy~\cite{Okamoto2019SciAdv,Zala2013PRL}. In pump-probe setups, direct transitions between excitons with different symmetries are allowed and have been observed by ultrafast THz optical spectroscopy in Sr$_2$IrO$_4$, showing a transfer of weight from the short-lived Drude response to a longer-lived intra-excitonic peak~\cite{mehio2023}. This type of exciton formation is not limited to spin pairing, but it can originate as well from a mixture of spin and orbital pairing, as was demonstrated in an analysis of the transient optical conductivity of LaVO$_3$ thin-films~\cite{lovinger2020}.

{\em Biexcitons ---}
There also exist bound objects that are composites of several excitons~\cite{jeckelmann2003}. The simplest example is the biexciton, which consist of two bound excitons.  Coulomb-bound biexcitons have been identified in a ED analysis of the square lattice $U$-$V$-$V'$ Hubbard model (nearest neighbor interaction $V$, next-nearest neighbor interaction $V'$), by simulating the optical conductivity in a laser excited system, where one-exciton states are produced by the pump~\cite{Shinjo2017}.  Experimental observations of biexcitons using pump-probe experiments have been reported for ET-F$_2$TCNQ~\cite{Miyamoto2019}. In addition, the in-gap states identified in photo-doped La$_2$CuO$_4$ and Nd$_2$CuO$_4$ \cite{Okamoto2010}, or the blue-shift of the Mott gap feature in the experiment by \onlinecite{Novelli2014} could be related to photo-induced biexcitons.

\subsection{Dynamical screening and bandgap renormalization}
\label{sec:dynamical}

The photo-doping of charge carriers into a Mott insulator almost instantaneously transforms the insulator into a conductor, which can have significant effects on the screening of the interactions. In the context of semiconductors, a similar screening due to the photo-induced electrons and holes has been discussed using the GW method~\cite{golez2022,perfetto2022,rossi2002,sayed1994,banyai1998}. In the case of Mott insulators, dynamical screening effects can be captured by EDMFT and GW+EDMFT. Also TDDFT+$U$ simulations based on the ACBN0 functional yield interactions which depend on the nonequilibrium population, and thus capture some form of screening~\cite{Tacogne2018}. Within EDMFT and GW+EDMFT, the screening modes, which originate from single-particle and collective charge excitations, are encoded in the frequency-dependent screened local interaction $W_{\text{loc}}(\omega,t)=-\frac{1}{\pi}\int_{t'}^\infty dt e^{i\omega(t-t')}W^R_\text{loc}(t,t')$, or in the effective local interaction $\mathcal{U}(\omega,t)$, which is related to the screened interaction and the local charge susceptibility $\chi_\text{loc}$ by $W_\text{loc}=\mathcal{U}-\mathcal{U}*\chi_\text{loc}*\mathcal{U}$. 

\begin{figure}[t]
\begin{center}
\includegraphics[angle=0, width=\columnwidth]{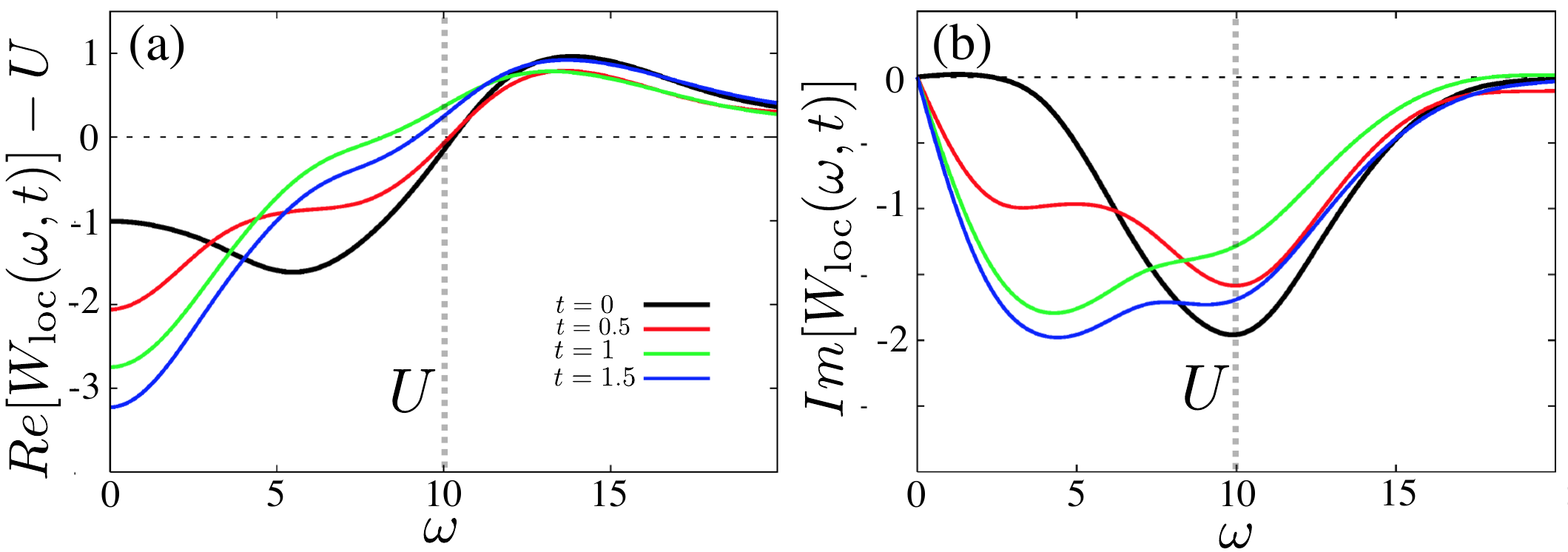}
\caption{
Change of~(a) the real part (shifted by $U$) and~(b) the imaginary part of the local screened interaction $W_{\text{loc}}(\omega,t)$ after a photo-excitation in the 2D extended Hubbard model. Calculations are based on EDMFT+NCA with $W=8$, $U=10$ and $V=2$.~(Adapted from \onlinecite{Golez16edmft}.) 
}
\label{fig_W}
\end{center}
\end{figure}

Figure~\ref{fig_W} illustrates the photo-doping-induced modifications in the real and imaginary parts of $W_{\text{loc}}(\omega,t)$, for an initially Mott insulating $U$-$V$ Hubbard model with bandwidth $W=8$, $U=10$, $V=2$ and inverse temperature $\beta=5$ (EDMFT results from \onlinecite{Golez16edmft}). The black line shows $W_\text{loc}(\omega)$ for the equilibrium Mott insulator. In the imaginary part, we see a broad peak centered at $\omega=U$ (dashed vertical line), which indicates that screening in the Mott state is associated with charge excitations across the Mott gap. Below this energy, $W_\text{loc}(\omega)$ is reduced, but the static value $W_\text{loc}(\omega=0)$ is still close to the bare interaction $U$.

%%%%%%%%%%%%%%%%%
%%%%%%%%%%%%%%%%%

Photo-doping introduces additional screening channels and results in substantial changes in $W_\text{loc}(\omega,t)$. In particular, as soon as doublons and holons are generated by the pulse, an additional low-energy peak appears in Im$[W_\text{loc}(\omega,t)]$, and grows with increasing density of photo-doped carriers. This peak is associated with excitations of doublons and holons within the Hubbard bands and represents low-energy photo-induced metallic screening. As seen in panel (a), this screening reduces the static value of $W_\text{loc}$ significantly. One consequence of the modified screening environment is a rapid shrinking of the Mott gap on the timescale of the electron hopping \cite{Golez16edmft}. 
Similar changes occur in the effective local interaction $\mathcal{U}(\omega,t)$. The dominant peaks in $\text{Im}\,\mathcal{U}(\omega,t)$ define the screening modes. From the integral over these modes (Eq.~\eqref{lambda_screening}) we can extract the effective electron-boson coupling strength, which is typically large. The coupling to charge fluctuations can thus lead to a substantial broadening of the Hubbard bands, and to the appearance of high-energy ``plasmon" satellites in the spectral function \cite{Golez2017}. 

\begin{figure}[t]
\begin{center}
\includegraphics[angle=0, width=\columnwidth]{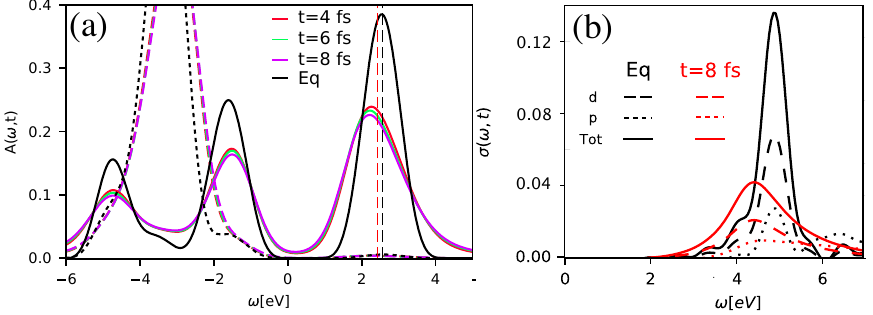}
\caption{
Time evolution  of the spectral function $A(\omega,t)$ (a) and optical conductivity $\sigma(\omega,t)$ (b) of the photo-doped $d$-$p$ model. The black lines show the results for the initial equilibrium state, and the colored lines the evolution during and after a photo-doping pulse with $\Omega=7.5$ eV, which creates about 5\% photodoping. Full lines in (a) are for the $d$ orbital and dashed lines for the $p$ orbitals. The vertical lines indicate the center of the upper Hubbard band.
(Adapted from \onlinecite{Golez2019}.) 
}
\label{fig_A_dp}
\end{center}
\end{figure}

In multi-band setups, where the photo-excitation transfers charge between different bands, there is a second pronounced effect which occurs on the timescale of the photo-doping pulse, namely a relative shift of the bands due to changes in the electrostatic interaction energy (Hartree shifts). These shifts were initially studied in the context of photo-induced insulator-metal transitions, see Sec.~\ref{sec:I_to_M_2}. The combined effect of Hartree shifts and dynamical screening has been studied in a photo-doped $d$-$p$ model using GW+EDMFT \cite{Golez2019}. The two effects cooperate to increase the bandgap renormalization and lead to substantial shifts in the optical conductivity and photo-emission spectrum, as shown in Fig.~\ref{fig_A_dp}. In addition to the broadening mentioned above, we see that the photo-doping results in a shift of the upper Hubbard band to lower energies (red vertical dashed line in panel (a)). This photo-induced shift also manifests itself as a red shift of the charge-transfer peak in the optical conductivity, see Fig.~\ref{fig_A_dp}(b). Similar conclusions were reached using TDDFT+$U$ based on the ACBN0 functional in the case of strong driving below the gap. In these simulations, incoherent processes lead to a modification of the optical gap and a reduced effective interaction~\cite{Tacogne2018}. 

%%%%%%%%%%%%%%%%%
%%%%%%%%%%%%%%%%%

The renormalization of the gap edge and position of the upper Hubbard band leads to shifts in experimental probes which are sensitive to the unoccupied states, like the optical conductivity or X-ray absorption. \onlinecite{Novelli2014} reported a red shift in the optical response after a photo-excitation of La$_2$CuO$_4$, which is naturally explained by ultra-fast changes in the screening. Recently, a bandgap renormalization was also observed with X-ray absorption spectroscopy on La$_{1.905}$Ba$_{0.095}$CuO$_4$ \cite{Baykusheva2022} and NiO~\cite{lojewski2023,wang2022}. The response highly depends on the frequency of the pump pulse. Above-gap excitations in NiO lead to a long-lived renormalization~\cite{lojewski2023}, while below-gap excitations can result in an observable~\cite{wang2022} or nonobservable~\cite{granas2022}  band gap renormalization. 
As these experiments were performed below the N\'{e}el temperature, the gap renormalization can be a combination of screening effects and a (partial) melting of the antiferromagnetic order. A clear separation of the two effects has not yet been achieved. 

Complementary information can be extracted from experiments which measure the occupied part of the spectrum, like photoemission spectroscopy. \onlinecite{cilento2018} reported a nonthermal renormalization and strong broadening of the oxygen band~(in the antinodal direction) in optimally doped Y-Bi2212, remarkably similar to the theoretical results shown in Fig.~\ref{fig_A_dp}.

\subsection{Recombination of charge carriers}
\label{sec_recombination}

In this subsection, we discuss the recombination of photo-doped carriers, see Fig.~\ref{fig_relaxation_processes}(c,e). These processes are also important for understanding the life-time of nontrivial quasi-stationary phases (Sec.~\ref{sec:noneq_phases}). As mentioned in Sec.~\ref{sec:general_thermalization}, for large-gap Mott systems, the recombination times associated with electron-electron and electron-boson scatterings are expected to be exponentially long, see Eq.~\eqref{eq_thermalization}. Still, compared to conventional band insulators, where the non-radiative recombination of electrons and holes via phonon emission can take ns or even $\mu$s, the recombination times in Mott insulators are typically short.

\begin{figure}[t]
\begin{center}
\includegraphics[angle=0, width=0.75\columnwidth]{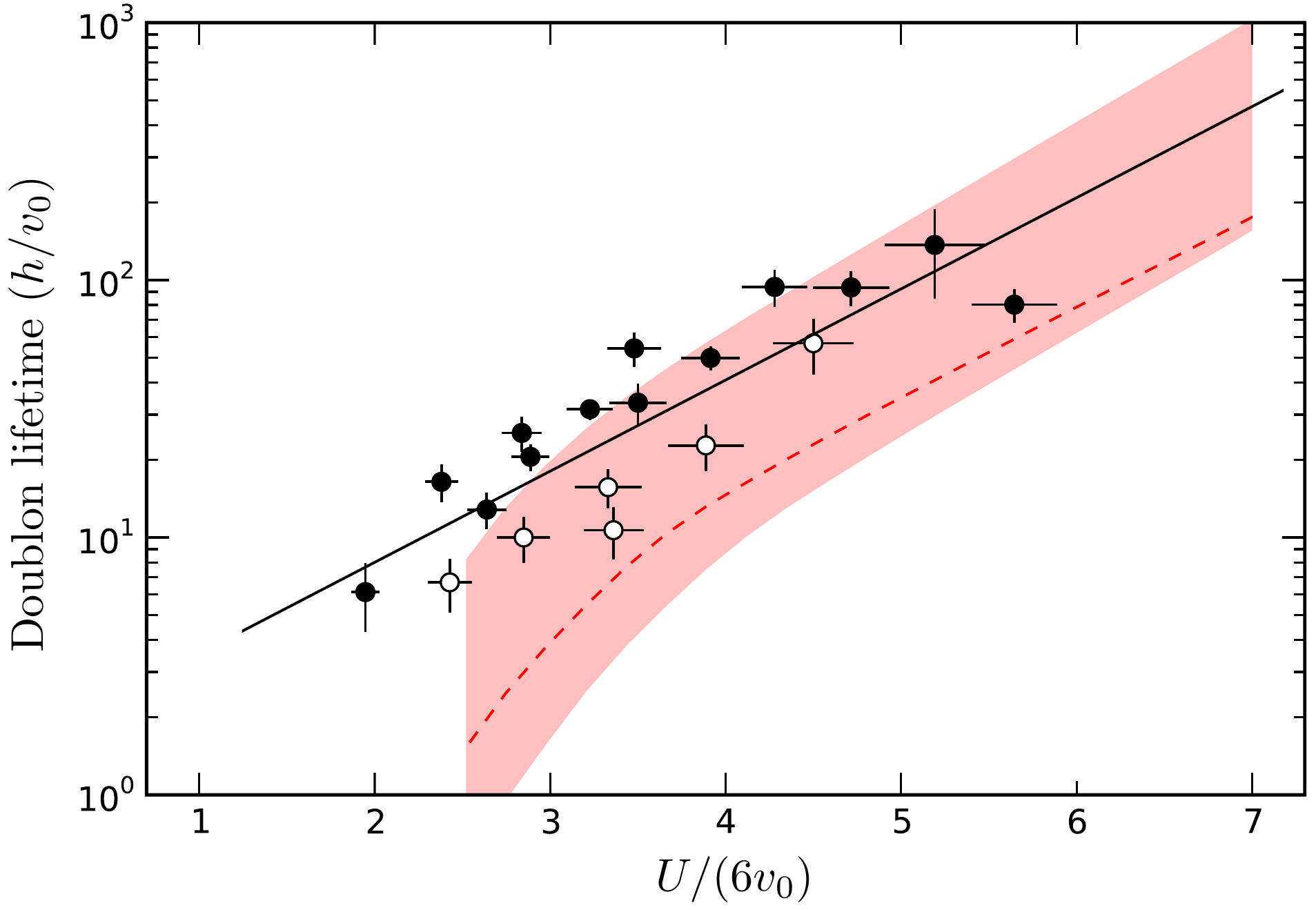}
\caption{
Lifetime of doublons measured in a cold atom simulation of the Hubbard model on a cubic lattice with hopping $v_0$, for two different mixtures of spin states (full and open circles). 
The black line indicates the asymptotic behavior \eqref{eq_thermalization} without the logarithmic correction. The shaded region was obtained by varying the filling factor in the theoretical calculations by 0.3. (Adapted from \cite{Strohmaier2010}.)}
\label{fig_cold_atom_d_time}
\end{center}
\end{figure}

{\it Electron-Electron scattering ---}
Photo-carriers can recombine as a result of electron-electron scattering processes, see Fig.~\ref{fig_relaxation_processes}(c). In this case, the released energy of order $U$ is transferred to the kinetic energy of the remaining charge carriers and $\epsilon_0$ in Eq.~\eqref{eq_thermalization} corresponds to the hopping $v_0$. The exponential scaling of the doublon lifetime has been confirmed in a cold atom experiment~\cite{Sensarma2010a,Strohmaier2010}, see Fig.~\ref{fig_cold_atom_d_time}. Here, a nonthermal state with excess doublons was created by a 10\% modulation in the depth of the lattice, with frequency close to $U$. The subsequent dynamics of the double occupancy follows an exponential relaxation, from which the doublon lifetime was extracted. 

%%%%%%%%%%%%%%%%%
%%%%%%%%%%%%%%%%%

\begin{figure}[t]
\begin{center}
\includegraphics[angle=0, width=1.0\columnwidth]{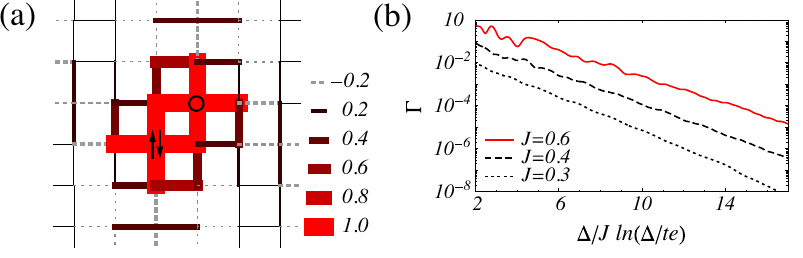}
\caption{(a) The most probable configuration leading to doublon-holon recombination in the 2D $t$-$J$ model with three-site term. The bond thickness represents the deviation of the bond energy relative to the AFM ground state for $J\equiv J_\text{ex}=0.4$.~(b)~Recombination rate $\Gamma$ versus the expected scaling $\Delta/J_{\text{ex}}\log(\Delta/v_0)$ for gap $\Delta$, different superexchanges $J_{\text{ex}}\equiv J$ and hopping $v_0\equiv t.$ (Adapted from~\cite{Zala2013PRL}.)
}
\label{fig_recom}
\end{center}
\end{figure}

{\it Electron-Boson scattering ---}
An alternative recombination pathway is provided by electron-boson scatterings, see Fig.~\ref{fig_relaxation_processes}(e). An explicit theoretical description of carrier recombination through electron-spin interactions has been developed for the decay of bound D-H pairs (excitons) in the strongly repulsive two-dimensional Hubbard model \cite{Zala2013PRL}. One starts from a Schrieffer-Wolff transformation to  separate the Hilbert subspaces with fixed numbers of D-H pairs~($t$-$J$ model) and focuses on the subspace with a single pair, see also Sec.~\ref{sec:simulation}.D. Within the $t$-$J$ model, the doublon-holon pair can form an exciton which is bound by spin fluctuations. The recombination rate of this pair can be described by Fermi's golden rule,  $\Gamma=2\pi \sum_{m}|\langle\Psi_{m}|H_3|\Psi_\text{hd}\rangle|^2 \delta(E_m-E_\text{dh})$, where $\Psi_\text{dh}$ is the initial state  ($s$-wave Mott exciton) with energy $E_\text{dh}$, and the $\Psi_{m}$ are final  states with energy $E_{m}$ and no doublons and holons. The relevant transition operator is the leading three-site correction to the $t$-$J$ Hamiltonian, $H_3=\frac{J}{2} \sum_{\langle ijk\rangle ss'}[h_{is} d_{ks'} {\boldsymbol\sigma_{ss'}} \cdot {\boldsymbol S_j} ]$, 
where $h_{is}$ $(d_{is})$ denotes the holon~(doublon) annihilation operator for spin $s$ at site $i$ and ${\boldsymbol\sigma}$ the Pauli matrices. Fermi's golden rule expression was evaluated using ED, see Fig.~\ref{fig_recom}(b). The decay rate $\Gamma$ essentially follows Eq.~\eqref{eq_thermalization}, where $\epsilon_0$ is now the exchange energy $J_\text{ex}$. Importantly, for realistic parameters for cuprates~($J_{\text{ex}}/v_0\approx 0.4$), the decay rate $\Gamma$ is large despite the large number of emitted magnons. This is a consequence of the strong spin-charge coupling in doped AFM insulators, which implies that the wave function of the initial exciton already contains configurations with many spin excitations (Fig.~\ref{fig_recom}(a)). The modeling was extended to charge-transfer insulators~\cite{zala2014} and a fairly good agreement was obtained with observed recombination times in the ps range for photo-doped cuprates, see Sec.~\ref{sec:exp_probes}.  A similar argument has been developed for 1D organic systems \cite{lenarcic2015}, where the decay of the exciton via spins is inefficient because of spin-charge separation, but a decay on the timescale of a few $100$~fs  is possible due to the large frequency of the relevant  local vibrations of the molecular crystal~\cite{mitrano2014}.

%%%%%%%%%%%%%%%%%
%%%%%%%%%%%%%%%%%

Besides spin fluctuations, nonlocal charge fluctuations can provide a high-energy bosonic mode which is relevant for the recombination. In models with nonlocal interactions, such as the $U$-$V$ Hubbard model (\ref{H_uvhubbard}), the coupling to single-particle and collective charge excitations leads to modifications in the local interaction $\mathcal{U}(\omega)$ (Sec.~\ref{sec:dynamical}). The frequency dependence can be interpreted as a coupling to a continuum of bosonic modes (``plasmons''),  with frequency $\omega$ and coupling strength $g_\omega^2=-\frac{1}{\pi}\text{Im}\,\mathcal{U}(\omega)$ \cite{Ayral2013,WernerCasula2016}. Already for moderate values of $V$, the effective electron-boson coupling can become very strong compared to typical electron-phonon couplings, and can therefore lead to a considerably faster recombination \cite{Golez2017}.

An interesting question concerns the dependence of the recombination time on the excitation density. While the decay of the exciton is exponential (similar to the approach of the  thermal state in Fig.~\ref{fig_thermal}), the recombination of two particles out of the continuum is a bimolecular process. It should therefore  be described by a nonlinear rate equation, where the scattering is determined by the probability at which a doublon encounters a holon. Specifically, $\frac{d}{dt}\mathcal{D} \propto -(\mathcal{D}-\mathcal{D}_\text{eq})^2$, where $\mathcal{D}_\text{eq}$ is the double occupancy in the equilibrium state and we have assumed that the doublons and holons have the same density. This equation, which implies a non-exponential decay, fits well the DMFT solution of  a Holstein-Hubbard model with strong electron-phonon coupling, where the relaxation occurs via phonon emission \cite{Werner15epl}. On the other hand, at least in the low photo-doping regime,  experiments seem compatible with an exponential relaxation~\cite{okamoto2011}. This hints at a two-stage process, where initially an exciton is formed, which then annihilates via a much slower recombination mechanism \cite{Zala2013PRL}. Deviations from the exponential behavior at higher photodoping in insulating cuprates \cite{sahota2019} have been interpreted as Auger-type processes (see Sec.~\ref{sec:photo_doped}.C). However, the recombination rate at high photo-doping concentration is not yet understood, and theoretical considerations should also take into account possible band-gap renormalizations due to photo-doping (Sec.~\ref{sec:photo_doped}.F).

Within DMFT simulations, the gap in the photo-doped state remains relatively robust, i.e., the Mott state can sustain a high density of photo-doped carriers, which can be important for the stabilization of new phases (Sec.~\ref{sec:noneq_phases}). In iTEBD simulations of 1D Mott insulators one observes a more pronounced shrinking and filling of the gap with photo-doping \cite{Murakami2022}, but nevertheless the state can sustain large photo-doping. The robustness of the Mott gap distinguishes this state from situations where the gap is opened by long-range orders, as in weak-coupling antiferromagnets \cite{tsuji2013} or excitonic insulators \cite{mor2017}.
If the gap is due to long range order, recombination can lead to a melting of the order and the closing of the gap, which potentially results in a highly nonlinear dynamics or even an avalanche breakdown of the insulator \cite{picano2021}.

%%%%%%%%%%%%%%%%%%%%%%%%%%%%%%%%%%%%%%%%%%%%%%%%%%%%%%%%%%%%%%%%%%%%%%%%
\section{Nonthermal phases and nonequilibrium phase transitions}
\label{sec:noneq_phases}

\subsection{Nonthermal and hidden phases: General remarks}
\label{sec:general_comments_hidden}

Mott insulators provide an intriguing platform to control the physical properties of materials by short laser pulses.  
The photo-generation of charge carriers can induce metallic transport and insulator-metal transitions on ultrafast timescales, influence competing or coexisting orders, and potentially lead to the emergence of 
{\em nonthermal phases} with exotic properties. 
For the purpose of this section, we 
define a nonthermal phase as a state which allows for an approximate steady-state description (quasi-steady state), but has properties distinct from thermal equilibrium states.
We also sometimes call such phases {\it hidden phases} to emphasize that they cannot be reached along thermodynamic pathways.
With this, we can identify three categories of nonthermal phases:

(i) Prethermal steady states: 
A prethemal quasi-steady state (or ``trapped'' state)  emerges  when the dynamics is 
constrained by approximate conservation laws. Prethermalization \cite{Berges2004} and its relation to almost conserved quantities has been widely discussed for systems close to integrability \cite{Moeckel2008a,Kollar2011a,Polkovnikov2011RMP,Langen2016}. In strongly correlated electron systems, the main constraint is imposed by the Mott gap (Sec.~\ref{sec:general_thermalization}), and the quasi-conserved quantities correspond to the density of photo-doped carriers, 
such as doublons and holons or more general local spin-orbital multiplets.
Within the prethermal category, nonthermal charge carriers can stabilize existing orders (termed ``stabilization  scenario'' below) or lead to the activation of new orders (``activation  scenario").  

(ii) Classical metastability: In this case, the quasi-steady state corresponds to a free energy minimum of emergent classical variables. The minimum can be global, as in the case of metastable phases close to a first order phase transition, or local, as in the presence of local defects in an ordered state.  
An example is the stabilization of photo-induced states by a lattice deformation.

(iii) Dissipative steady states 
associated with
energy and particle flows:  
Such a state arises for example if 
a  flow of energy from the electronic sector transiently supports a dissipative driven state within some subset of low-energy degrees of freedom. 
 Such emergent nonthermal steady states can also be related to nonthermal criticality 
\cite{Berges2008,Nowak2011,MikheevArxiv2023}  
and turbulence, with energy and particle flow between different length scales.
In photo-excited correlated systems, this scenario has been investigated much less than prethermal or metastable states.
Theoretical predictions include a nonequilibrium phase transition in a quantum antiferromagnet \cite{Kalthoff2022}, or a 
population inversion in the charge sector which is maintained by doublon-holon (D-H) recombination energy \cite{Werner2015hirsch,Golez2017}. 

From a phenomenological perspective, both prethermal and metastable phases fit the traditional description of photo-induced phase transitions~\cite{NasuPPT} in terms of a nonthermal free energy surface for a small set of nonequilibrium order parameters. However, in the case of strongly correlated systems, it remains an outstanding challenge for microscopic theory to determine such nonthermal free 
energies or even to  identify the relevant coarse-grained degrees of freedom. 
The quasi-steady description of photo-doped states (Sec.~\ref{sec:quasiseq}) provides a possible starting point.
In this section we review nonthermal phases in Mott insulators which belong to the prethermal and metastable category, 
pathways to reach such states via the ultrafast melting of order, 
and protocols to switch between different phases.

\subsection{Photo-induced insulator-metal transitions}

\subsubsection{Metallic properties of photo-doped states}

Time-resolved optical pump-probe experiments generally 
lead to an almost instantaneous appearance of a low energy conductivity signal (Drude peak) in photo-doped Mott insulators;
see, e.g., \onlinecite{Iwai2003, Okamoto2010, mitrano2014, Okamoto07, Giannetti2016}, and Fig.~\ref{fig_okamoto}. 
The Drude peak  indicates that the photo-induced charge carriers (e.g.~holons and doublons) are highly mobile and support a transient metallic state.
In large-gap Mott insulators, due to the thermalization bottleneck
(Sec.~\ref{sec:general_thermalization}), such photo-doped metallic phases fall into the category of trapped states.  

%%%%%%%%%%%%%%%%%
\begin{figure}[t]
\begin{center}
\includegraphics[angle=0, width=\columnwidth]{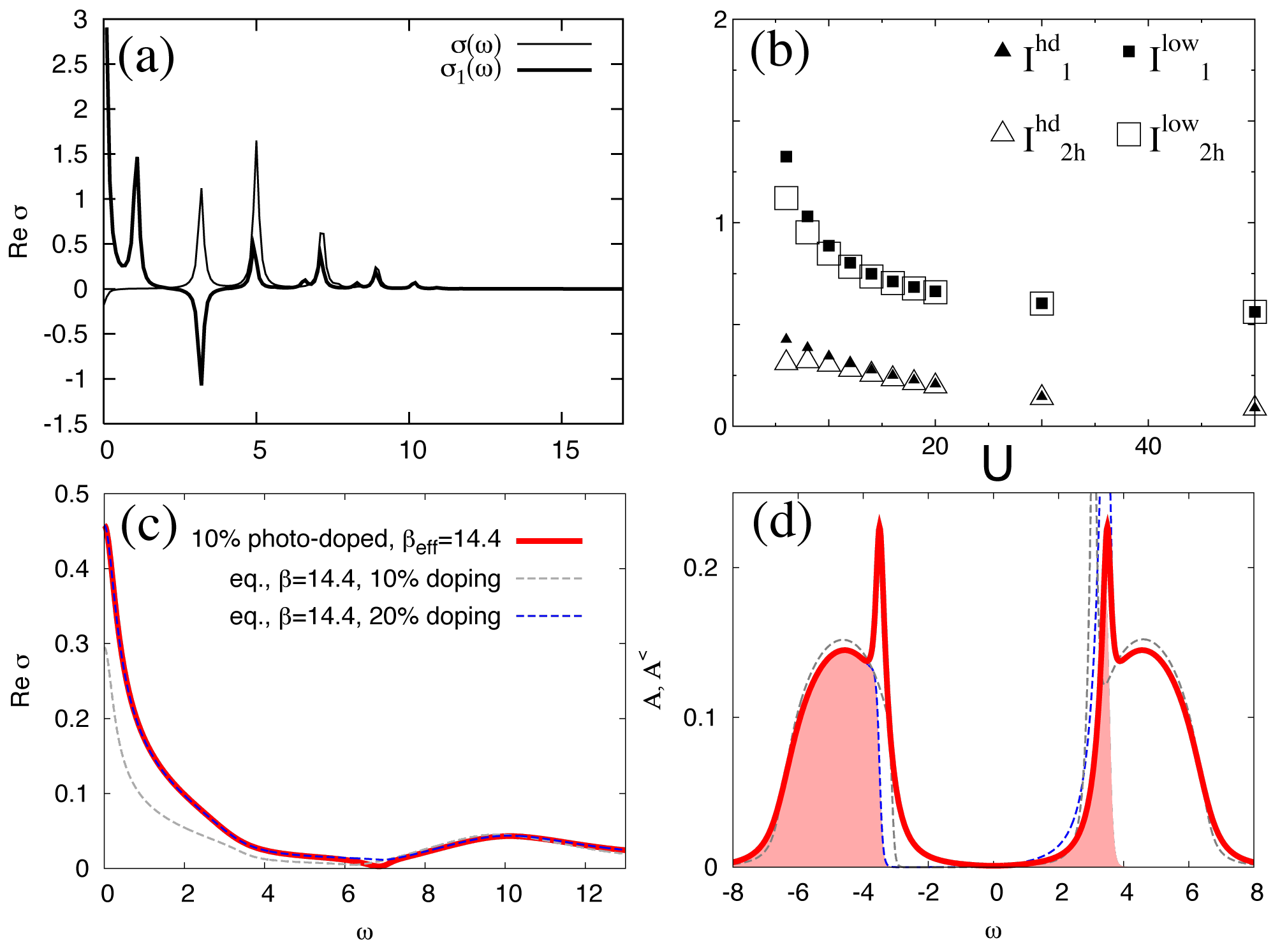}
\caption{Optical conductivity and spectral function of photo-doped Mott insulating Hubbard models.
(a) 1D Hubbard model with $U=6$ and bandwidth $W=4$: $\sigma(\omega)$ and  $\sigma_1(\omega)$ show the conductivity in equilibrium and for a photo-doped state, respectively. (b) Weight of the Drude peak (squares) and of the continuum (triangles), for the photo-doped state (solid symbols) and an equilibrium system with two holons (empty symbols). Lower panels: 
optical conductivity (c) and spectral function (d) of the photo-doped Hubbard model on the Bethe lattice with $U = 9$, $W=4$
(10\% doublons and holons, effective inverse temperature $\beta_{\rm eff}=14.4$). 
The gray (blue) dashed lines show equilibrium results for the indicated 
chemical doping. In (d),  the equilibrium spectra are shifted to match  the positions of the Hubbard bands in nonequilibrium. (Adapted from \onlinecite{maeshima2005} and \onlinecite{Werner2019b}).}
\label{fig_cond}
\end{center}
\end{figure}
%%%%%%%%%%%%%%%%%

The experimental observation of a Drude peak is in agreement with theoretical simulations for Hubbard  type models based on ED \cite{maeshima2005, Shao2016, Shinjo2017}, DMFT~\cite{Eckstein2013, Werner2019b} and DMRG~\cite{rincon2021}. As an illustration, Fig.~\ref{fig_cond}(a) shows results from the ED study by \onlinecite{maeshima2005} on the one-dimensional (1D) Hubbard. 
In equilibrium,  the real part of the conductivity (thin line) exhibits a gap and a ``continuum" 
of excitations (up to finite-size effects) in a range 
$U-W\lesssim \omega\lesssim U+W$ set by the bandwidth $W$. 
In the photo-doped state (mimicked here by the first optically allowed excited state), the conductivity has a Drude peak (bold line). In addition, negative spectral weight at the lower edge of the continuum indicates 
energy gain in optical deexcitation processes due to a partially inverted population. Qualitatively similar results were also found for the infinite-dimensional Hubbard model within DMFT~\cite{Werner2019b}), as exemplified in Fig.~\ref{fig_cond}(c).
In the conductivity of the photo-doped state, one again observes a dip near the lower edge of the continuum (which would develop into negative weight for larger photo-doping), and a prominent photo-induced Drude peak. The continuum of excitations across the Mott gap (near $U\approx 9$) is hardly modified by the photodoping,  apart from a slight bleaching. 

A basic question is to what extent the photo-doped system with $x\%$ doublons and $x\%$ holons resembles a chemically doped Mott insulator with doping concentration $2x\%$. 
\onlinecite{maeshima2005} compared the weight of the Drude peak ($I^\text{low}$) and the continuum feature ($I^\text{hd}$) for a  photo-doped system with a single D-H pair and a chemically-doped system with two holons and found a good agreement (Fig.~\ref{fig_cond}(b)).  
Also in the DMFT simulations, the Drude peak of the photo-doped state corresponding to $x=10\%$ is well reproduced by $2x\%$ chemical doping, which means that doublons and holons independently contribute to the low-energy conductivity (Fig.~\ref{fig_cond}(c)). In contrast, the spectral function of the photo-doped system shows a quasiparticle peak for both holons and doublons 
and cannot be directly compared to that of the chemically doped state (Fig.~\ref{fig_cond}(d)).

Another relevant question is the lifetime of the Drude peak and the dynamics of its buildup. Consistent with the sum rule 
$\int d\omega \text{Re}\sigma(\omega,t)=-E_\text{kin}(t)$ for a single-band model,  the buildup of the Drude peak requires a 
cooling of the photo-induced carriers.
Consequently, in the DMFT simulations, 
the Drude peak grows as the photo-doped carriers dissipate energy to bosonic excitations (Sec.~\ref{sec:intraband_relax}),
while it  remains broad and of small weight if the coupling to the bath is inefficient~\cite{Eckstein2013}. 
In experiments, the Drude peak can emerge within a few femtoseconds, such as in the cuprate Mott insulator Nd$_2$CuO$_4$ \cite{miyamoto2018}. On the other hand, 
at longer times, the Drude peak can disappear 
even before the recombination of the charge carriers (Fig.~\ref{fig_okamoto}), as the latter may get trapped by impurity states \cite{Okamoto2010}, dressed into lattice polarons~\cite{matsueda2011, tohyama2013, yonemitsu2009, Golez2012b, vitalone2022} and spin-polarons~\cite{lenarcic2014}, or form (bi)-excitons (Sec.~\ref{sec:MottExciton}).

\subsubsection{Photo-induced changes of the electronic structure}
\label{sec:I_to_M_2}

In correlated electron systems, an insulator-metal transition cannot only 
be induced through the generation of mobile carriers (transfer of population across the gap), but also by changing the electronic structure,
i.e., the generation of new states within the gap or closing of the gap.  
In particular, photo-induced changes of the electronic structure may occur due to dynamical screening effects (Sec.~\ref{sec:dynamical}) and the electrostatic interaction energy between electrons in different bands (Hartree shifts).  The latter can already have interesting consequences in metals, such as a change of the Fermi surface topology \cite{Beaulieu2021}. A gap closing resulting from changes in the screening environment has been proposed to explain the observation of a short-lived metallic phase in VO$_2$ \cite{Wegkamp2014}, which was also detected with time-resolved electron diffraction \cite{Morrison2014}. Especially in multi-band systems, a restructuring of the density of states provides a promising pathway for photo-induced insulator-metal transitions. Studies within the Hartree-Fock approximation~\cite{he2016} 
predict that a carrier redistribution between bands can lead to metastable metallic states. 
A photo-induced metastable state has also been found within the time-dependent Gutzwiller approximation for a quarter-filled two-band Hubbard model~\cite{Sandri2015}
in the vicinity of a first order insulator-metal transition: 
For certain intermediate excitations the charge gap can be closed,
while spin correlations survive. Similar features were observed in time-resolved X-ray and photoemission studies on V$_2$O$_3$, which reported an induced metallic state with a lifetime of several picoseconds~\cite{lantz2017}, although its nonthermal nature is still under debate~\cite{moreno2019}.

\subsubsection{Metastable  phases and inhomogeneities}

In many systems the Mott transition is intertwined with a lattice distortion and can become first order, with profound effects on the  dynamics.  
A paradigmatic example is the photo-induced insulator-metal transition in VO$_2$ \cite{Cavalleri2001a,Kuebler2007a,Hilton2007}. 
The entangled motion of the electrons and lattice makes it difficult to decide whether transitions in VO$_2$ are  electronically- or lattice-driven. Experiments indicate an ultrafast  transition from the insulator to an  
isostructural metallic state, which can be reached by optical excitation \cite{Wall2013a,Wegkamp2014,Morrison2014}, strong-field tunneling breakdown 
(Sec.~\ref{sec:breakdown})  \cite{Gray2018}, or short voltage pulses \cite{Sood2021},
but has also been interpreted within a thermal scenario \cite{vidas2020}.
A metastable photo-induced metallic phase  
was also reported in Ca$_2$RuO$_4$~\cite{verma2023}. 
Supported by DMFT simulations, the authors 
proposed that photo-doped charge carriers can quickly reduce the orbital order that stabilizes the equilibrium Mott phase. 
The coupling of the orbital order to the lattice leads to a slow evolution into a nonthermal metallic phase,
with a distinct lattice configuration compared to the high-temperature metallic phase (revealed by time-resolved X-ray diffraction).

In the conventional picture of ultrafast structural phase transitions, atoms move in a slowly varying free energy landscape determined by the fast electrons, 
as qualitatively captured by a time-dependent Ginzburg Landau theory for few global lattice displacement coordinates.  However, heterogeneities (both of extrinsic  \cite{Callahan2015} and intrinsic nature) can have a profound effect on the dynamics of the transient state: On a timescale of tens of picoseconds, nucleation and growth dynamics  of metallic domains  has been reported for thin films  of ${\mathrm{V}}_{2}{\mathrm{O}}_{3}$ \cite{Abreu2015} and Nickelates \cite{Abreu2020}. Also the long-lived hidden metallic states in 1$T$-TaS$_2$ \cite{Stojchevska2014} have distinct nanoscale patterns \cite{Gerasimenko2019,Ravnik2021}. Intrinsic disorder can emerge on shorter timescales. Within the so-called ultrafast disordering scenario, as proposed for the photo-induced transition in VO$_2$ \cite{Wall2018,Munoz2023} and in a layered manganite \cite{Perez-Salinas2022}, the nonequilibrium dynamics leads to strong disorder on the atomic scale, such that the mean value of the order parameter is no longer representative of the local environment. Theoretically,  ultrafast disordering has been studied in the Holstein model \cite{Picano2023,sous2021}, while it is still an open problem for  strongly correlated electron systems.

Recent experimental progress is opening new avenues for studying this spatio-temporal dynamics in correlated systems with femtosecond temporal resolution and nanometer spatial resolution:  Diffuse electron scattering can map out phonon occupations \cite{Seiler2021}, ultrafast electron diffraction has revealed defects and transient subdominant orders in a charge density wave (CDW) phase \cite{Zong2019,Zong2021}, while FEL-based ultrafast X-ray imaging has been used to study inhomogeneities in the photo-induced transition of VO$_2$ with a time resolution of 150 fs  \cite{Johnson2023}. Electron microscopy \cite{Vogelgesang2018, Danz2021} has revealed the evolution of CDW domains and topological defects in the unconventional CDW in 1$T$-TaS$_2$.  Finally, an intriguing development is ultrafast scanning tunneling microscopy \cite{Plankl2021}. Applications of these techniques to correlated insulators should further clarify the role of spatial patterns in nonequilibrium transitions. 

\subsection{Melting and trapping of electronic orders}
\label{sec:melting}

Due to the delayed thermalization, electronic orders in large-gap Mott insulators can persist for extended periods even if the injected energy is 
sufficient to disorder the thermalized state \cite{Werner2012afm}. Nevertheless, the injected mobile carriers can disrupt the ordered state at early times after the excitation. This ultrafast melting dynamics is  important for understanding relaxation pathways towards nonthermal hidden phases.

\begin{figure}[tbp]
\begin{center}
\includegraphics[angle=0, width=0.8\columnwidth]{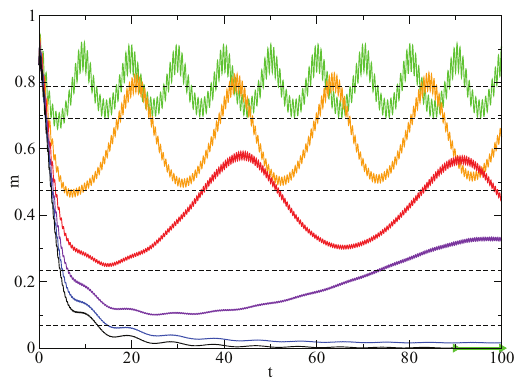}
\caption{
Time evolution  of the staggered magnetization $m$ in the Hubbard model with bandwidth $W=4$ for interaction quenches $U_i=4\rightarrow U_f=12,14,16,18,20,22$ (top to bottom). 
For all cases, the thermalized state would be disordered ($m=0$). 
(From \onlinecite{Sandri2013}.) 
}
\label{fig_afm_gutzwiller}
\end{center}
\end{figure}

As a simple example, let us discuss the dynamics of antiferromagnetic (AFM) order in the Hubbard model after an interaction quench. (For the purpose of this discussion, the quench acts similar to a photo-doping pulse, as it produces a state with a nonthermal doublon/holon population.) As was shown in Fig.~\ref{fig_afm_semi}, 
the generation of doublons and holons by the quench does not always lead to a melting of the staggered magnetization on numerically accessible timescales, even when the fully thermalized state is expected to be paramagnetic. The behavior of the order changes from trapped to exponentially decaying at a sharply defined excitation threshold.  Similar physics has also been studied by \onlinecite{Sandri2013} using the time-dependent Gutzwiller variational approach (Sec.~\ref{sec_gutzwiller}).
Figure~\ref{fig_afm_gutzwiller} shows the time evolution of the staggered magnetization $m$ in a Hubbard model for interaction quenches from an initial 
interaction $U_i$ in the AFM phase to larger final interactions.
Already for the smallest quench, the thermalized system would be paramagnetic,
but the staggered magnetization oscillates around 
a large value $m\approx 0.8$. The frequency $\omega_2$  
shows a critical slowdown with increasing $U_f$, and vanishes at $U_f^c\approx 21$, 
where the dynamics switches to an exponential decay. 

%%%%%%%%%%%%%%%%%
%%%%%%%%%%%%%%%%%

These transitions are examples of dynamical phase transitions between ordered and disordered phases which are both prethermal.
Such transitions have been widely discussed for systems close to integrability \cite{Marino2022}, including the quench dynamics of superconducting and antiferromagnetic states in the weakly-correlated regime~\cite{Barankov2006,Yuzbashyan2006,tsuji2013}.
A critical slow-down, as predicted for such dynamical phase transitions (see also Fig.~\ref{fig_afm_semi}(c)), has been observed for melting of charge density wave (CDW) \cite{Zong2019a}.  
A  dynamical phase transition in a strongly correlated electron system has been reported for the photo-induced melting of AFM order in Sr$_2$IrO$_4$ \cite{delaTorre2022}. In this case, one finds separate fluence thresholds for the ultrafast melting of AFM order and for the critical slowdown of the recovery dynamics.
This indicates that the 
lower threshold,  which is for the melting of AFM order, corresponds to a dynamical phase transition, consistent with the 
numerical predictions above.

In the presence of coexisting order parameters, the ultrafast partial melting of orders can provide an interesting pathway leading to hidden states. 
An important aspect is that an order parameter which is dominant in equilibrium can be {\em less} robust in the nonequilibrium setting.
For example, the signatures of light-induced  superconductivity appearing within the stripe phase of high-$T_c$ cuprates \cite{Fausti11,Nicoletti2014} may be explained by the melting of 
the initially dominant stripe order. From a more general perspective, \onlinecite{Sun2020} proposed that in a system with intertwined orders, after a  partial melting of both orders, the system can preferentially relax into a subdominant metastable free energy minimum. This rather generic pathway to reach hidden states was demonstrated within a time-dependent Ginzburg-Landau theory for a multi-component order parameter (assuming an over-damped, so-called ``model A'' dynamics \cite{HohenbergHalperin1977}).  A somewhat related theoretical prediction is discussed in Sec.~\ref{hidden_magnetic_orbital}, where one order parameter (orbital order) is more robust against thermal fluctuations than another (magnetic order), but nevertheless melts more efficiently upon photo-doping \cite{Li2018}.

\subsection{Switching on electronic timescales}
\label{sec_switching}

While the previous section pointed out pathways to hidden states through the incoherent melting of orders, one can also ask whether more coherent switching pathways can be realized. 
This is particularly interesting for systems with a discrete manifold of long-lived states, 
including (i) phases close to a first order transition, and (ii) equivalent realizations of the order under the breaking of a discrete symmetry. 
Such switching pathways  may also be relevant for possible ultrafast memory devices. In this section, we review mainly proposals where the switching relies on electronic mechanisms  and happens on correspondingly fast timescales.

%%%%%%%%%%%%%%%%
\begin{figure}[t]
\begin{center}
\includegraphics[angle=0, width=1.0\columnwidth]{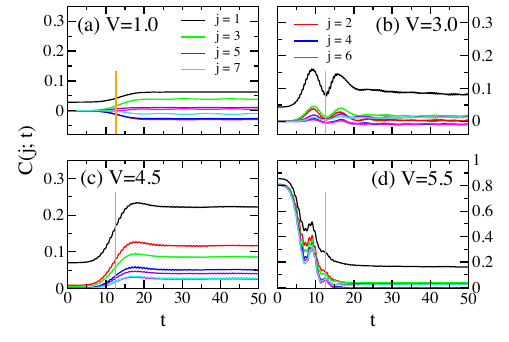}
\caption{ED results for the time evolution  of the 
charge correlations $C(j,t)$ at distance $j$ in a photo-excited half-filled extended Hubbard model on a $14$-site chain.
$V$ is the nearest-neighbor interaction and the Gaussian laser pulse is centered at $t=12.5$ (vertical lines). 
The pump frequencies match the resonance peaks of the optical absorption spectrum. Parameters: $U=10$, (a)~$\omega_{\rm pump}=7.1,E_0=0.10$, (b)~$\omega_{\rm pump}=6.1,E_0=0.30$, (c)~$\omega_{\rm pump}=4.0,E_0=0.07$, (d)~$\omega_{\rm pump}=4.1,E_0=0.60$.~(From \onlinecite{Tohyama2012PRL}.) 
}
\label{Tohyama_CDW}
\end{center}
\end{figure}  
%%%%%%%%%%%%%%%%

{\em Optical access to competing states ---}
Close to a first order transition, states featuring a competing order can appear as low-lying excited states 
which may be reached by  optical excitation. 
A proposal for controlled optical switching based on this scenario was made for the extended Hubbard model (\ref{H_uvhubbard}) on a 
1D chain \cite{Tohyama2012PRL}. 
At half filling, a first-order phase transition occurs around $U\simeq 2V$ between a conventional Mott insulator characterized by singly occupied sites and AFM spin correlations, and a CDW phase. As shown in Fig.~\ref{Tohyama_CDW}, the charge order correlations are enhanced by a few-cycle electric field pulse when the system is in the Mott insulating phase but close to the phase boundary. This photo-induced charge order can be realized only if the excitation pulse is resonant with an excited state with large CDW correlations. 
A related switching mechanism has been predicted for the competing charge orders in the extended Hubbard model on a frustrated lattice~\cite{Hashimoto2014}.
Experimentally, such an optical switching was demonstrated in a bromine-bridged Pd-chain compound~\cite{Okamoto2014PRL}, although 
the electron-phonon coupling also matters there. 
The material shows a first-order transition from a CDW phase to a Mott insulator with decreasing temperature.
Exciting  the low-temperature phase resonantly with the Mott exciton leads to a transition to the CDW phase, while above-gap excitations induce a metallic state.
Somewhat related transitions have been induced by strong THz pulses in organic molecular compounds~\cite{Yamakawa2021}.
The opposite effect was observed in the iodine-bridged platinum compound, where the equilibrium CDW is transformed into the Mott insulating phase~\cite{kimura2009}.

{\em Switching by periodic driving --- }
As an alternative to resonant optical transitions between competing states, one can try to use the effective (time averaged) forces arising from a periodic drive to transfer a system between different free energy minima. Experimentally, the most established pathway along these lines is a switching via non-linear phononics (Sec.~\ref{sec:nonlinear_phononics}), as recently demonstrated 
for ferromagnetic order in YTiO$_3$~\cite{Disa2023}, and ferroelectric order in LiNbO$_3$~\cite{mankowsky2017}.  
Theoretical proposals include  a switching of the lattice structure in ErFeO$_3$ \cite{Juraschek2017}, and a modification of the spin 
configuration in Cr$_2$O$_3$ via the transient modification of the exchange interaction \cite{Fechner2018}. An obvious question is whether related switching protocols can be realized using a transient Floquet engineering of the electronic structure and interactions. A promising class of materials for this physics are systems with orbital order of electronic origin, which feature a discrete set of equivalent orders. An example is the switching of composite order, which  has been theoretically discussed in the context of the fulleride compounds A$_3$C$_{60}$~\cite{Werner2017}. 
These half-filled three-orbital systems exhibit a Mott transition, and a nearby peculiar Jahn-Teller metal state with coexisting metallic and Mott insulating orbitals \cite{Zadik2015}, see Fig.~\ref{fig_expmott}. This state has been identified as a spontaneous orbital-selective Mott phase \cite{Hoshino2017}, where two orbitals are in a paired Mott state and the third one is metallic.  There are therefore three equivalent ordered states with different metallic orbitals. 
Due to the orbital-dependent spatial anisotropy of the hopping, one can selectively modify the hoppings 
for the three orbitals  through dynamical localization by a suitable choice of the polarization of an applied laser field (Sec.~\ref{sec_bandwidth_ren}). This enables the controlled switching 
between the three order realizations on electronic timescales. 

Similarly, also conventional orbital order due to the Kugel-Khomskii orbital exchange mechanisms has discrete domains, and laser-control of the orbital exchange (Sec.~\ref{sec:effective_j_ac}) along selected crystallographic directions can be used to rotate the order parameter \cite{Grandi2021b}. In such switching protocols, relevant forces can also arise due to anharmonic interactions and a transient modification of the order parameter fluctuations.

%%%%%%%%%%%%%%%%%%%%%%%%%%%%%%%%%%%%%%%%%%%%%%%%%%%%%%%%%%%%%%%%%%%%%%%%
\subsection{Hidden phases}
\label{sec:hidden_phases}
%%%%%%%%%%%%%%%%%%%%%%%%%%%%%%%%%%%%%%%%%%%%%%%%%%%%%%%%%%%%%%%%%%%%%%%%

In this section we discuss examples of hidden states in Mott insulators, with a focus on those supported by the presence of long-lived charge carriers. 
We try to classify the physics using the concepts introduced in Sec.~\ref{sec:general_comments_hidden} (melting, trapping, stabilization, activation).

\subsubsection{Magnetic and orbital order}
\label{hidden_magnetic_orbital}

Strongly correlated multi-orbital systems with spin and orbital degrees of freedom display competition or coexistence between magnetic and orbital orders, which provides a rich playground for the study of hidden phases. 
Photo-generated charge carriers can trigger transitions in these systems through the (partial) melting of spin and orbital orders.
Relevant examples include the metal-insulator transition in a perovskite manganite \cite{Miyano1997}, and transient hidden orders in a manganite \cite{Ichikawa2011}. 
\onlinecite{Beaud2014}  identified the energy density as the main nonequilibrium order parameter for a transition in a manganite.
 In combination with strain in thin films, even a single optical pulse can initiate a transition to a long-lived metastable hidden metallic phase \cite{zhang2016,Teitelbaum2019}. 
Because the excitation of orbital order is often strongly coupled to the lattice dynamics \cite{Wall2009}, a microscopic theory of these photo-induced transitions is challenging. However, if the initial step of the switching towards a hidden state involves the perturbation of orbital orders by the photo-doped carriers, typically on the fs timescale \cite{Singla2013a}, relevant aspects of the switching can already be captured in simple theoretical models.

%%%%%%%%%%%%%%%%%
\begin{figure}[t]
\begin{center}
\includegraphics[angle=0, width=1.0\columnwidth]{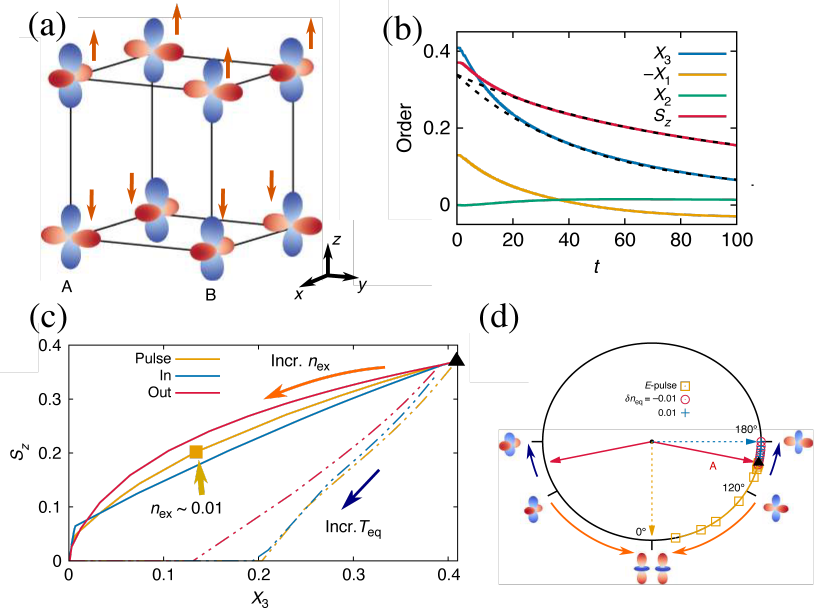}
\caption{(a) Spin and orbital order of the two-orbital Hubbard model with three electrons per site
in equilibrium.  The unoccupied orbital is shown, as well as the total electron spin moment (arrows). 
(b)~DMFT evolution of spin and orbital order after an electric field pulse which creates a photo-doping $n_{\rm ex}=0.015$. 
(c) Order parameter in the $(S_z,X_3)$ plane, for the 
prethermal photo-doped state at different $n_\text{ex}$ (solid lines), in equilibrium with increasing temperature at integer filling (yellow dotted line), and for chemically doped systems with $\delta n = \pm 0.01$ (red/blue dotted lines).
(d) Normalized orbital-order components $X_1$ and $X_3$ in the trapped state shown on the pseudospin compass. (From \onlinecite{Li2018}.) }
\label{fig:Li2018}
\end{center}
\end{figure} 
%%%%%%%%%%%%%%%%%

A model which allows to realize and explain a
transient hidden state with intertwined spin and orbital orders is the three-quarter filled two-band Hubbard model on a cubic lattice \cite{Li2018}. The two orbitals could represent partially filled $e_g$ orbitals ($d_{y^2-z^2}$ and $d_{3x^2-r^2}$), as in KCuF$_3$ \cite{Pavarini2004}.
The low-temperature equilibrium state features A-type AFM spin order (antiferromagnetically aligned ferromagnetic planes) and G-type antiferro-orbital order (alternating orbital occupations in all directions), see Fig.~\ref{fig:Li2018}(a). 
The order is characterized by the ordered magnetic moment $S_z$, and the orbital 
pseudospin-$1/2$ moment $\mathbf{X}=(X_1,X_2,X_3)$ in the basis  $(d_{y^2-z^2},d_{3x^2-r^2})$. 
The excitation of the system by an electric field pulse induces a partial melting of the spin and orbital order, see Fig.~\ref{fig:Li2018}(b). The prethermal (trapped) photo-doped state is obtained by extrapolating the curves to long times. Figure~\ref{fig:Li2018}(c) locates the  corresponding order in the $(X_3,S_z)$ plane for different photo-doping densities $n_\text{ex}$ and various excitation protocols (solid lines), and compares it to the equilibrium behavior for different temperatures (dashed lines). Upon photo-doping, one can realize a state with dominant magnetic order, while in equilibrium, with increasing temperature,  the magnetic order melts before the orbital order.
Also the change in the direction of the orbital pseudo-spin is opposite in equilibrium and in the photo-doped case (Fig.~\ref{fig:Li2018}(d)). 
Hence, the photo-doped state realizes a nonthermal hidden order. This can be understood by considering how the photo-doping 
melts the two orders.  
The efficient melting of G-type orbital order originates from the motion of the injected charge carriers, which leave behind strings of defects in the ordered background (analogous to the case of G-type AFM order  in Sec.~\ref{electron-spin}). The magnetic order, on the other hand, is ferromagnetic in the planes, and therefore perturbed only by the out-of plane motion of carriers.
This difference in the effect of the charge motion explains the faster melting of the orbital order, in spite of the fact that this is the more robust order in equilibrium.

A related theoretical proposal for a hidden state realized by the partial melting of orders in multi-band systems has been made for the half-filled three-band Hubbard model with negative Hund coupling~\cite{Werner2021}, which describes the alkali-doped fullerides A$_3$C$_{60}$ \cite{Fabrizio1997,Capone2009}. At low temperatures, the strongly-correlated system shows an AFM spin order and a composite ferro-type orbital order~\cite{Hoshino2017}. As in the previous example, photo-excitation of electrons across the gap rapidly melts the AFM order, while the ferro-type order is only moderately perturbed. This allows to realize a pure composite orbital order that does not occur in equilibrium upon increasing temperature. After an initial relaxation, the composite order is favored by the presence of photo-doped doublons/holons, similar to the stabilization of ferromagnetic order by itinerant carriers in the double exchange model \cite{Zener1951,deGennes1960}, see also Sec.~\ref{sec:spin_state}.

%%%%%%%%%%%%%%%%%
\begin{figure}[t]
\begin{center}
\includegraphics[angle=0, width=0.8\columnwidth]{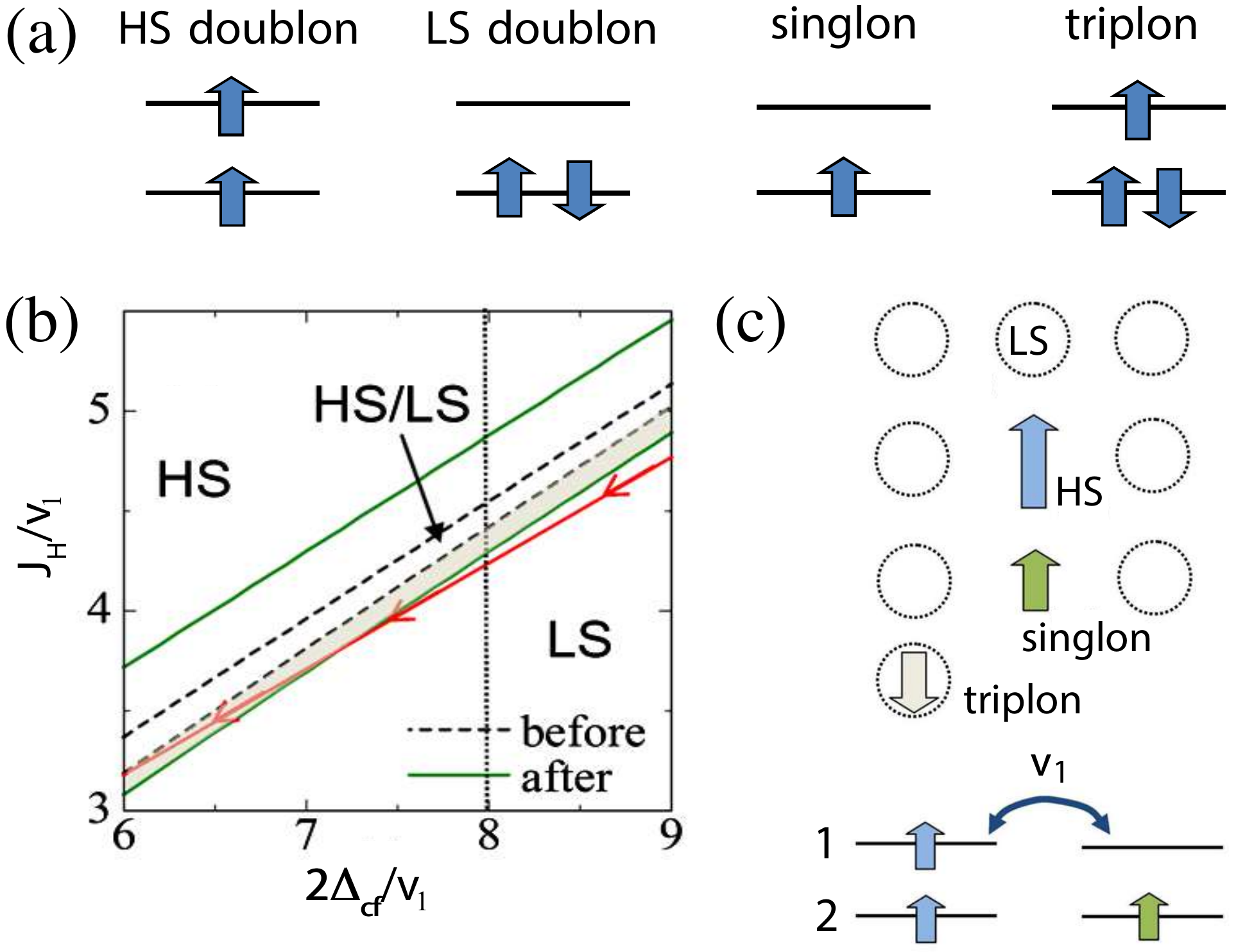}
\caption{
Two-orbital model with crystal field splitting $\Delta_\text{cf}$, Hund coupling $J_\text{H}$ and nearest-neighbor orbital-dependent hopping $v_1,v_2$ ($v_1>v_2$). 
(a) Sketch of the HS doublon, LS doublon, singlon and triplon configurations.~(b) Phase diagrams before and after photo-doping, obtained using ED,
showing the boundaries of the HS-LS coexistence region in equilibrium (dashed lines) and in the photo-doped state with one additional singlon-triplon pair (bold green lines). 
(c) Illustration of the singlon-HS bound state and the double-exchange mechanism which stabilizes this state.
(Adapted from \onlinecite{Ishihara2011PRL}.)
}
\label{fig:NESS_eff_Ishihara}
\end{center}
\end{figure}  
%%%%%%%%%%%%%%%%% 

\subsubsection{Hidden phases near a spin state transition} 
\label{sec:spin_state}

 Another interesting playground for hidden states exists in the vicinity of spin state transitions, which arise from the competition between crystal field splitting and Hund coupling.   The dynamics of the photo-induced high-spin (HS) to low-spin (LS) transitions has been experimentally studied in molecular spin crossover compounds~\cite{Bertoni2015} and spin crossover complexes~\cite{Ogawa2000,Tayagaki2001,Ohkoshi2011}. Related physics in Mott systems can be described by the two-band Hubbard model with Hund coupling (see Eq.~\eqref{H_multiorbital}) and crystal field splitting $H_\text{cf}=\Delta_\text{cf}(n_1-n_2)\label{eq_cf}$ \cite{Werner2007cf}. For $\Delta_\text{cf}\gtrsim \sqrt{2}J_H$, the system favors the low-spin~(LS) doublon configuration, while for $\Delta\lesssim \sqrt{2}J_H$, the high-spin~(HS) doublon configuration is dominant (see Fig.~\ref{fig:NESS_eff_Ishihara}(a) for a representation of the local states).

{\em Photo-induced high-spin states ---} 
The photo-doped two-orbital Hubbard model with orbital-dependent hopping allows to realize a hidden phase stabilized by a nonthermal population of local multiplets.
Using ED, \onlinecite{Ishihara2011PRL} studied the nonequilibrium phase diagram of the half-filled Mott system by fixing the photo-doping density (number of singlons and triplons) in a generalized strong coupling model, as discussed in Sec.~\ref{sec:quasiseq}. They found that the LS-HS coexistence region expands in the photo-doped case (compare green lines and dashed black lines in Fig.~\ref{fig:NESS_eff_Ishihara}(b)). The additional stabilization of the HS doublon state can be understood within a double-exchange mechanism, which can give rise to a bound state between a singlon and a HS state, see the bottom part of Fig.~\ref{fig:NESS_eff_Ishihara}(c).  The singlon-HS bound state induces a new absorption peak at $\omega\simeq 2v_1$, which corresponds to the excitation of the electron in orbital 1 from a bonding to an anti-bonding states. 
These results explain the experimental observations from fs pump-probe spectroscopy on a series of cobaltites $R$BaCo$_2$O$_{6-\delta}$ ($R=$ Sm, Gd and Tb)~\cite{Okimoto2011PRB}, which show distinct in-gap spectra in the photo-doped state and the high-temperature equilibrium state, and an increase of the in-gap spectral weight with increasing ionic radius of $R$.

\begin{figure}[t]
\begin{center}
\includegraphics[angle=0, width=1.0\columnwidth]{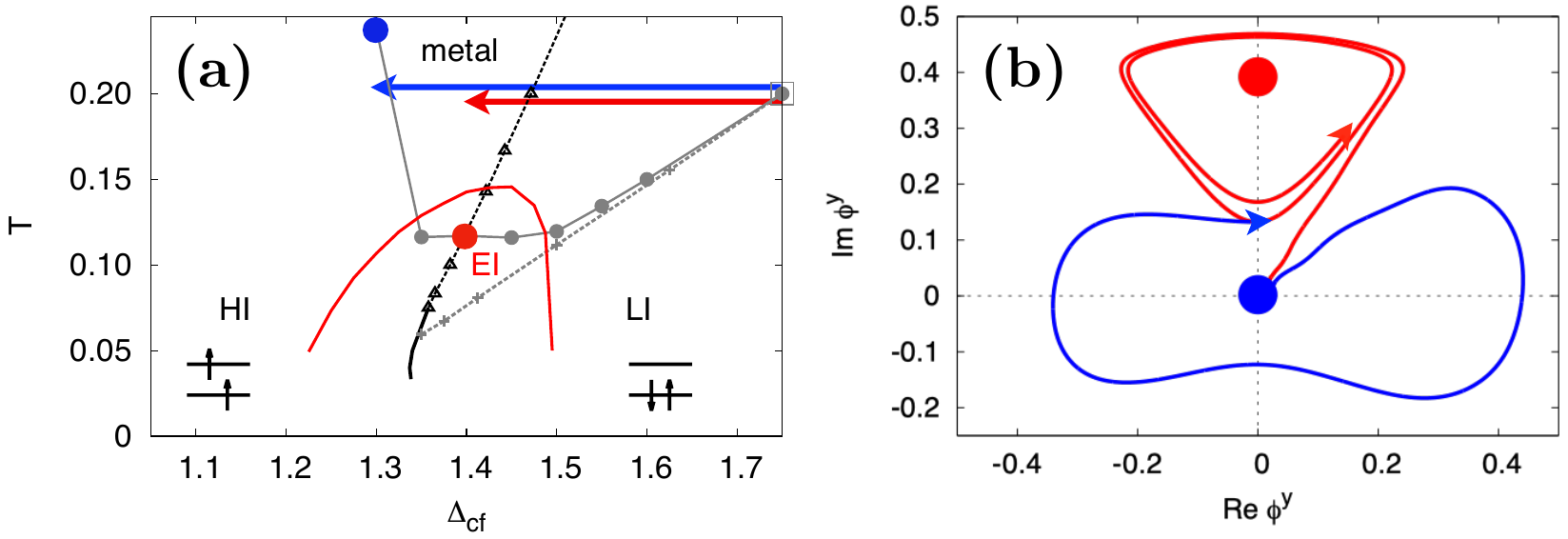}\\
\caption{(a)~Phase diagram of the two-orbital Hubbard model with  black solid (dashed) lines indicating the high-spin insulator (HI) to low-spin insulator (LI) transition (crossover). 
The boundary of the excitonic insulator (EI) phase is shown by the red line. The gray dashed line marks the isentropic line and the gray solid line with dots the thermalized results after quenches starting from the open square. (b) Order parameter trajectories after crystal field quenches from $\Delta_\text{cf}=1.75$, $T=0.2$ to $\Delta_\text{cf}=1.4$ (red) and $1.3$ (blue). A small seed field is applied in the imaginary $\phi^Y$ direction. Parameters: $U=6$, $J_H=1$, $W=4$.  (From \onlinecite{Werner2020}.)
}
\label{fig_2orbitalEI}
\end{center}
\end{figure}   

{\em Transient excitonic order ---}
The initial dynamics after an excitation can sometimes be qualitatively described by the mean-field dynamics. Mean-field theory has many conserved quantities, 
which results in prethermal (``trapped'') orders, and phenomena related to this trapping can survive in more accurate treatments.  
Near a spin state transition, one finds an interesting example of a hidden phase associated with such a trapping.
 In the two-orbital Hubbard model (\ref{H_multiorbital}) with orbital-independent hopping, the system near the HS/LS transition is susceptible to excitonic (spin-orbital) order \cite{Kunes2015}, see the DMFT+NCA phase diagram in Fig.~\ref{fig_2orbitalEI}(a). The spin-triplet excitonic order parameters can be defined as $\phi^\lambda = \sum_{\sigma\sigma'} \sigma^\lambda_{\sigma\sigma'}\langle c^\dagger_{1\sigma}c_{2\sigma'}\rangle$, where $\sigma^\lambda$ denotes the Pauli matrix for $\lambda = X, Y, Z$. (The $Y$ component is selected for the realization of the order below.) \onlinecite{Werner2020} studied the dynamics after a quench of the crystal field  $\Delta_\text{cf}$, starting from a normal metallic state in the 
low-spin regime above the highest excitonic transition temperature. In an adiabatic ramp of $\Delta_\text{cf}$, the system would follow the isentropic line (gray dashed line in Fig.~\ref{fig_2orbitalEI}(a)). Along this line, the formation of disordered HS moments leads to entropy reshuffling from the electronic to the spin sector, and the electronic  temperature can drop below the excitonic transition temperature. The expected thermalized temperature after a sudden quench of  $\Delta_\text{cf}$ is higher than after an adiabatic ramp (see the filled dots), but it can still drop below the transition temperature. Interestingly, the dynamics after such quenches shows a dynamical phase transition, as seen from the trace of the order parameter $\phi^Y$ in Fig.~\ref{fig_2orbitalEI}(b). For quenches to the ordered regime ($\Delta_\text{cf}\ge 1.35$), the order parameter oscillates counterclockwise around the expected equilibrium value indicated by the colored dot. For quenches to $\Delta_\text{cf}\le 1.3$, where the thermalized state is disordered, one observes the transient emergence of order with trajectories that rotate clockwise around zero. This is a distinctly nonthermal type of symmetry-breaking. The existence of the two qualitatively different dynamical regimes can be reproduced by a strong-coupling mean-field theory, whose dynamics is restricted by the conservation of the local entropy. The model thus exhibits a dynamical phase transition \cite{Marino2022} between a thermal-like and nonthermal symmetry breaking.

\subsubsection{Superconductivity and $\eta$-pairing states}
\label{sec:super_eta}

While the enhancement of the superconducting gap by microwave radiation (Eliashberg effect) has long ago been predicted \cite{Eliashberg1970} and experimentally confirmed even in cuprates \cite{Vedeneev2008}, recent research on nonequilibrium superconducting orders has been strongly motivated by experimental works reporting superconducting-like phases induced far above the equilibrium $T_c$ by strong mid-IR or THz pulses~\cite{Fausti11,Kaiser2014,Mitrano2015,Buzzi2020}. 
On the theory side, significant efforts have been made in the exploration of nonequilibrium superconducting states.
In the context of nonlinear phononics, various scenarios for the enhancement of 
pairing interactions have been discussed, e.g. parametric phonon excitations~\cite{knap2016,Komnik2016,babadi2017, murakami2017}, nonlinear electron-phonon couplings~\cite{kennes2017, sentef2017} and excitations of phonons coupled to interband transitions~\cite{Eckhardt2023,Chattopadhyay2023}. 
As in the Eliashberg effect, the possibility of an effective cooling of the system has been discussed~\cite{nava2018,Grunwald2023}.
In Mott insulators, photo-doping leads to a metallization of the system and naturally enhances the correlations for Cooper pairs of various types \cite{Wang2018PRL,Werner2018PRB,Bittner2019JPSJ}.

%%%%%%%%%%%%%%%%%
\begin{figure}[t]
\begin{center}
\includegraphics[angle=0, width=1.0\columnwidth]{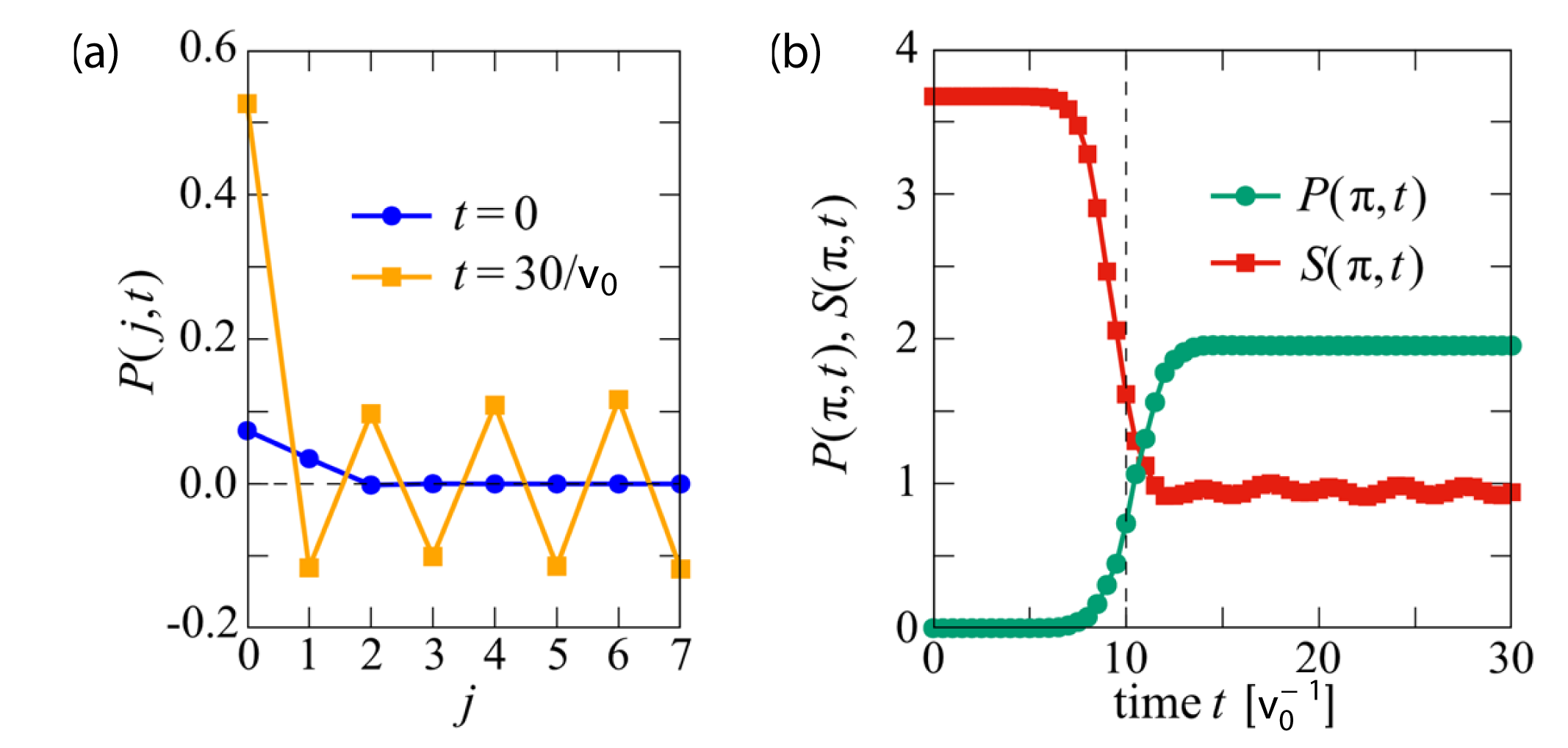}
\caption{(a) Real space profile of the pair correlations $P(j,t)$ before~($t=0$) and after~($t=30/v_0$) an electric field excitation. (b) {T}ime evolution  of the pair and spin correlations at $q=\pi$. These ED results are for a Hubbard chain with $N=14$ sites and $U=8v_0.$  (From \onlinecite{Kaneko2019}.) 
}	
\label{Kaneko_eta_pair}
\end{center}
\end{figure} 
%%%%%%%%%%%%%%%%%

In the following, we review mechanisms which are unique to photo-doped states and arise from activated correlations between photo-excited carriers.
To begin, let us consider the single-band Hubbard model on a bipartite lattice.
By analyzing the time evolution  of the 1D Hubbard model using ED, \onlinecite{Kaneko2019} showed that electric field excitations strongly enhance the $\eta$-pairing correlations. Specifically, 
they measured the pair-correlation function $P(j,t)=\frac{1}{N} \sum_i\langle \Psi(t)|(\hat{\Delta}_{i+j}^\dagger \hat{\Delta}_{i} + h.c.)|\Psi(t)\rangle$ and the spin correlation function $S(j,t)=\frac{1}{N}\sum_i \langle \Psi(t)|\hat{m}^z_{i+j} \hat{m}^z_i|\Psi(t)\rangle$, where $\hat{\Delta}_i=\hat{c}_{i,\uparrow}\hat{c}_{i,\downarrow}$ and $\hat{m}_i^z = \hat{n}_{i,\uparrow}-\hat{n}_{i,\downarrow}$, 
and the corresponding Fourier transforms $P(q,t)$ and $S(q,t)$. After an electric field excitation, the staggered component of the pair-correlation function
$P(q=\pi,t)$ is strongly enhanced, while the spin correlations $S(q=\pi,t)$ are suppressed, see Fig.~\ref{Kaneko_eta_pair}. The pump frequency dependence of the enhancement of the pair correlations matches the (linear) optical spectrum. The staggered nature of the photo-induced $P(j,t)$ indicates that the momentum of the corresponding pairs is peaked at $\pi$, in contrast to conventional $s$-wave condensation of zero-momentum pairs. 
The photo-excited state reported by \onlinecite{Kaneko2019,Kaneko2020PRR} can 
therefore be associated with the $\eta$-pairing states originally introduced by \onlinecite{Yang1989PRL} as exact eigenstates 
of the Hubbard model on a bipartite lattice. This model has a 
SU$(2)$ symmetry with respect to the $\eta$ spins ($\hat{\eta}^+ = \sum_i (-)^i c^\dagger_{i\downarrow} c^\dagger_{i\uparrow}$,  $\hat{\eta}^- = \sum_i(-)^i  c_{i\uparrow} c_{i\downarrow}$ and $\hat{\eta}^z = \frac{1}{2}\sum_i (n_i-1)$). 
Energy eigenstates can be expressed as simultaneous 
eigenstates $|\eta,\eta_z\rangle$ of $\hat{\eta}^2$ and $\hat{\eta}^z$,  
with a direct relation to the pair correlation function: $P(q=\pi)=2\langle \eta,\eta_z|\hat{\eta}^+\hat{\eta}^-|  \eta,\eta_z\rangle = 2(\eta(\eta+1)-\eta^z(\eta^z-1))/N$. 
The equilibrium states are $|\eta,\eta_z=-\eta\rangle$ and exhibit no $P(q=\pi)$ correlations~\cite{Essler1992,Essler2005}.  On the other hand, the electric field breaks the SU(2) symmetry and 
states $|\eta,\eta_z\neq-\eta\rangle$ are created during the excitation~\cite{Kaneko2019}.
(This follows from the Wigner-Eckart theorem and the fact that the current operator is a rank-one operator in terms of the $\hat{\eta}$ operators.) 
After the pulse, the evolution is again constrained by the SU(2) symmetry, so that the enhancement of $P(q=\pi)$ due to  the photo-generated admixture of $|\eta,\eta_z\neq-\eta\rangle$ states is long lived.
While the results reported in Fig.~\ref{Kaneko_eta_pair} are for a finite chain, an enhancement of $\eta$ pairing correlations was also reported in the thermodynamic limit and at nonzero temperatures using iTEBD~\cite{Ejima2020PRR}.
A related enhancement of pairing was found in a steady-state nonequilibrium system realized by the selective excitation of spin degrees of freedom~\cite{Tindall2019PRL}.
Due to the $\eta$-SU(2) symmetry, the pairing correlations were enhanced despite strong heating. 
A similar phenomenon was also reported in the Kondo lattice model~\cite{shirakawa2020}.

%%%%%%%%%%%%%%%%%
\begin{figure}[t]
\begin{center}
\includegraphics[angle=0, width=0.8\columnwidth]{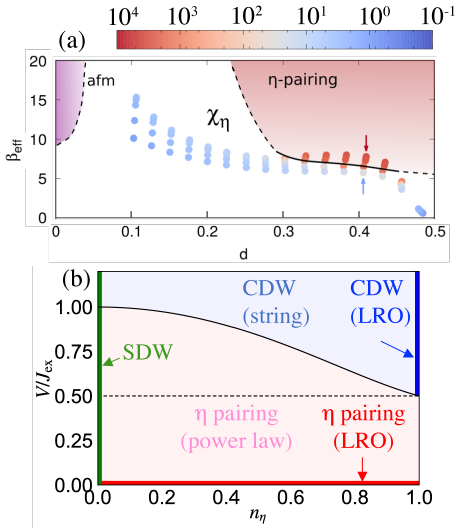}
\caption{(a)  Nonequilibrium phase diagram of the photo-doped steady state of a Mott insulator with doublon density $d$ and effective inverse temperature $\beta_{\text{eff}}$, obtained using the steady-state DMFT formalism (infinitely coordinated Bethe lattice, $W=4$, $U=8$).
The data points show the $\eta$-pairing susceptibility $\chi_\eta$, while the phase boundary is a guide to the eye (defined by $\chi_\eta\simeq 10^3$). 
(b) Ground state ($T_{\rm eff}=0$) phase diagram of the photo-doped 1D Mott insulator described by Eq.~\eqref{eq:Heff} 
in the limit $J_{\rm ex}\rightarrow 0$~(keeping $V/J_{\rm ex}$ constant) 
at half-filling, for D-H density $n_{\eta}$ and nearest-neighbor interaction $V$. (From \onlinecite{Li2020} and  \onlinecite{Murakami2023PRL}.) 
}
\label{eta_phase}
\end{center}
\end{figure} 
%%%%%%%%%%%%%%%%%

The emergence of the $\eta$-pairing phase was also predicted for quasi-steady states of photo-doped Mott insulators (large-gap Hubbard model).
Motivated by cold atom experiments, 
\onlinecite{Rosch2008} studied the effective model obtained from the Schrieffer-Wolff (SW) transformation for a system consisting of only doublons and holons, i.e., the SU$(2)$ ferromagnetic Heisenberg model for $\eta$ spins, see $\hH_{\rm dh,ex}$ in Eq.~\eqref{eq:Heff}.
The ground state of this model 
corresponds to the $\eta$-pairing state characterized by staggered off-diagonal long range order. This result raised the question 
whether long-lived $\eta$-pairing can also be realized away from this fully photo-doped limit. For the photo-doped Hubbard model, this has been numerically studied using nonequilibrium DMFT:
In~\onlinecite{Li2020}, a quasi-steady state with some given photo-doping density was prepared and stabilized by attaching electron baths (NESS approach discussed in Sec.~\ref{sec:quasiseq}). 
The $\eta$-pairing phase was  found to be stable over a wide range of photo-doping densities and up to inverse effective temperatures $\beta_{\text{eff}}$ comparable to the N\'eel temperature in the undoped Mott phase 
(Fig.~\ref{eta_phase}(a)). 
The same phase was also realized with the entropy-cooling protocol discussed in Sec.~\ref{sec_entropy} \cite{Werner2019b}. 
The imaginary part of the optical conductivity in the $\eta$-pairing phase has the $1/\omega$ behavior which is characteristic of superconducting states, and the behavior of the superfluid density is well explained within a strong-coupling model derived from the SW transformation. 
The real part of the conductivity generically features a Drude peak (instead of a gap), since the condensed doublons and holons coexist with normal singlons \cite{Li2020}.

For 1D systems with on-site interaction $U$ and nearest-neighbor interaction $V$, further physical insights into the photo-doped states have been obtained 
 in an analysis of the effective model~\eqref{eq:Heff} based on the quasi-equilibrium approach~\cite{Murakami2022,Murakami2023PRL}. 
For these systems, the wave function of the metastable photo-doped state at large $U$ can be rigorously factorized as 
\eqq{
|\Psi\rangle = |\Psi_{\rm charge}\rangle |\Psi_{\rm spin}\rangle |\Psi_{\rm \eta-spin}\rangle. \label{eq:MTKLP_state}
}
Here, $|\Psi_{\rm charge}\rangle$ represents the configuration of the singlons and is ruled by the free Hamiltonian of spinless fermions.
$|\Psi_{\rm spin}\rangle$ represents the configuration of spins and is ruled by the isotropic Heisenberg model in the
``squeezed spin space'' where the doubly occupied or empty sites are eliminated. $|\Psi_{\rm \eta-spin}\rangle$ represents the configuration of the doublons and holons and is ruled by the XXZ model of the $\eta$-spins in the squeezed space where the singly occupied sites are eliminated. 
(For example, for a state $|\Psi\rangle = |\!\!\uparrow,d,h,\downarrow,d\rangle$, we have $ |\Psi_{\rm charge}\rangle=|1,0,0,1,0\rangle$,  $|\Psi_{\rm spin}\rangle=|\!\!\uparrow,\downarrow\rangle$, and $ |\Psi_{\rm \eta-spin}\rangle = |d,h,d\rangle$.)
The wave function \eqref{eq:MTKLP_state} is a natural extension of the Ogata-Shiba state for the doped Hubbard model in equilibrium~\cite{Ogata1990PRB}, and indicates the presence of spin, charge and $\eta$-spin separation in the 1D metastable state. In particular, the XXZ model of the $\eta$-spins in the squeezed space is described as
$\hH^{(SQ)}_{\eta\text{-spin}}= -J^{\eta}_{X} \sum_j (\hat{\eta}^x_{j+1}\hat{\eta}^x_{j} + \hat{\eta}^y_{j+1}\hat{\eta}^y_{j}) +  J^{\eta}_Z\sum_j \hat{\eta}^z_{j+1}\hat{\eta}^z_{j}$,
where $J^\eta_X$ and $J^\eta_Z$ are functions of the density of $\eta$-spins, $V$ and $J_{\rm ex}$. The XXZ model explains the origin and nature of the photo-doped metastable states: The gapless state of the XXZ model for $J^\eta_X>|J^\eta_Z|$ corresponds to the $\eta$-pairing phase, while the gapful state of the XXZ model for $J^\eta_X<J^\eta_Z$ corresponds to the CDW phase, see Fig.~\ref{eta_phase}(b). The photo-induced CDW phase exhibits a hidden long-range order in the squeezed space (string order), similar to  the Haldane phase of the spin chain.

The development of pairing correlations associated with $\eta$ spins has also been studied for geometrically frustrated systems and multiorbital systems. In an analysis of a Hubbard model on a chain with tunable frustration, the time evolution of the system has been simulated using ED~\cite{Tindall2020}. While the presence of frustration removes the $\eta$-SU(2) symmetry, a transient enhancement of $\eta$ pairing correlations has nevertheless been observed for frustrations below a critical value. \onlinecite{Li2023} studied photo-doped Mott insulators on triangular, Kagome and similar three-colorable lattices using nonequilibrium DMFT and ED simulations. They found metastable superconducting phases with a 120$^\circ$ phase twist between the different sublattices. These are chiral superconducting states which are stabilized by the frustrated $\eta$-spin correlations. Because of the phase twist, there are persistent loop currents in the symmetry broken phase, which lead to a peculiar transverse supercurrent response to external fields applied in certain directions. 
In multi-orbital systems, the existence of $\eta$-pairing type phases of the spin/orbital triplet type (depending on the sign of $J_\text{H}$) has been demonstrated in the strongly photo-doped regime~\cite{Ray2023}.

%%%%%%%%%%%%%%%%%%%%%%%%%%%%%%%%%%%%%%%%%%%%%%%%%%%%%%%%%%%%%%%%%%%%%%%%
\section{Outlook}
\label{sec:summary}
%%%%%%%%%%%%%%%%%%%%%%%%%%%%%%%%%%%%%%%%%%%%%%%%%%%%%%%%%%%%%%%%%%%%%%%%

The intricate nature of strongly correlated systems, in particular states in proximity to a Mott phase, results in remarkable physical phenomena whose underlying mechanisms are often not yet fully understood. 
Driving such systems out of equilibrium can provide new perspectives on correlation phenomena and reveal effects or properties which are not accessible in equilibrium. This review discussed the responses of Mott insulators to various driving protocols, including AC/DC fields and photodoping excitations, and the 
resulting transient or long-lived nonthermal phases.
The significant recent progress in the development of experimental and theoretical tools 
has created a situation where the cross-fertilization between theory and experiment is driving further progress in this field. 
As experimental techniques evolve and more subtle effects are revealed, the comparison 
to reliable theoretical calculations becomes essential for understanding the microscopic mechanisms. At the same time, model studies and theoretical insights into nonlinear responses or nonthermal phases can suggest further experiments with refined control over the relevant miscroscopic degrees of freedom.

For a closer handshake between theory and experiment, it is important to further improve the theoretical approaches to enable a realistic modeling of the complex dynamics of strongly correlated materials.
This will likely necessitate the development of hierarchical schemes, which allow to consistently link the descriptions on different scales.
Thus, a future perspective is the development of a systematic approach which starts from a highly accurate local short-time description based on {\em accurate and efficient many-body solvers}, and expands this description to encompass {\em nonlocal correlations}, and/or {\em longer temporal scales} as well as {\em spatial inhomogeneities}. 
The ultimate goal is to develop an {\em ab initio framework} for nonequilibrium strongly-correlated systems.

{\em Many-body solvers---} 
Reliable and realistic nonequilibrium simulations require both accurate and efficient many-body solvers. Within the context of DMFT, various highly accurate impurity solvers have recently emerged, including inchworm, quasi Monte Carlo, and tensor-train techniques. Applying these solvers to real-world problems beyond the reach of perturbative solvers, and extending them to multiband systems 
remains an important challenge. Beyond DMFT, numerical methods based on  diagrammatic Monte Carlo, tensor networks and machine learning are 
being actively developed. Although their application to real-time simulations is currently limited to relatively short times, these methods hold promise for nonequilibrium steady-state investigations.

In the nonequilibrium Green's function context, several time propagation schemes with significantly reduced computational cost have recently been proposed, based on memory truncation, memory compression, or approximate quantum kinetic equations.
Furthermore, in realistic setups, one often must deal with correlation functions with multi-dimensional indices, including time, space, spins and orbitals. 
Applying modern tensor compression techniques in this context may enable new and more efficient implementations of many-body calculations.

{\em Nonlocal correlations---}
In equilibrium, nonlocal correlations can cause intriguing many-body phenomena, such as non-Fermi liquid behavior and $d$-wave superconductivity, and they should also play an essential role in the nonequilibrium dynamics. Understanding this physics becomes even more important in light of recent experimental developments such as time-resolved X-ray spectroscopies (XAS and RIXS), which allow to measure the evolution of nonlocal correlations and fluctuations in real time. However, most current theories treat only short-range fluctuations, which are expected to primarily influence the short-time dynamics. To predict the evolution at longer times, it is essential to incorporate the effects of 
long-range spin, charge and orbital fluctuations and their feedback on the electronic degrees of freedom. In higher-dimensional systems, diagrammatic extensions of DMFT provide a possible route for capturing long-range fluctuations. Numerical approaches based on diagrammatic Monte Carlo and tensor-network techniques are also promising options.

{\em Longer time dynamics---} Another important future task is to connect the short-time dynamics, controlled by the coupling between strongly correlated degrees of freedom, to the more slowly evolving (collective) modes associated with the lattice or symmetry-broken states. Throughout this review, we have focused on high-energy bosonic modes, which provide an efficient dissipation channel for the 
electrons. However, low-energy bosonic modes, which influence the electronic states in a quasi-static fashion, control the long-time dynamics. A similar challenge arises in the description of quasi-steady states, where the inclusion of terms which break the approximate conservation laws, such as doublon-holon recombination processes, can lead to slow electronic dynamics. A 
possible strategy is to combine nonequilibrium steady-state simulations with a quantum Boltzmann description of the slow population dynamics. In the case of symmetry-broken states, a related question is how to combine explicit simulations of nonequilibrium strongly correlated systems with time-dependent 
Ginzburg-Landau theory. 
Here, the relevant issue is how the feedback from the correlated nonequilibrium state influences the free energy landscape. 
Since Ginzburg-Landau theory is often used to interpret experimental results, a systematically extended formalism would enable a closer comparison between theory and experiment and deeper insights into the slow dynamics in materials like vanadates, ruthenates, and 1$T$-TaS$_2$.

{\em Inhomogeneity---} 
Recent advances in experimental techniques, like near-field optical spectroscopy, scanning tunnelling spectroscopy, or time-resolved electron microscopy, allow to resolve both the temporal and spatial profiles of electronic and lattice patterns. Similarly, the progress in material preparation enables the stacking of individual layers into heterostructures. 
Theoretical modeling  beyond the usual homogeneous assumption by a direct real-space formalism becomes computationally expensive for many sites or layers, and one will eventually need a suitable hierarchical multi-scale approach. 
The latter should provide an atomic resolution within some small region and a mean-field-type description at larger distances. Ideally, such formalisms should be able to describe the emergence of various patterns, like domains, filaments, or jammed structures, and reveal their impact on the material properties. The rapid development of nanoscale experimental techniques should enable systematic tests of theoretical ideas and the development of new concepts which extend beyond the phenomena described in this review.

{\em Ab initio description---} 
Material-specific simulations of strongly correlated systems have been achieved in equilibrium by combining downfolding schemes with appropriate many-body solvers. Prominent examples include DFT+DMFT and GW+EDMFT. In principle, the same procedure can be applied to study material-specific nonequilibrium dynamics, but we still lack powerful and versatile many-body solvers which can handle the complexity of realistic models.
In this review, we discussed the nonequilibrium extension of GW+EDMFT for the $dp$ model of cuprates, which represents a first step in the direction of material-specific nonequilibrium simulations.
Ultimately, these calculations need to be embedded into an ab initio framework using an appropriate multi-tier downfolding procedure. This allows to obtain a hierarchical description in orbital space, where higher-lying orbitals are treated within a simpler and numerically more efficient approximation.  An alternative route is the development of new functionals within TDDFT, which are capable of capturing strong correlation effects. First steps in this direction are the ACBN0 functional, which provides a DFT+$U$ like description, and attempts to derive functionals for Mott systems from equilibrium DMFT solutions.

Finally, the holy grail in the field is the nonequilibrium control of material properties, which means controlled access to or dynamical stabilization of nonthermal phases. We mentioned several experimental examples, which are not yet fully understood, and discussed theoretical ideas, which currently lack experimental realizations. At present, there still exists a gap between the complexity of real material systems, and the relative simplicity of the numerically accessible models. The future developments mentioned in this outlook will narrow this gap and enable more specific predictions and interpretations of nonthermal phases.

\begin{acknowledgments}
All authors contributed to this review equally.
We thank H. Aoki, R. Arita, N. Bittner, L. Boehnke, J. Bon\v{c}a, U. Bovensiepen, J. Chen, D. Choi, N. Dasari, T. Esslinger, N. Gedik, A. Georges, F. Grandi, S. Hoshino, K. Held, K. Inayoshi, S. Ishihara, S. Iwai, M. Ivanov, S. Johnson, T. Kaneko, J. Kaye, A. Kim, S. Kitamura, A. Koga, A. L\"auchli, Z. Lenar\v{c}i\v{c}, J. Li, M. Mierzejewski, D. Mihailovi\`{c}, M. Mitrano, A. Millis, M. M\"uller, C. Monney, C. Nicholson, T. Oka, H. Okamoto, O. Parcollet, F. Petocchi, A. Picano, P. Prelov\v{s}ek, S. Ray, G. Refael, K. Sandholzer, M. Sch\"uler, M. Sentef, K. Shinjo, H. Shinaoka, C. Stahl, U. Staub, H. Strand,  K. Sugimoto, Z. Sun, S. Takayoshi, K. Tanaka, T. Tohyama, N. Tsuji, K. Uchida, L. Vidmar, W. Widdra, and K. Yonemitsu for stimulating discussions and collaborations on topics related to this review.
Y. M. was supported by Grant-in-Aid for Scientific Research from JSPS, KAKENHI Grant Nos. JP20K14412, JP21H05017, and JST CREST Grant No. JPMJCR1901. D. G. acknowledges the support by the program No. P1-0044 and No. J1-2455 of the Slovenian Research Agency (ARRS). 
P. W. was supported by ERC Consolidator Grant No. 724103 and the Swiss National Science Foundation through Grant No. 200021-19696 and NCCR Marvel. 
D. G. acknowledges the hospitality of College de France, and P. W. the hospitality of the Aspen Center for Physics. 
\end{acknowledgments}

\clearpage

\section*{List of Symbols}

\begin{table}[b]
\begin{ruledtabular}
\begin{tabular}{ll} 
$\mathbf{A}$&Vector potential\\
A($\omega$,t)&Time-dependent spectral function\\
A($\omega$)&Spectral function\\ 
$c_{i\sigma}$&Annihilation operator at site $i$ for spin $\sigma$\\
$\mathbf{d}_{ab}$&Dipolar matrix element ($a,b$ Wannier orbitals) \\
$\mathcal{D}$& Double occupation \\
$D_{\mathbf{k}}(t,t')$&Two-time bosonic propagator for momentum $\mathbf{k}$ \\
$\mathbf{E}$&Electric field~(component along the field direction)\\
$E_0$&Electric field amplitude\\
$g$&Electron-boson (phonon) interaction\\
$G_i(t,t')$&Two-time Greens function for flavor $i$\\
$H$&Hamiltonian\\
$\mathbf{J}$& current $-\delta H[\mathbf{A}]/\delta \mathbf{A}$\\
$\mathbf{j}$ & microscopic current density $\mathbf{j}=\mathbf{J} + \partial_t \mathbf{P}$\\
$J_H$&Hund coupling\\
$J_{\text{ex}}$&Superexchange interaction\\
$N(\omega,t)$&Time-dependent occupation function\\
$\mathbf{P}$& Polarization density \\
$P_{\mathbf{k}}(t,t')$&Two-time Polarization for momentum $\mathbf{k}$\\
$q$&Electric charge\\
$\mathbf{S}$&Vector of spin operators\\
$S$&Entropy\\
$\mathcal{S}$&Action\\
$\mathcal{T}_\mathcal{C}$ & Time-ordering operator on the contour $\mathcal{C}$\\
$U$&Hubbard interaction\\
$\mathcal{U}$&Bare impurity interaction\\
$v_{ij}$&Hopping integral between site $i$ and $j$\\
$v_{0}$&Nearest neighbor hopping integral\\
$v_*$&Rescaled hopping integral on hypercubic lattice\\
$W$&Bandwidth\\
$W_{\mathbf{k}}(t,t')$&Two-time retarded interaction for momentum $\mathbf{k}$ \\ 
$X$ $(P)$&Lattice distortion~(momentum)\\
$\Gamma$&Relaxation rate\\
$\Delta(t,t')$&Two-time hybridization function\\
$\Delta_{\text{cf}}$&Crystal field splitting\\
$\epsilon_{\mathbf{k}}$&Noninteracting dispersion\\
$\boldsymbol{\eta}$&Vector of $\eta$ operators\\
$\lambda$&Effective electron-boson coupling\\
$\mu$&Chemical potential\\
$\sigma(t,t')$&Two-time optical conductivity\\
$\Sigma_{\mathbf{k}}(t,t')$&Two-time selfenergy for momentum $\mathbf{k}$\\
$\chi(t,t')$&Two-time charge susceptibility\\
$\omega_0$&Phonon frequency\\
\end{tabular}
\end{ruledtabular}
\end{table}
Different components of the propagators are marked by superscripts, namely $R$ for the retarded component, $<$ for the lesser component and $M$ for the Matsubara component.
\bibliographystyle{apsrmp}
\bibliography{rmpbib}

\end{document}